\documentclass[aps,prd,showpacs,nofootinbib,twocolumn]{revtex4}
\usepackage{latexsym}
\usepackage{amsmath,amsfonts}
\usepackage{amsbsy}
\usepackage{mathrsfs}
\usepackage{color}
\usepackage{float}
\usepackage{dsfont}
\usepackage{psfrag}
\usepackage{enumerate}
\usepackage{soul}
\usepackage{url}
\usepackage{hyperref}

\usepackage{amsmath,amssymb,calc,amsfonts}
\usepackage{latexsym}
\usepackage{hyperref}
\usepackage{color}
\usepackage{graphicx}
\usepackage{ulem}
\begin{document}
\author{Gast\' on Briozzo$^{1}$ and Emanuel Gallo$^1$ }

\newcommand{\red}[1]{\textcolor{red}{#1}}
\newcommand{\blue}[1]{\textcolor{blue}{#1}}

\affiliation{$^1$FaMAF, UNC; Instituto de Física Enrique Gaviola (IFEG), CONICET, \\
Ciudad Universitaria, (5000) C\'ordoba, Argentina. }

\title{Analytical expressions for pulse profile of neutron stars in plasma environments}

\begin{abstract}

We present an analytical study of light curves of slowly rotating radio pulsars with emphasis on the chromatic effects derived from the presence of a plasma environment; analyzing the effects of the compactness, the metric model, and the electronic plasma density profile.
After doing a numerical integration of the trajectories and luminosity curves of pulsars for different spherically symmetric metrics representing the exterior region of the pulsar, we  generalize the approximate Beloborodov formula in order to include plasma corrections, obtaining simple analytical expressions for the trajectories and the observed flux and significantly simplifying the calculation of the pulse profiles by a drastic reduction of their computational cost. We study the errors committed by our approximation, comparing the numerical and analytical procedures.
We also show how to use the new formalism to model the flux coming from different emission caps, not necessarily circular or antipodal and including the case of ring-shaped hot spots.
Finally, we extend the classification introduced by Beloborodov to the case of two distinguishable, non-antipodal, finite size emission caps, showing the respective classification maps and some of the characteristic pulse profiles. 
\end{abstract}


\maketitle

\section{Introduction}
\label{c2s1}

    Neutron stars represent one of the most relativistic stellar objects accessible to observation \cite{Shapiro_1983, Harding_2013}. Due to the extreme conditions to which they are subjected, most of their properties remain uncertain, making the determination of their equation of state an open problem in nuclear astrophysics \cite{Ozel_2016, annala2018gravitational, Lattimer_2021}. However, this difficulty can be overcome by observing the neutron stars themselves and the phenomena associated with them, such as their luminosity or their lensing power. 
    At the same time, neutron stars are one of the best natural candidates to probe physics in extreme environments and test the theory of General Relativity in the strong field regime~\cite{ Berti:2015itd}.
    Indeed, the discoveries of neutron stars of around two solar masses \cite{demorest2010two}, as well as the observation of gravitational waves resulting from mergers of binary systems \cite{abbott2016observation}, impose severe constraints on the equation of state of these objects, in the high-density regime. More recently, the mass and radius of neutron stars have been estimated by fitting the rotating hot spot patterns using data of Neutron Star Interior Composition Explorer (NICER) and X-ray Multi-Mirror 
    (XMM-Newton) X-ray observations~\cite{Riley:2019yda,Miller:2021qha,Bogdanov:2021yip}. 
    
    Most of the neutron stars emit highly collimated jets of electromagnetic radiation \cite{petri2016theory} coming from highly localized ``hot spots'' that emit significant quantities of observable radiation.
    Neutron stars exhibiting these ``emission caps'' are called pulsars, and because of its rotation, they produce observable luminosity curves, similar to a lighthouse.
    
    These pulsar pulse profiles encode information about the physical properties of the emission caps, such as their size, geometry, temperature distribution and spectra, as well as containing information about the geometry of the spacetime itself around the star

    \cite{Pechenick_1983, sotani2017pulse, Sotani_2017, Sotani_2020, Silva_2019,      Dabrowski_1995, Silva:2019leq, Xu:2020vbs, Hu:2021tyw}.
    
    The observation of these periodic electromagnetic pulses offers another way to investigate the properties of these extremely compact objects.
    Due to their strong gravitational field, the radiation in the vicinity of neutron stars experiments a strong deflection, generating a gravitational lensing effect, whose magnitude is intimately related to the mass and radius of the star, so that their observation allows in principle to infer these parameters, or at least to constrain their values, providing information about the stellar structure that facilitates the determination of the equation of state \cite{psaltis2014prospects, bogdanov2016prospects}.
    
    The problem of modeling pulse profiles has been extensively studied in the literature. Studies have been done considering the Doppler effect, aberration; time delay \cite{poutanen2006pulse} and fast rotation or stellar oblongation \cite{2007ApJ...654..458C,psaltis2014pulse,2010MNRAS.409..481N,Sotani:2018oad}. In particular, In \cite{Pechenick_1983} it was studied gravitational lensing properties in slowly rotating neutron stars described by the Schwarzschild metric, consisting of one or two antipodal, uniform, circular polar caps on the surface that emit photons to the observer at infinity. In general, the treatment of photon propagation in a spherically symmetric and stationary spacetime leads to elliptic-type integrals or even more complicated expressions that need to be solved by numerical evaluation. The implementation of numerical methods is often unavoidable when considering a continuous surface temperature distribution, anisotropic emission or an arbitrary shape of the emission regions. While these models may be able to describe the pulse profile of stars of any compactness, it requires a complicated and computationally expensive numerical treatment. In this context, Beloborodov found in \cite{Beloborodov_2002} a simple analytical expression for the deflection angle in a Schwarzschild metric background, that replaces the elliptic integral, approximating with high accuracy the pulse profiles of point emission caps, assuming that the compactness of the star is not too big. In turn, Turolla and Nobili \cite{Turolla_2013} generalized this analysis to extended, uniform, and circular caps, giving simple analytical expressions that closely approximates the total observed flux. 
    The Beloborodov formula not only has been used to study pulse profiles of NS under different situations~\cite{Hu:2022ehk,Zyuzin:2021teu,Petri:2019oqa,Giraud:2019iba,Hu:2019zqy,Gotthelf:2010um,Viironen:2004ze} but also to study polarimetric images of accretion disks around black holes \cite{Melia:2011pq, Loktev:2021nhk,EventHorizonTelescope:2021btj}.
    
    It is to be expected that neutron stars are immersed in a dense, plasma-rich magnetospheres \cite{GJ_1969, Petri:2016tqe}. In the visible spectrum, the modification in photon trajectories and pulse profiles due to the presence of plasma are negligible, so generally gravitational lensing theory deals with the propagation of light rays in vacuum, where the trajectories and deflection angles are independent of the photon frequency, and thus the effects are achromatic. However, the effects of the optical medium cannot be safely neglected in the radio frequency range, where the refraction index of the plasma causes strong frequency-dependent modifications of the usual gravitational lensing behavior. In this range, photon propagation becomes frequency dependent, giving rise to chromatic phenomena. 
    These effects have been studied in the context of black holes and its shadows \cite{Perlick:2021aok,Perlick:2017fio,Kimpson:2019mji,Huang:2018rfn,Zhang:2022osx,Badia:2021kpk}, in weak fields under the weak and strong lensing regime \cite{CG_2018, CGR_2019, CGV_2019,Crisnejo:2019ril, Bisnovatyi-Kogan:2010flt,Bisnovatyi-Kogan:2015dxa,Er:2019jkg,Er:2017lue, Er:2021shs,Er:2013efa}, in studies of microlensing \cite{Tsupko:2019axo}, and on its effects on the fast radio burst \cite{Er_2020,Er:2021pjc} to cite some examples.
    The influence of plasma in light propagation and associated light curves produced by radio pulsars have been numerically studied in the past by \cite{Rogers_2015,Rogers:2017ofq, Rogers:2016xcc,Battye:2021xvt} assuming a Schwarzschild model for the exterior of the metric and a cold non-magnetized plasma (see also \cite{Witte:2021arp} where a Minkowski metric is assumed, but the magnetic field influence is taken into account). From the observational point of view, in \cite{Main:2018kfc} the pulsar emission regions were resolved and amplified by plasma lensing effect in an eclipse binary. Similar results have been carried out by other pulsars \cite{Lin:2021epe,Wang:2021cqk}.
    
    In this work, using numerical evaluations and introducing an analytical approach that generalize the Beloborodov's formula, we study the gravitational lensing phenomena in slowly rotating radio pulsars taking into account the dispersive effect of the pulsar's magnetosphere on the propagation of light rays. We will make use of alternative metrics to Schwarzschild to describe the exterior region in order to analyze the difference in their influence on the luminosity curves. More precisely, 
    we will focus on the Schwarzschild (Sch) {and the so called Reissner-Nordström like metrics consisting in solutions from alternatives gravitational theories which are functionally similar to the RN metric but with a parameter $Q$ unrelated to the electrical charge; for example associated to tidal charges in brane theories \cite{BW_2005}, associated to parameters of the theory, as in Horndeski theories \cite{Babichev:2017guv}  or being proportional to the gravitational mass of the star in Modified Gravity \cite{Moffat:2014aja}.} 
    The particular choice of these metrics is only to demonstrate the differences that appear in the properties of compact objects in the strong field regime and that can potentially be observed. 
    We will consider circular, homogeneous and isotropic emission caps (either thermal emission hot spots or electromagnetic emission poles) of finite size, in highly compact slowly rotating radio pulsars, so we will not take into account Doppler effect, aberration, time delay, fast rotation or stellar oblongation.
    As in \cite{Rogers_2015,Rogers:2016xcc,Rogers:2017ofq,Battye:2021xvt} the neutron star will be embedded in a pressureless and non-magnetized plasma environment, whose electron density will be given by a power law inversely proportional to the radial coordinate $r$, resulting in a chromatic analysis of the problem\footnote{
    A strong magnetic field in a plasma environment can affect the propagation of light rays by behaving as a birefringent dispersive medium, affect their polarization through Faraday rotation, or interfere through phenomena such as pair creation. However, in this work we will not consider this type of effects, which should be taken into account in a more detailed model.
    }. 
    
    This paper is organized as follows. In Sec.(\ref{c2s2}) we introduce the family of metrics that we will consider in the rest of the work. In Sec.(\ref{c2s3}) we study numerically the photon trajectories considering the different metric models and plasma density and distribution, emphasizing the chromatic effects derived from the presence of the plasma environment, whose interaction with the light will depend on its wavelength. In Sec.(\ref{c2s4}) we discuss the luminosity curve resulting from a single circular, homogeneous and isotropic emission cap on the stellar surface and how this is affected by the spacetime model, the distribution and density of the plasma, the angles between the cap and the pulsar rotation axis and between the rotation axis and the observer or the compactness and charge of the star. This analysis will be extended to the case of two identical and antipodal caps. In Sec.(\ref{c2s5}), motivated by Beloborodov's approximate analytical model, we modify the well known analytical formula in order to introduce corrections that take into account plasma environments, including now analytical expressions to describe uniform and circular extended caps immersed in a plasma environment. Then we apply the new formalism to the different metric models used above, obtaining for each of them simple analytical expressions for the photon trajectories (\ref{c2s5ss1}) and the observed flux (\ref{c2s5ss2}) that considerably simplify the calculation of the pulse profiles, drastically reducing their computational cost. We will compare the numerical results with the analytical approximations, studying the relative errors committed by the approach and its dependence on different magnitudes of interest according to the characteristics of the problem. Once the validity ranges of our model have been established, in Sec.(\ref{c2s6}) we will show how to use the resulting approximations to easily model more realistic, non-antipodal, homogeneous or circular emission caps. In particular, we will focus on the case of ring-shaped caps. Under this new formalism it will be possible to examine in detail the effects of the shape and size of individual hot spots, without any great loss of accuracy or generality, thus facilitating the study and understanding of these objects. In Sec.(\ref{c2s7}) we expand the classification system introduced by Beloborodov for the case of two non-antipodal, distinguishable caps, which can be also applied even in the case that the influence of plasma on the propagation of light rays can be neglected. {Finally, in Appendix (\ref{AppendixA}) (see supplementary material) we include some plots obtained for different plasma distributions in order to show how sensible are the photon trajectories and pulse profiles to the behavior of the electron density profiles, expanding the results of Secs. (\ref{c2s3}) to (\ref{c2s5}).} 

\section{Metric}
\label{c2s2}

    Let us use normalized fundamental units, $c=1$, $G=1$ and $\hbar=1$, being $c$ the speed of light in vacuum, $G$ the universal gravitational constant and $\hbar$ Planck's constant, and the metric signature $(-,+,+,+)$. 
    For the exterior region of the star, we employ simple metric models, consisting of asymptotically flat and static spacetimes with spherical symmetry.  These assumptions allow us to express the metric as follows\footnote{ Even when it is always possible to choose a suitable coordinate system for which the spacetime is characterized by only two metric functions $A$ and $B$, in what follows we will prefer the form \eqref{eq:elementodelinea} because it allows to write general expressions that remain valid for a vast family of coordinate systems.}
    \begin{equation}
        ds^2=-A(r)dt^2+B(r)dr^2+C(r)d\Omega^2,
        \label{eq:elementodelinea}
    \end{equation}
    where the functions $A(r)$, $B(r)$ and $C(r)$ depend only on the radial coordinate $r$, satisfying also the asymptotically flat conditions, i.e.,
    \begin{equation}
    \begin{split}
        \lim_{r \to \infty} A(r)=1, \\
        \lim_{r \to \infty} B(r)=1, \\ 
        \lim_{r \to \infty} \frac{C(r)}{r^2}=1. 
    \end{split}
    \label{eq:lims}
    \end{equation}

    The simplest metric in GR describing a static, spherically symmetric and asymptotically flat spacetime with a gravitating $M$ mass is the Schwarzschild (Sch) metric.
    This solution is a useful approximation for describing slowly rotating astronomical objects. 
    
    A generalization of the Schwarzschild metric, that describes a spherically symmetric metric of a massive body with non-zero net electrical charge $Q$ is the Reissner-Nordström (RN) metric, which is a solution of the Einstein-Maxwell equations \cite{Reissner_1916, Nordstrom1918}. 
    
    As mentioned in the introduction there exist solutions of alternative gravity's theories which are functionally similar to the RN but with a charge parameter $q^*$ unrelated to the electrical charge (See Table \ref{tab:NS_Metricas}). In particular, in Brane world (BW) theories, a negative $q^*$ is theoretically preferred \cite{Dadhich:2000am}. Studies of NS in braneworld theory can be found in \cite{Germani:2001du} (where it is shown that a uniform density star can be matched to a RN-like solution for the exterior of the star, see also the discussion in \cite{Kotrlova:2008xs}) and observational constraints on the value of $q^*$ in NS are presented. For a study of the plasma magnetosphere around NS in BW theories we refer to \cite{Morozova:2010gg}.

    {
    Given the similarities between the models for alternative theories when $q^*\neq 0$, when $q^*<0$ we will refer to them as RN-like metrics, 
    associating the parameter $Q^2=-q^*$, 
    while when $q^*>0$ we will refer to them as RN metrics, associating the parameter $Q^2=q^*$, so that both models turn out structurally identical and can be described by a single parameter $q^*$.}
    
    {
    Even though in the metrics used in this work $C(r)=r^2$, in the following, we prefer to keep the coefficient $C(r)$ in order to obtain general expressions that hold valid for alternative metrics to those used here.
    }
    
    \begin{table}
        \centering
        \begin{tabular}{|l|l|l|l|l|}
        \hline 
        Metric & $A(r)$                            & $B(r)$            & $C(r)$ & Restriction          \\ \hline
        Sch    & $1-\frac{2M}{r}$                  & $\frac{1}{A(r)}$  & $r^2$  &                      \\ \hline
        RN-like     & $1-\frac{2M}{r}+\frac{q^*}{r^2}$  & $\frac{1}{A(r)}$  & $r^2$  & $q^*\leq M^2$         \\ \hline
        \end{tabular}
        \caption{\label{tab:NS_Metricas} Family of metrics around compact objects with a gravitating mass $M$ and a charge $q^*$, in static, spherically symmetric and asymptotically flat spacetimes.}
    \end{table}

\section{Ray Tracing}
\label{c2s3}

\begin{figure}
        \centering
        \includegraphics[scale=0.24,trim=0 0 0 0]{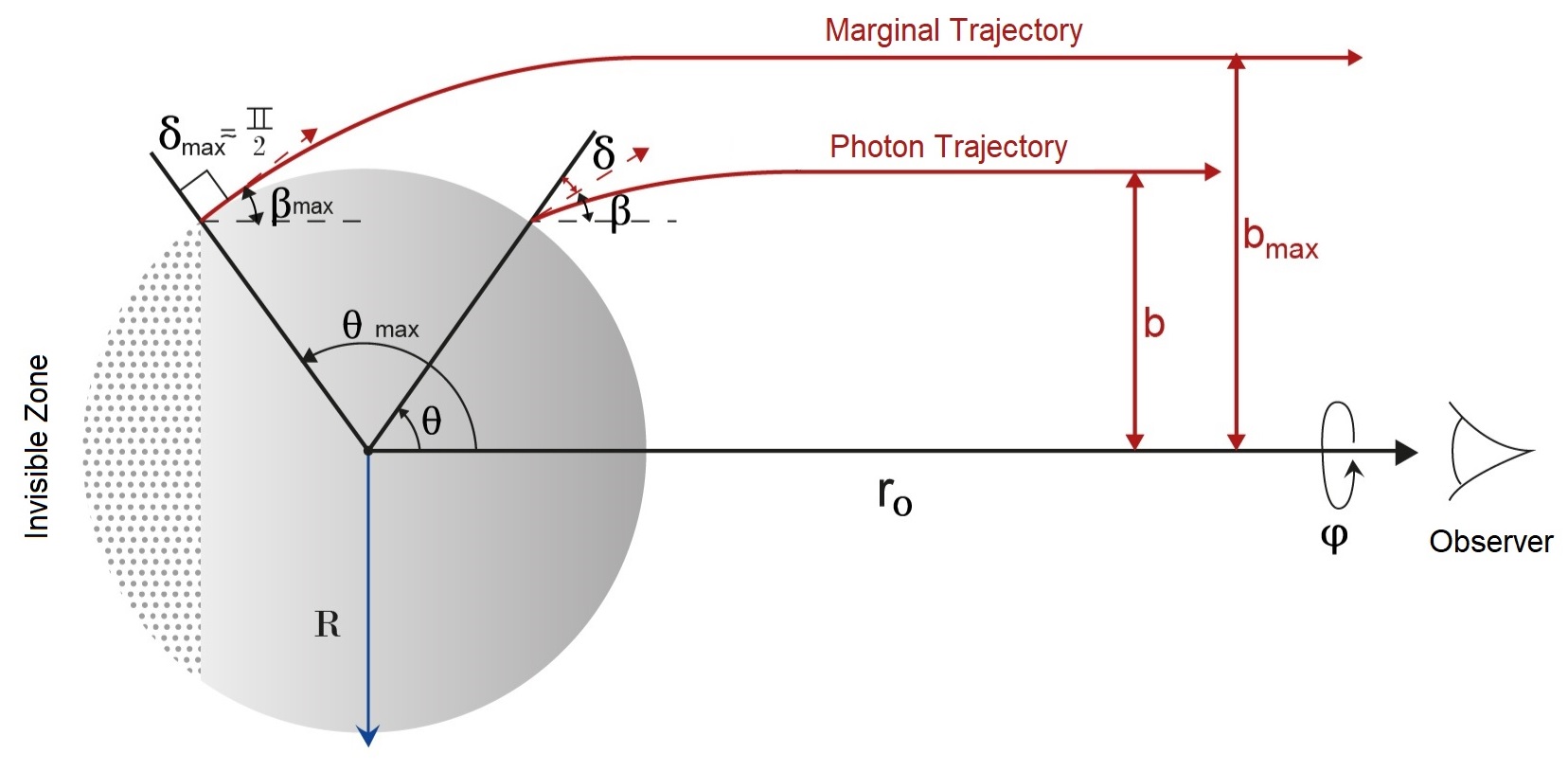}
        \caption{A photon’s trajectory, with impact parameter $b$, from the stellar surface at $R$ to the observer at $r_O$. The deflection angle $\beta$ is calculated as $\theta-\delta$.}
        \label{fig:TrazadodeRayos}
    \end{figure}

    The stellar surface will be placed at the radial coordinate $r=R$, the region $r<R$ will be totally opaque since it corresponds to the stellar interior. The light emission will originate from this surface, with an intensity $I$ dependent on the angle $\delta$ between the normal direction to the stellar surface and the photon emission direction (see Fig.\eqref{fig:TrazadodeRayos}). The observer will be considered static at $r=r_O$ with $r_O\rightarrow\infty$. 
    We will employ Greek indices to denote spacetime coordinates, while we will reserve Latin indices for spatial coordinates only.
    
    In this section we will follow the work presented in \cite{Rogers_2015}, recovering his results for the Schwarzschild metric and extending its application to the different spacetimes mentioned in Table \eqref{tab:NS_Metricas}. 
    Consider a spacetime of the form given by Eq. \eqref{eq:elementodelinea}
    where the metric functions $A$, $B$ and $C$ depend parametrically on the mass $M$ and charge $q^*$ of the pulsar.
    In the geometrical optic limit, photon trajectories within a plasma medium are described by the Hamiltonian \cite{Synge_1960}, 
    \begin{equation}
        H(x^\alpha,p_\alpha)=\frac{1}{2}[g^{\alpha\beta}p_\alpha p_\beta-(n^2-1)(p_\alpha V^\alpha)^2],
        \label{eq:ham1}
    \end{equation}
    being
    \begin{equation}
        n^2=1+\frac{p_\alpha p^\alpha}{(p_\beta V^\beta)^2}
        \label{eq:refindSynge}
    \end{equation}
    the refractive index of a general dispersive medium not necessarily coming from a plasma, while $p^\alpha$ is the linear photon $4-$momentum, and $V^\alpha$ the plasma $4$-velocity.
    The light rays are obtained as solutions to the equations of motion
     \begin{equation}
        \label{eq:dxpdlgen}
        \begin{split}
        \frac{dx^\alpha}{d\lambda} &=\;\;\frac{\partial H}{\partial p_\alpha}, \\
        \frac{dp_\alpha}{d\lambda} &=-\frac{\partial H}{\partial x^\alpha},\\
        H(x^\alpha,p_\alpha)       &=0.
        \end{split}
    \end{equation}

    We will assume a spherically symmetric plasma distribution around the pulsar. When light passes through it, the plasma will act as a dispersive medium, with a refractive index $n=n(x^\alpha,\omega)$ which will depend on the frequency $\omega=-p_\alpha V^\alpha$ of the photon. 
    Assuming that the plasma is a static medium such that $V^t=\sqrt{-g^{tt}}$ and $V^i=0$, it follows that
    \begin{equation}
        p_t\sqrt{-g^{tt}}=-p^t\sqrt{-g_{tt}}=-\omega(x^i),
        \label{eq:ptgtt}
    \end{equation}
    so that the Hamiltonian does not depend on $t$ or $\phi$, so $p_t$ and $p_\phi$ are conserved. We set the timelike component of the momentum as the asymptotic frequency of the photon as measured by an asymptotic observer at rest
    \begin{equation}
        p_t=-\omega_\infty.
        \label{eq:pt}
    \end{equation}
    From Eqs. \eqref{eq:ptgtt} and \eqref{eq:pt} we can find the relation for the effective redshift,
    \begin{equation}
        \omega(r)=\frac{\omega_\infty}{A(r)^{1/2}},
        \label{eq:redshift}
    \end{equation}
    where $\omega(r)$ is the frequency of light measured by a static observer at the radial coordinate $r$.
    
    In the rest of this work, we will assume a non-magnetized pressureless plasma with the index of refraction taking the form
    \begin{equation}
        n^2=1-\frac{\omega_e^2(r)}{\omega^2(r)},
        \label{eq:refind}
    \end{equation}
    where
    \begin{equation}
        \omega_e^2=\frac{ e^2}{\epsilon_0 m_e} N(r), 
        \label{eq:W}
    \end{equation}
    being $\omega_e(r)$ the plasma frequency \cite{IntroPlasmaMHD_2003}, $e$ and $m_e$ the electron charge and mass respectively, $\epsilon_0$ the vacuum permittivity and $N(r)$ the number electron density of the plasma\footnote{In Sec.\eqref{c2s4ss2} we will consider a modification of this expression, which takes into account the relativistic motion of electron-positron pairs around the magnetic fields lines, producing an effective plasma frequency several order of magnitude less than the frequency given by \eqref{eq:W}.}. Together with Eqs. \eqref{eq:ham1} and \eqref{eq:refind}, the above expression allows us to obtain the dispersion relation for photons,
    \begin{equation}
        H(x^\alpha,p_\alpha)=\frac{1}{2}[g^{\alpha\beta}p_\alpha p_\beta+\omega_e^2].
        \label{eq:ham2}
    \end{equation}

    Given the spherical symmetry of the problem, we can always choose a referential such that $p_\phi=0$, corresponding to planar meridian trajectories ($\phi$=constant)\footnote{The reason to orientate the reference frame in this way instead of the standard orientation where equatorial orbits are analyzed, is because of, as observed by \cite{Turolla_2013}, it allows explicit analytical integrations in the Beloborodov approximation as discussed in Sec.\eqref{c2s5}.}.
    For these trajectories, it follows that $p_\theta$ is also a constant of motion.
    
    To construct the ray tracing, i.e., the spatial trajectory of photons, we can get rid of the parameter $\lambda$ of Eq. \eqref{eq:dxpdlgen} as follows,
    \begin{equation}
        \frac{d\theta}{dr}=\frac{d\theta}{d\lambda} \left[\frac{dr}{d\lambda}\right]^{-1} =\frac{p_\theta}{p_r}\frac{B(r)}{C(r)}.
        \label{eq:dpdr1}
    \end{equation}
    
    At the same time, from Eqs. \eqref{eq:ham2} and \eqref{eq:dxpdlgen} it can be deduced that
    \begin{equation}
        p_r=\pm \sqrt{\frac{B}{A}p_t^2-\frac{B}{C}p_\theta^2-B\omega_e^2}
        \label{eq:pr}
    \end{equation}
    Here the positive solution corresponds to the case where both $r$ and $\theta$ are increasing, while the negative solution otherwise. We now introduce the impact parameter $b$ and the asymptotic group velocity $n_0$ to express the constant $p_\theta$, which can be written as
    \begin{equation}
        p_\theta= \omega_\infty n_0 b.
        \label{eq:pf}
    \end{equation}
    The group velocity usually tends asymptotically to $1$, except in cases where one has a constant, or non-zero, plasma distribution around the observer, in which case we have
    \begin{equation}
        n_0^2=1-\frac{\omega_e^2(r_O)}{\omega^2(r_O)},
        \label{eq:n0}
    \end{equation}
    with $r_O$ is the radial coordinate of the observer. 
    
    Thus, the trajectory equation becomes \cite{CGV_2019}
    \begin{equation}
        \frac{d\theta}{dr}= \frac{1}{C}\left[ \frac{1}{AB} \left(\frac{n^2}{n_0^2}\frac{1}{b^2}-\frac{A}{C} \right)   \right]^{-1/2}.
        \label{eq:dpdr2}
    \end{equation}
    This equation allows us to describe the photon trajectory in a meridian plane in the general case of strong deflection, which is appropriate for orbits near the surface of the star. The refractive index of the plasma in this expression introduces the frequency dependence of the orbits. In order to facilitate the numerical integration, we introduce the variable $u=1/r$, for which it turns out that
    \begin{equation}
        \frac{d\theta}{du}= \frac{1}{u^2C}\left[ \frac{1}{AB} \left(\frac{n^2}{n_0^2}\frac{1}{b^2}-\frac{A}{C} \right)   \right]^{-1/2}, 
        \label{eq:dpdu}
    \end{equation}
    where both the metric elements and the refractive index $n$, must be written in terms of $u$.
    We now would like to make use of Eq. \eqref{eq:dpdu} to study the outgoing ray tracing of a neutron star with a cold, nonmagnetic plasma atmosphere modeled by \eqref{eq:refind}, where the photons move with $\phi=const$.

    For this, we will place the observer on the $\theta=0$ axis of the coordinate system, at a large distance from the star, at $r=r_O\rightarrow\infty$, such that the line of sight connecting the center of the star with the observer corresponds to the impact parameter $b=0$. We are interested in describing rays that leave the surface of the star at $u_R=1/R$, $\theta_R=\theta(R)$ and reach the observer at $u_O=0$, $\theta_O=0$. The angle $\theta(R)$ of the photon at the instant it leaves the pulsar can be calculated as
    \begin{equation}
        \theta(R)=\int_0^{u_R} \frac{1}{u^2C}\left[ \frac{1}{AB} \left(\frac{n^2}{n_0^2}\frac{1}{b^2}-\frac{A}{C} \right)   \right]^{-1/2}du.
        \label{eq:D}
    \end{equation}

    Now, if the neutron star is sufficiently compact, it is possible that the value of $\theta$ given by Eq. \eqref{eq:D} may be larger than $\pi$, so this angle will not be exactly the photon's colatitude coordinate at the moment of leaving the stellar surface, but rather the angle from the observer's position vector to the photon's position vector measured counterclockwise, as shown in Figure \ref{fig:TrazadodeRayos}. From now on, when we refer to the $\theta$ angle of the photon we will mean the angle we have just defined, while we will use the symbol $\Theta$ to talk about its colatitude.
    
    At this point it only remains for us to find an expression for the impact parameter of the rays whose trajectory connects the surface of the star with the observer. For this, we define $\delta$ as the angle between the radial and angular components of the photon momentum. At the surface of the star $r=R$, $\delta$ is the angle at which the ray is emitted with respect to the normal. 
    
    Under these assumptions, it is possible to express the impact parameter $b$ in terms of $\delta$ as follows
    \begin{equation}
        b=\frac{n(R)}{n_0}\sqrt{\frac{C(R)}{A(R)}}\sin(\delta),
        \label{eq:b}
    \end{equation}
    where we see that the maximum impact parameter $b_{max}$ occurs for $\delta=\pi/2$. Thus, photons coming from the stellar surface will arrive at the observer with an impact parameter less than or equal to $b_{max}$. The presence of the refractive index in Eq.\eqref{eq:b} predicts that the apparent size of the pulsar for a distant observer vanishes when $n(R)\rightarrow0$.
    
    From these expressions we can obtain the maximum angle $\theta_{max}(b_{max})$ of the stellar surface that is visible to the observer. Thus, the observer has access to all points on the surface with $\Theta\leq\theta_{max}(b_{max})$, corresponding to $b\leq b_{max}$, while the points with $\Theta\geq\theta_{max}(b_{max})$ will form an invisible region. As in the case of light propagation without plasma, the resulting fraction $\delta f$ of the visible stellar surface is given by $\delta f =(1-\cos[\theta_{max}(b_{max})])/2$ (cf. \cite{Silva_2019}, Eq. 26).

    The expressions shown up to this point are valid for a generic plasma distribution, with a refractive index $n=n(x^\alpha, \omega)$ that depends on both photon frequency and position. For the case where $\omega_e\ll\omega$, $n\rightarrow1$, thus recovering the case of pure gravity. The propagation of a photon through this medium requires that $\omega_e(r)<\omega(r)$ at every point along the trajectory. The necessary condition for the rays to propagate from the star's surface to infinity then results
    \begin{equation}
        \omega_e(r)\sqrt{A(r)}<\omega_\infty
        \label{eq:winf}
    \end{equation}
    for all $r\geq R$.
    
    Numerical ray tracing requires a specific plasma distribution. Following \cite{Rogers_2015}, we choose some number density profiles that contain as particular case the Goldreich-Julian density \cite{GJ_1969}. Normally, the pulsar magnetosphere contains plasma with densities that are sustained by Goldreich-Julian currents, transferring charge carriers from the neutron star surface to the magnetosphere. Therefore, these currents provide a lower limit for the concentration of charge in a standard magnetosphere. Using the plasma frequency presented in Eq.~\eqref{eq:W}, let us consider a radial power law
    \begin{equation}
        N(r)=\frac{N_0}{r^h},
        \label{eq:Nr}
    \end{equation}
    with $h\geq0$. In order to simplify the numerical integration and to get rid of the largest number of trivial constants, we now introduce the parameter $\epsilon$, defined as the ratio between the plasma frequency and the photon frequency over the stellar surface,
    \begin{equation}
        \epsilon=\frac{\omega_e(R)}{\omega(R)},
        \label{eq:eps}
    \end{equation}
    so that the index of refraction can be simply rewritten as
    \begin{equation}
        n^2(r)=1-\frac{A(r)}{A(R)}\left(\frac{R}{r}\right)^h\epsilon^2,
        \label{eq:refindnum}
    \end{equation}
    while the propagation condition is now expressed as
    \begin{equation}
        \epsilon^2 < \frac{A(R)}{A(r)} \left( \frac{r}{R} \right)^h
        \label{eq:propagationcondition}
    \end{equation}
    for all $r\geq R$.
    For the metric models and plasma distributions that we will use in this paper, the above condition can be reduced to
    \begin{equation}
        \epsilon^2<1
        \label{eq:conpro}
    \end{equation}
    in the cases where the plasma density decays to zero faster than $A(r)$ grows as $r$ goes from $R$ to $\infty$, or to
    \begin{equation}
        \epsilon^2<A(R)
        \label{eq:conpro1}
    \end{equation}
    when we consider a constant plasma density that does not vanish around the observer.
  
    Coming back to Eq. \eqref{eq:b}, if we assume a vanishing electronic density at the neighborhood of the asymptotic observer, it follows that independently of the plasma profile model under consideration, the maximum impact parameter $b_{max}$ is expressed in terms of $\epsilon$ as
    \begin{equation}
        b_{max}={(1-\epsilon^2)}^{1/2}\sqrt{\frac{C(R)}{A(R)}}.
        \label{eq:bmax22}
    \end{equation}
    Hence, varying the observation frequency (and therefore the value of $\epsilon$ for a given system), we obtain information of the value of the metric quotient $A(R)/C(R)$ at the surface of the star. Of course, $b_{max}$ is not a direct observable, however, due to the relation between $b_{max}$ and $\theta_{max}$ we can infer that observation of pulse profiles of a given NS at different radiofrequencies codifies information on the geometric properties of the spacetime at the neighborhood of the star.
    
\subsection{Summary}
\label{c2s3ss1}
    
    We already have all the necessary tools to perform ray tracing from a highly compact and relativistic object immersed in a pressureless non-magnetized plasma medium to an observer located at infinity. Below, for the reader's convenience, we list the steps to be followed to carry out the ray tracing.
    
    \begin{enumerate}
        \item Select a metric. Some examples are shown in the Table \ref{tab:NS_Metricas}.
        \item Set the parameters star radius $R$ ($R>2M$), charge $q^*$ (Table \ref{tab:NS_Metricas}), frequency ratio $\epsilon$ ($0\leq\epsilon<1$) and plasma distribution $h$ ($h\geq$0).
        \item Calculate the asymptotic group velocity $n_0$ (at infinity) from Eq.~\eqref{eq:n0} and check that $\epsilon$ satisfies the pertinent propagation condition.
        \item Obtain the maximum impact parameter $b_{max}$ from Eq.~\eqref{eq:b} and choose an impact parameter $b$ in the range $0<b<b_{max}$. 
        \item Calculate the angle $\theta(b)$ from which the ray will leave the stellar surface according to Eq. \eqref{eq:D}.
        \item Starting from the initial position $(u_R,\theta_R)=(1/R,\theta(R,b))$, numerically integrate the ray path from Eq. \eqref{eq:dpdu}. 
    \end{enumerate}

\subsection{Results}
\label{c2s3ss2}

    The ray orbits obtained for the Schwarzschild, RN and RN like metrics are shown in Figures \ref{tab:TR_h=3} and \ref{tab:TR_h=0} below. 
    A highly relativistic star with radius $R/r_S=1.60$, with $r_S=2M$ the Schwarzschild radius, was considered. The traced rays count with positive impact parameters in the range from $0$ to $b_{max}$, whose values correspond to integer multiples of $b_{max}/5$. The absolute value of the charge in each figure is 
    $|q^*|=0.25M^2$.

    We chose this value since it allows us to distinguish the different trajectories obtained for each metric model. 
    In turn, each figure corresponds to a particular plasma distribution, and compares the results obtained for different $\epsilon$ ratios of frequencies.     
    
    For this, we wrote a code in Fortran 90 that uses the Runge-Kutta 4th order method to solve the differential Eq. \eqref{eq:dpdu} and a Gaussian quadrature rule to integrate Eq. \eqref{eq:D}.
    
    We will start considering the plasma profile suggested in \cite{Rogers_2015}, where $h=3$ was used as it considers the Goldreich-Julian density and its dependence on the polar magnetic field strength. Thus, the index of refraction is expressed in the whole space as
    \begin{equation}
        n_3^2(r)=1-\frac{A(r)}{A(R)}\left(\frac{R}{r}\right)^3\epsilon^2.
        \label{eq:NS_n_h=3}
    \end{equation}
    The results obtained for this distribution of plasma are shown in Figure \ref{tab:TR_h=3}.
    
    Moreover, we shall consider a homogeneous plasma distribution with constant density, so that the refractive index is expressed as
    \begin{equation}
        n_0^2(r)=1-\frac{A(r)}{A(R)}\epsilon^2.
        \label{eq:NS_n_h=0}
    \end{equation}
    As it is well known, there is a correspondence between the dynamics of light rays of frequency $\omega$ in a homogeneous, non-magnetized, pressureless plasma, with frequency $\omega_e$, and the timelike geodesic motion of test massive particles of mass $\mu$ and energy $E_\infty$ measured by an asymptotic observer in the same gravitational field \cite{kulsrud1992dynamics}. In particular, given the transformation $\omega_e\rightarrow\mu$ and $\omega_\infty\rightarrow E_\infty$ it is possible to use the same Hamiltonian to describe both phenomena. Thus, a constant plasma density is a model of great interest, where photons behave as if they have an effective inertial mass. {In fact, it has been suggested in the literature \cite{Dey:2016psn,Dey:2021mwb} that neutrinos from magnetosphere caps in the PeV regime could be detected by various neutrino detectors such as the IceCube. Their dynamics can then be equivalently studied from the study of photons in a homogeneous plasma.} The plots obtained for this distribution are shown in Fig. \ref{tab:TR_h=0}. 
    In addition, in order to see how sensible are the photon orbits to the behavior of the electron density profiles, two others plasma densities are shown in the Appendix \ref{AppendixA}, one of them of the form of Eq.\eqref{eq:Nr}, with $h=2$ and another one with an exponential decay profile given by $n_e^2(r)=1-\frac{A(r)}{A(R)}e^{R-r}\epsilon^2$. 
    
    In general, as in \cite{Rogers_2015}, we observe that plasma concentrations produce a divergent lensing effect that competes with the convergent effect of gravitational lensing. Thus, while the gravitational potential tries to deflect the light rays towards the star, increasing the values of $\theta_{max}$ and $b_{max}$, the diffraction produced by the plasma deflects the rays in the opposite direction, moving them away from the star and reducing the aforementioned values. This produces the photon trajectories to have a $S$ shape, where the convergent effect of the gravitational lens dominates in the vicinity of the surface, being overcome by the divergence of the plasma lens as we move away from it. These effects become more pronounced as the plasma density increases {(or alternatively the observation frequency decreases)}, i.e., for $\epsilon\rightarrow1$ the trajectories converge
more and more at $y = 0$ axis in Fig.\eqref{tab:TR_h=3}, that is,  $\theta_{max}\rightarrow0$ and $b_{max}\rightarrow0$, so that the star is no longer visible.

     \begin{figure}
        \centering
        \begin{tabular}{ccc}
            {\rotatebox{90}{\hspace{0.8cm}$\epsilon=0.00$}} &
            {\includegraphics[scale=0.20,trim=0 0 0 0]{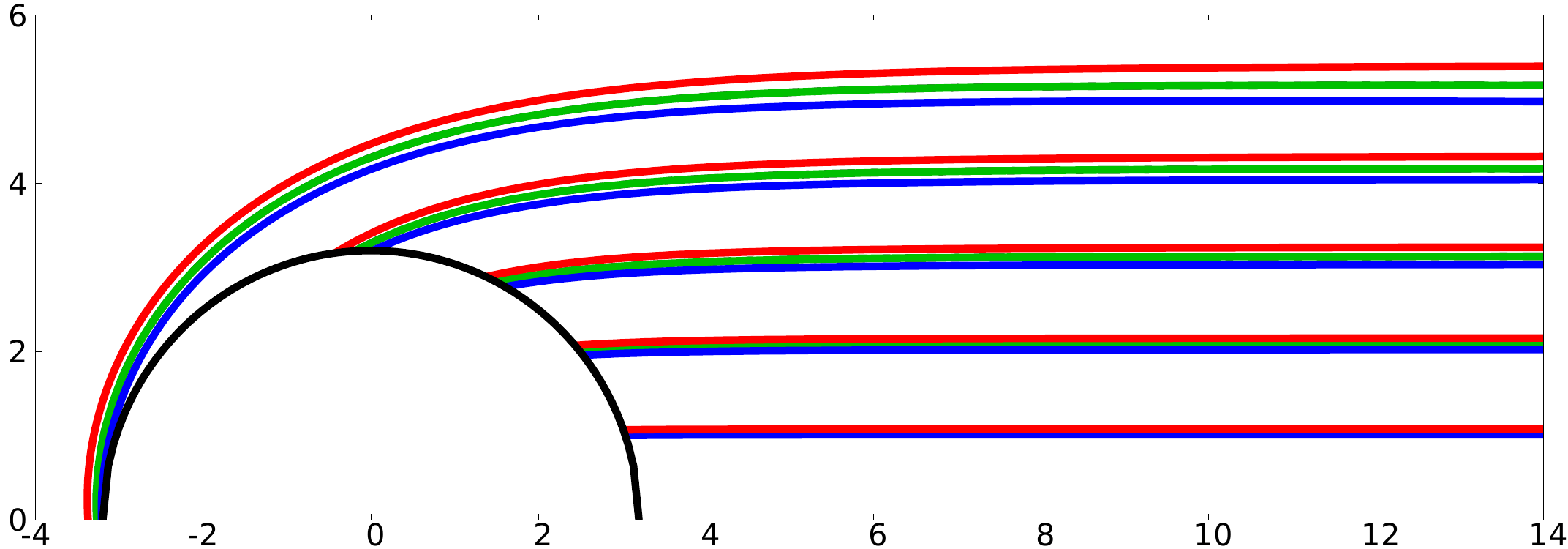}}\\
            {\rotatebox{90}{\hspace{0.8cm}$\epsilon=0.30$}} &
            {\includegraphics[scale=0.20,trim=0 0 0 0]{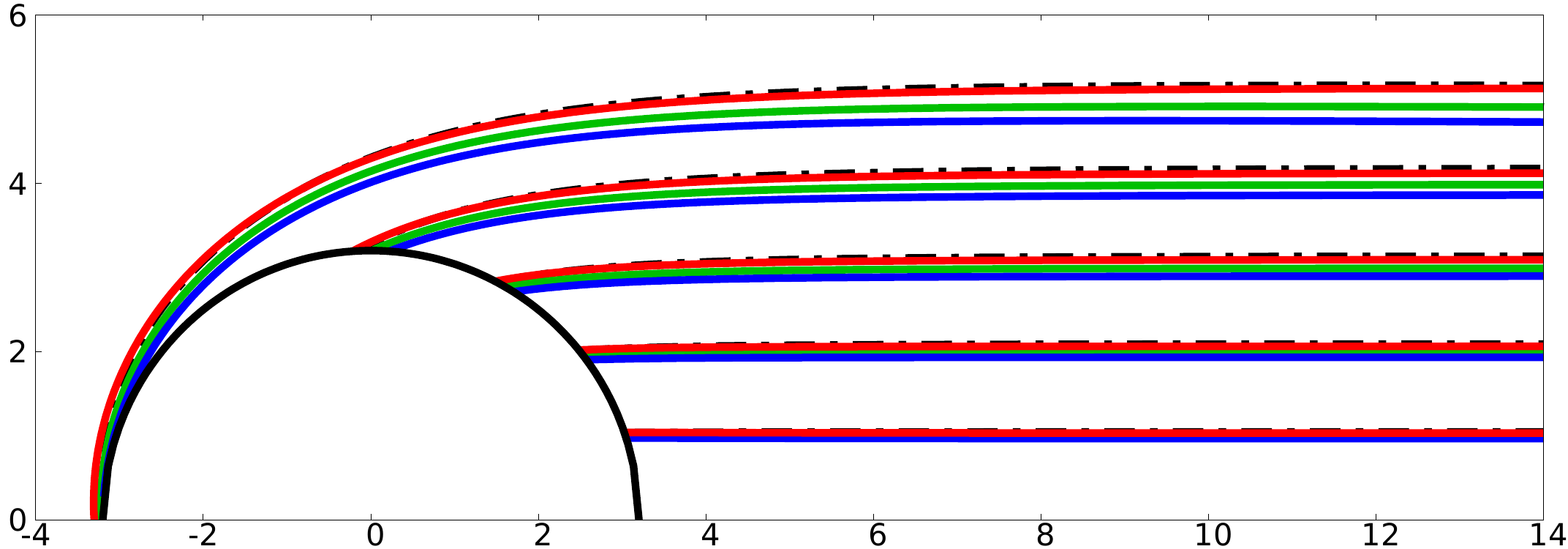}}\\
            {\rotatebox{90}{\hspace{0.8cm}$\epsilon=0.60$}} &
            {\includegraphics[scale=0.20,trim=0 0 0 0]{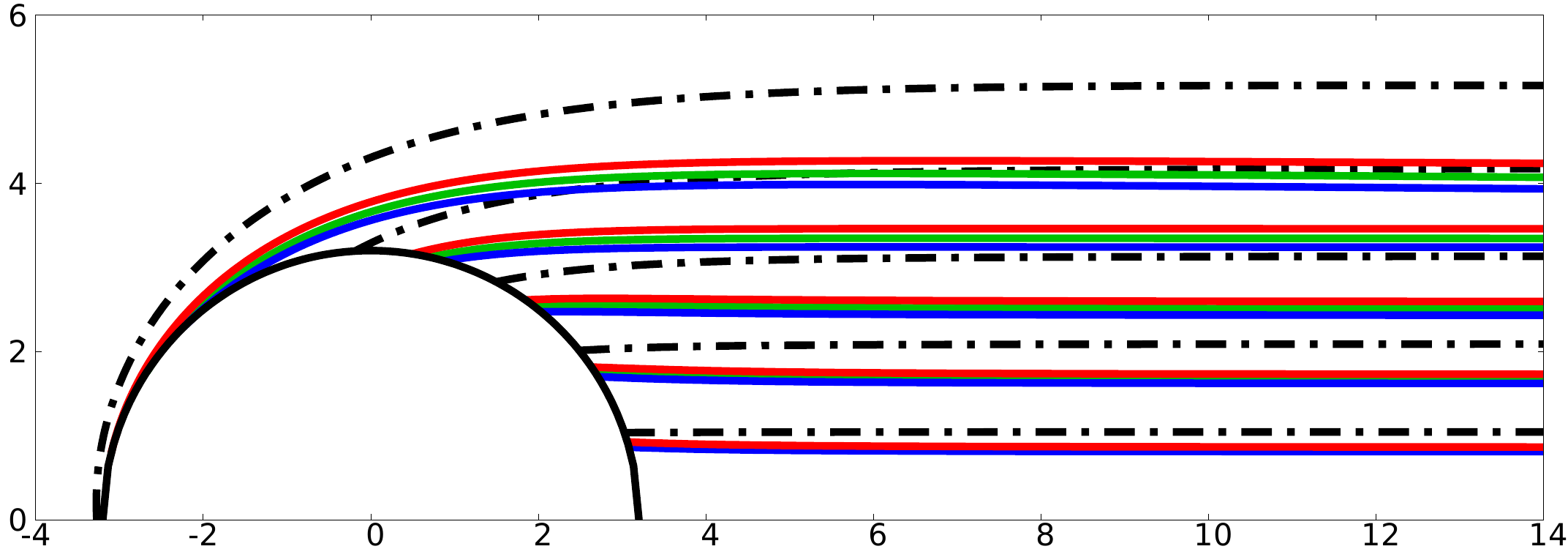}}\\
            {\rotatebox{90}{\hspace{0.8cm}$\epsilon=0.90$}} &
            {\includegraphics[scale=0.20,trim=0 0 0 0]{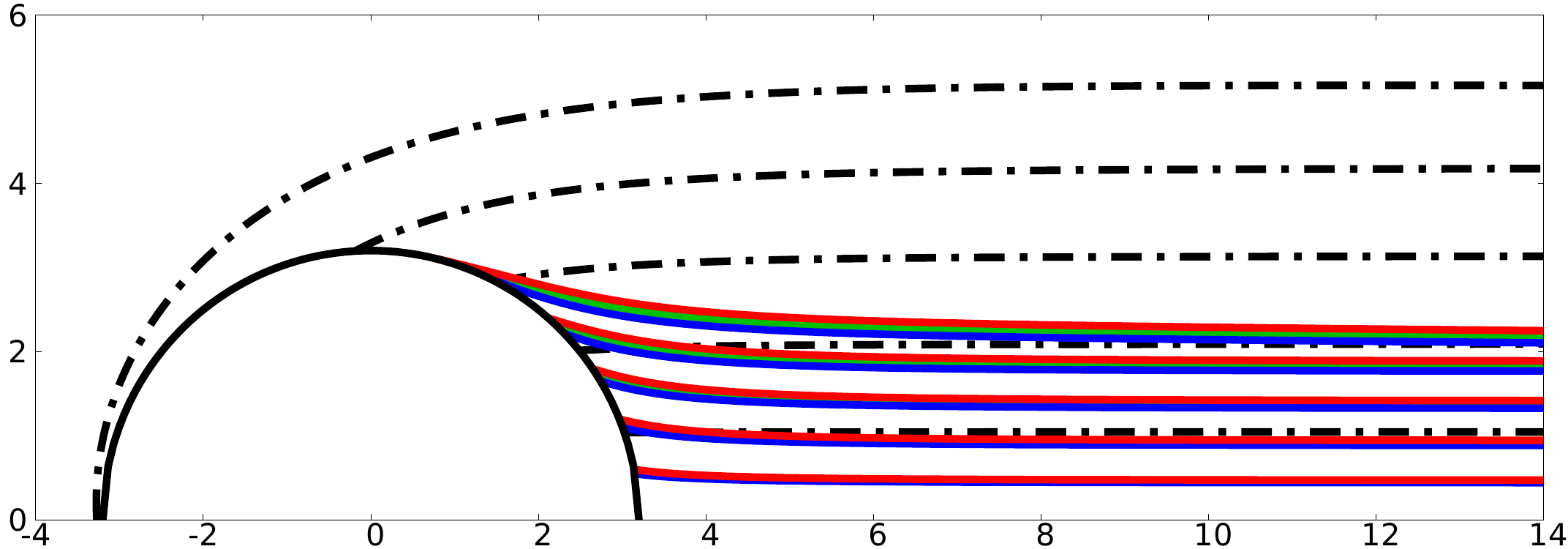}}
        \end{tabular}
        \caption{Ray tracing. $R=3.2M$, $|q^*|=0.25M^2$, $h=3$. The dotted line corresponds to Schwarzschild without plasma, green to Schwarzschild, blue to $q^*>0$ {and red to $q^*<0$}. As the ratio $\epsilon=\omega_e(R)/\omega(R)$ increases, both the apparent size of the star $b_{max}$ and the maximum angle of visibility $\theta_{max}$ decrease. 
        We see that $b_{max}$ and $\theta_{max}$ increases for $q^*<0$ and decreases for $q^*>0$, with Schwarzschild metric in between.}
        \label{tab:TR_h=3}
    \end{figure}
    
    From these plots, and those shown in Appendix \eqref{AppendixA} we can see that the divergent lensing effects produced by the plasma distribution are magnified as the plasma concentration decay rate increases (i.e. higher $h$ or exponential decay). Thus, the plasma profile with a decay given by a power law exhibits marked divergent lensing properties, while a homogeneous plasma distribution does not produce any divergent effect. This is in accordance with the basic notions of optics in refractive media, which indicate that the greater the variation of the refractive index at an interface, the greater its deflection power.
    
    \begin{figure}
        \centering
        \begin{tabular}{ccc}
            {\rotatebox{90}{\hspace{0.8cm}$\epsilon=0.00$}} &
            {\includegraphics[scale=0.20,trim=0 0 0 0]{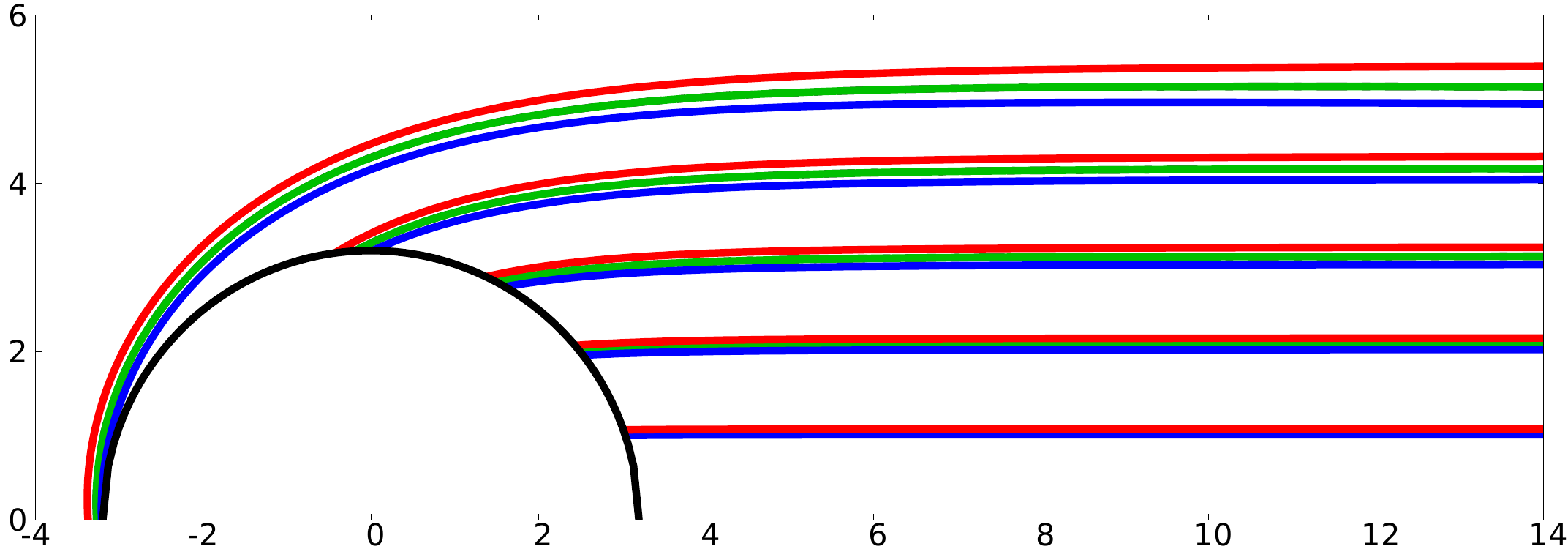}}\\
            {\rotatebox{90}{\hspace{0.8cm}$\epsilon=0.30$}} &
            {\includegraphics[scale=0.20,trim=0 0 0 0]{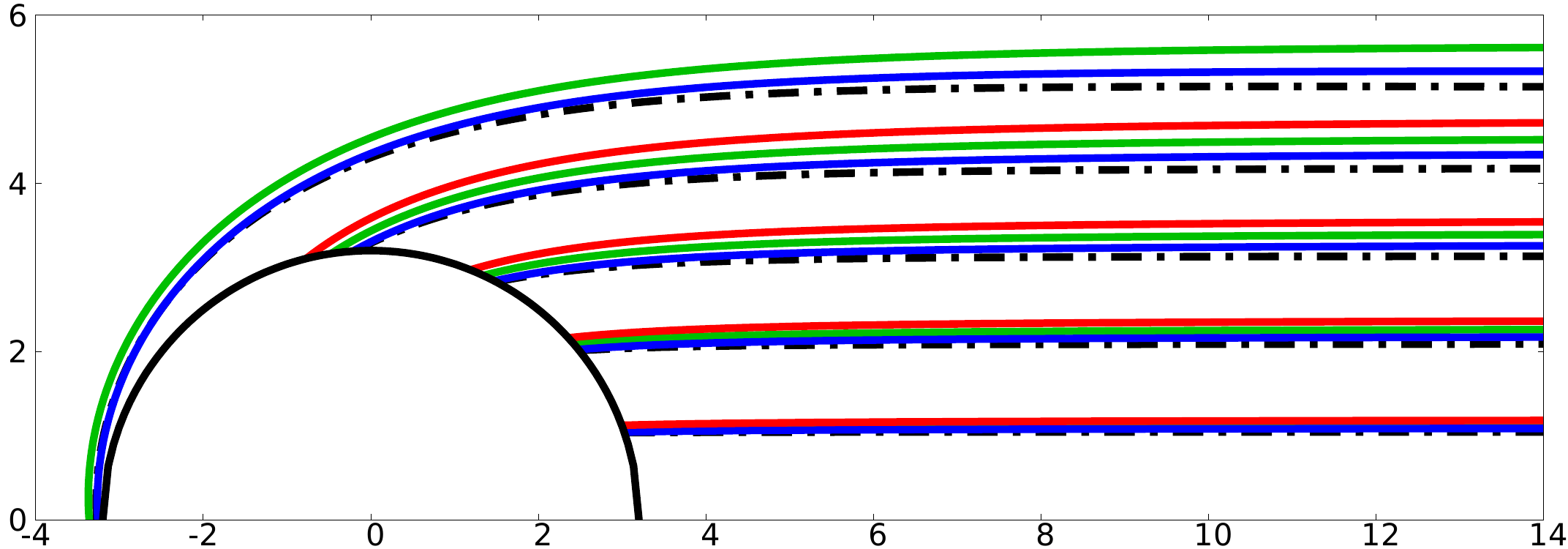}}\\
            {\rotatebox{90}{\hspace{0.8cm}$\epsilon=0.60$}} &
            {\includegraphics[scale=0.20,trim=0 0 0 0]{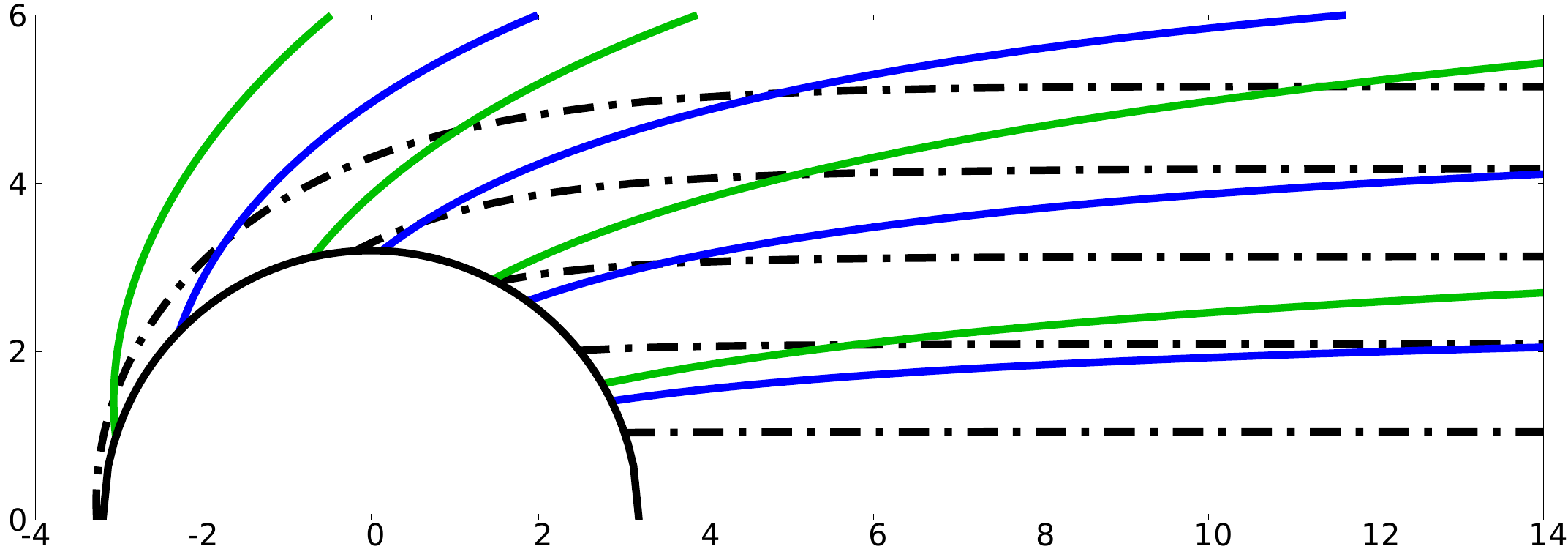}}\\
        \end{tabular}
        \caption{Same as in Fig. \ref{tab:TR_h=3} but for $h=0$. As the ratio $\epsilon={\omega_e(R)}/{\omega(R)}$ increases, both the apparent size of the star $b_{max}$ and the maximum visibility angle $\theta_{max}$ diverge. Note there is not plot at $\epsilon=0.9$, because it does not satisfy the propagation condition \eqref{eq:conpro1}.
        In addition, for $q^*<0$, as the value of $\epsilon$ increases so does the radius of the photon sphere, eventually exceeding the stellar radius, so that trajectories with $b \leq b_{max}$ do not reach the observer. That is why we do not show the $q^*<0$ example in the bottom panel.
        }
        \label{tab:TR_h=0}
    \end{figure}
    
    A curious result happens when considering a homogeneous plasma distribution. As we already said, the divergent lensing effect of the plasma is completely cancelled out in this profile. In Fig. \ref{tab:TR_h=0} it is clear that both the impact parameter and the visible surface of the star increase at larger values of $\epsilon$. If we analyze Eqs. \eqref{eq:n0} and \eqref{eq:b}, we see that this is because, given a constant plasma density, $n_0\rightarrow0$ as $\epsilon \rightarrow \sqrt{A(R)}<1$,
    and the impact parameter $b$ diverges as $1/n_0$. 
    However, as we will shown in Eq. \eqref{eq:Iobs2}, the observed light intensity goes to zero as $n_0^2$. 
    We add that, the quotient $\epsilon=0.90$ does not satisfy the propagation conditions for this profile, so it was not plotted.
    
   We note that for $h>0$ increasing the charge $q^*$ of the star tends to decrease both the observable surface area by reducing the value of $\theta_{max}$, and the apparent size of the star by reducing the value of $b_{max}$. 
    {The opposite effect results for decreasing $q^*$.  For the particular case of the RN metric ($q^*>0$) similar conclusions were observed by \cite{Dabrowski_1995} and \cite{Sotani_2017}.
    }
    
\section{Pulse Profile}
\label{c2s4}

    Consider a radio pulsar with bright polar regions of emission on its surface that will produce periodic pulses in the luminosity curve.
    The observed flux $F$ will be given by the integral of the intensity times the solid angle measured by the observer on his celestial sphere. In the case that the same emission region produces multiple images, all of them must be considered when integrating the flux. While this approach is slightly idealized, it allows a first approximation to describe qualitatively the effects associated with the various parameters of the star, its emission caps, the plasma environment and the spacetime itself.
    
    The geometry of this model is detailed in \cite{Pechenick_1983} for the gravity pure situation.
    Again, in this section we will follow the work \cite{Rogers_2015}, reproducing its results for the Schwarzschild metric and expanding the procedure to other spacetimes.
    
    The surface of the star, at $r=R$, is described by the angular spherical coordinates $\Theta$ and $\phi$ that we used above. In general, we will use $\theta$ instead of $\Theta$.
    We will place the observer on the positive semi-axis $z$ ($\Theta=0$), at $r=r_O\rightarrow\infty$,
    defining its coordinate system with respect to a plane perpendicular to the line of sight (the ``detector"). The detector surface registers the flux received from the object (the image) over the angles of its celestial sphere $\theta'$ and $\phi'$. 
    Since the star is spherically symmetric, the trajectories will remain in the same meridian plane, so we can assume $\phi=\phi'$.
    
    Hence, we can express the solid angle element in the observer's coordinate system as follows
    \begin{equation}
        d\Omega'=\sin{\theta'}d\theta'd\phi, 
        \label{eq:EAS1}
    \end{equation}
    where the angle $\theta'$ is assumed to be sufficiently small, so
    \begin{equation}
        d\Omega'=\theta'd\theta'd\phi.
        \label{eq:EAS12}
    \end{equation}
   This allows us to express the latitudinal angle $\theta'$ in terms of the impact parameter $b$, where $b\approx r_O\theta'$, resulting in
    \begin{equation}
        d\Omega'=\frac{1}{r_O^2}bdbd\phi,
        \label{eq:EAS2}
    \end{equation}
    where $b$ is given by Eq.~\eqref{eq:b} and varies between $0$ and $b_{max}$ for trajectories starting from the stellar surface. Thus, employing the approximation $b\approx r_O\theta'$ and Eq. \eqref{eq:D} we can map each observed image point in $(\theta'(b),\phi')$ to its location on the surface of the star in $(\theta(b),\phi)$, recalling that $\phi=\phi'$. 
    That is, we can map the observer's two-dimensional solid angle element onto the corresponding surface area element on the star. To find the pulse profile in question, we must find the flux of the emission region on the stellar surface and calculate the projection of this area on the observer's sky.
    
    \begin{figure}
        \centering
        \includegraphics[scale=0.33,trim=0 0 0 0]{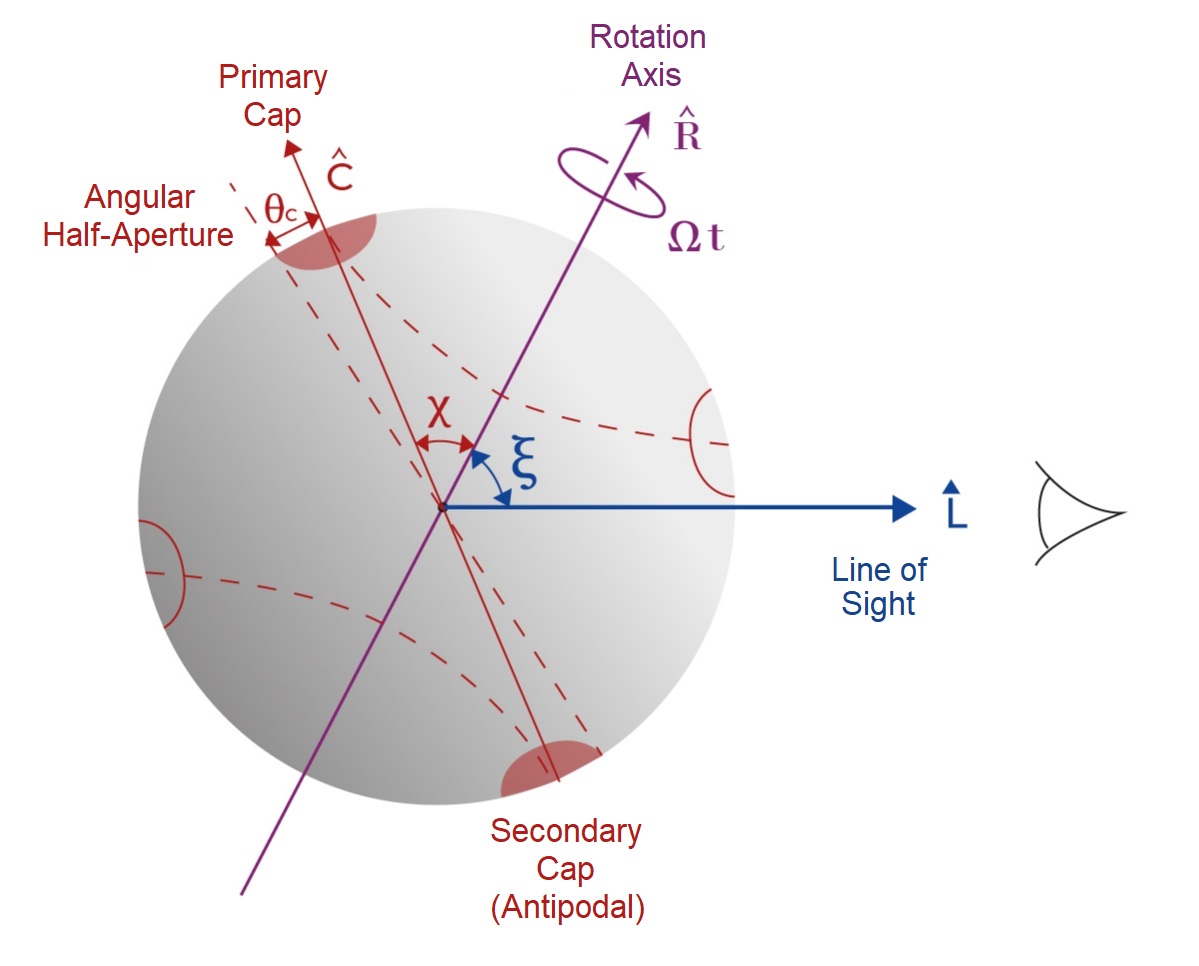} 
        \caption{Angular configuration between the observer $\hat{L}$, the rotation axis of the star $\hat{R}$ and the center of the emitting cap $\hat{C}$, of angular half-aperture $\theta_c$.}
        \label{fig:ConfiguraciónAngular}
    \end{figure}
    
    Let us consider for now a single polar emission cap, also called hot spot, of circular shape and angular half-aperture $\theta_c$, whose position with respect to the center of the star is indicated by the versor $\hat{C}$, being centered at $\theta=\theta_0$. We further define the versor along the line of sight $\hat{L}$ in the outgoing direction from the pulsar to the observer, resulting in $\theta_0=\cos^{-1}\hat{C}\cdot\hat{L}$ the angle between both versors.
    In general, $\theta_0$ will be a time-dependent quantity since the star rotates about an $\hat{R}$ axis that need not coincide with the line of sight $\hat{L}$. The angle between the rotation axis $\hat{R}$ and the line of sight is defined as $\xi=\cos^{-1}{\hat{R}\cdot\hat{L}}$, while the angle between the rotation axis and the center of the cap is given by $\chi=\cos^{-1}{\hat{R}\cdot\hat{C}}$. Both $\xi$ and $\chi$ are constants.
    These angles are represented in Fig. \ref{fig:ConfiguraciónAngular}.
    
    To find the angular position of the cap $\theta_0$ as a function of time $t$, we must express the phase of the pulsar $\gamma_p(t)=\Omega t$, where $\Omega=2\pi/P$ is the angular velocity, with $P$ being the rotation period of the star.
    Thus, the latitudinal orientation of the cap can be expressed as
    \begin{equation}
        \theta_0(t)=\cos^{-1}\left[\cos{\xi}\cos{\chi}-\sin{\xi}\sin{\chi}\cos{\gamma_p(t)}\right],
        \label{eq:t0(t)}
    \end{equation}
    where it can be seen that $\gamma_p(t)=2n\pi$ corresponds to the maximum value of $\theta_0(t)$. For a single cap, is conventionally taken $0\leq\xi\leq\pi/2$, $0\leq\chi\leq\pi$ and $0\leq\theta_0\leq\pi$.
    
    In a spherical coordinate system fixed to the cap, whose pole is given by the versor $\hat{C}$, an arbitrary point on the edge of the cap is expressed as $(\theta_c,\Phi)$, while the same point is expressed as $(\theta,\phi_b)$ in the object coordinates, marking $\phi_b$ the boundary of the cap in the parallel $\theta$. 
    Using the transformation between the Cartesian components of the two systems, we arrive at the expression
    \begin{equation}
        \phi_b(\theta)=\cos^{-1}\left(\frac{\cos{\theta_c}-\cos{\theta_0}\cos{\theta}}{\sin{\theta_0}\sin{\theta}}\right),
        \label{eq:pb(t)}
    \end{equation}
    where $2\phi_b(\theta)$ should be understood as the angular length of the circle segment in the observer's sky with the same impact parameter $b(\theta)$, corresponding to the one-dimensional intersection between the $\theta$ parallel and the emitting cap on the surface of the star. In this way, we can determine the boundaries of the cap at the observer's coordinates along the $\theta$ parallel for a given $\theta_0$ orientation and $\theta_c$ half-aperture. Over such parallel, the cap extends in the range $[-\phi_b,\phi_b]$, so we will define the function $h(\theta,\theta_c,\theta_0)=2\phi_b$ as the range of values of $\phi$ that belong to the one-dimensional intersection between the $\theta$ parallel and the emitting cap, 
    \begin{equation}
        h(\theta,\theta_c,\theta_0) = \begin{cases} 
        2\pi,             & ~~~~~~~\Delta\leq-1 \\
        2\arccos(\Delta), & -1<\Delta<+1 \\
        0,                & ~~~+1\leq\Delta \end{cases}
        \label{eq:h2}
    \end{equation}
    where $\Delta$ is the argument of the function $\cos^{-1}$ in Eq. \eqref{eq:pb(t)}.
    
    \begin{figure}
        \centering
        \includegraphics[scale=0.2,trim=0 0 0 0]{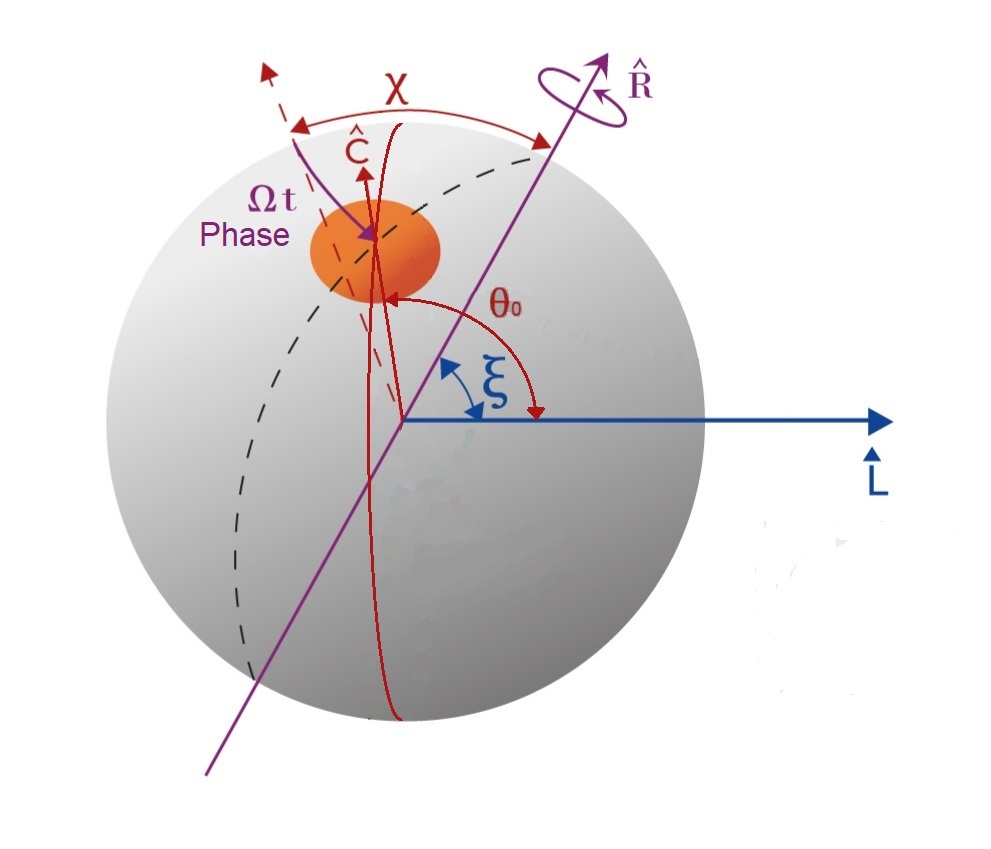} 
        \includegraphics[scale=0.2,trim=0 0 0 0]{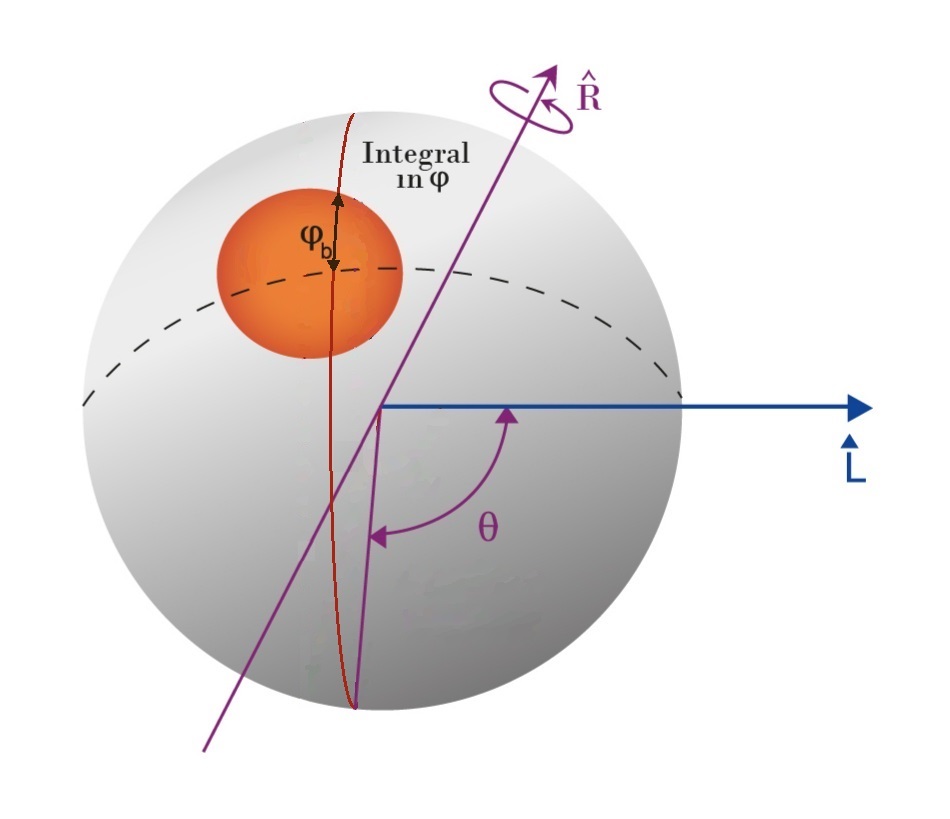}
        \caption{The figure on the left shows the dependence of the observed position of the cap $\theta_0$ on the phase of the pulsar $\Omega t$. In the figure on the right we see a graphical representation of the $\phi$ integral, $\phi_b$, at a given $\theta$.}
        \label{fig:IntegralenPhi}
    \end{figure}
    
    Actually, the definition of $\phi_b$ is more complicated than the one described above, since it is not well defined for any $\theta_0$ orientation. This point is discussed in more detail in \cite{Pechenick_1983} and \cite{Dabrowski_1995}. However, for the purpose of this paper, it is sufficient for us with Eq. \eqref{eq:h2}, which will allow us to simplify the solid angle element.
    
    Following to \cite{Rogers_2015}, we introduce the normalized impact parameter $x=b/M$ ($M=1$).
    The solid angle segment results
    \begin{equation}
        d\Omega'=\frac{1}{r_O^2}h(\theta(x),\theta_c,\theta_0)xdx,
        \label{eq:EAS3}
    \end{equation}
    so that, to determine the observable flux of an emission cap, only a one-dimensional integral over the impact parameter $x$ is required.
    In order to contemplate the occurrence of multiple images, we must allow $\theta$ to take all values in the interval $[0,\theta_{max}]$. For convenience, we rewrite Eq. \eqref{eq:D} in terms of $x$ as
    \begin{equation}
        \theta=\int_0^{u_R} \frac{1}{u^2C}\left[ \frac{1}{AB} \left(\frac{n^2}{n_0^2}\frac{1}{x^2}-\frac{A}{C} \right)   \right]^{-1/2}du.
        \label{eq:t(b)}
    \end{equation}
    
    The emitted intensity $I_{em}$ of a bright spot on the stellar surface at $r=R$ in a dispersive medium is related to the observed intensity $I_{obs}$ of that spot by a detector at $r=r_O$ by \cite{BH_1975}
    \begin{equation}
        \frac{I_{obs}}{\omega(r_O)^3n(r_O)^2}=\frac{I_{em}}{\omega(R)^3n(R)^2}=\text{constant}.
        \label{eq:Iobs1}
    \end{equation}
     The observed intensity is then 
    \begin{equation}
        I_{obs}=\left(\frac{A(R)}{A(r_O)}\right)^{3/2}\left(\frac{n(r_O)}{n(R)}\right)^{2}I_{em}.
        \label{eq:Iobs2}
    \end{equation}

    We will now consider the possibility that the emitted intensity of the polar cap depends on the emission angle $\delta$ with respect to the stellar surface normal. Let us take for convenience
    \begin{equation}
        I_{em}=r_O^2\frac{A(r_O)^{3/2}}{C(R)}\left(\frac{n(R)}{n(r_O)}\right)^{2}I_0f_B(\delta),
        \label{eq:Iem}
    \end{equation}
    where $I_0$ is a parametric constant and $f_B$ describes the emission anisotropy. The factor $r_O^2$ has been included so that it can be cancelled out when joining this expression with Eq. \eqref{eq:EAS3}, and has been divided by $C(R)$ to normalize the flux by the emission area of the star. 
    In this way, the expression for the observed flux can be rewritten as
    \begin{equation}
        I_{obs}=r_O^2\frac{A(R)^{3/2}}{C(R)}I_0f_B(\delta).
        \label{eq:Iobs3}
    \end{equation}
    On the other hand, the emission angle at the surface is a function of $x$ since, as can be deduced from Eq.~\eqref{eq:b}, it turns out to be
    \begin{equation}
        \delta=\arcsin{\left(\frac{x}{x_{max}}\right)},
        \label{eq:d(x)}
    \end{equation}
    so that we can express the observed flux differential $dF$ in the detector's solid angle differential $d\Omega'$ as
    \begin{equation}
        dF = I_{obs}d\Omega'= \frac{A(R)^{3/2}}{C(R)}I_0f_B[\delta(x)]h[\theta(x),\theta_c,\theta_0]xdx.
        \label{eq:dF}
    \end{equation}
    
    To obtain the total observed flux $F$, it only remains to integrate the previous expression in $x$, obtaining
    \begin{equation}
        F=\frac{A(R)^{3/2}}{C(R)}I_0\int_0^{x_{max}}f_B[\delta(x)]h[\theta(x),\theta_c,\theta_0]xdx.
        \label{eq:flux}
    \end{equation}
    This result, derived in \cite{Rogers_2015} for the particular case of a Schwarzschild metric, is similar to that obtained in \cite{Pechenick_1983}. However, the values of $\theta$ and $x_{max}$ now depend on the frequencies $\omega$ and $\omega_e$, thus introducing the effects produced by the presence of plasma and its distribution. 
    
    The existence of a second antipodal cap identical to the first simply requires the addition of a second component with $\theta_{0,2}=\pi-\theta_{0,1}$,
    \begin{equation}
        F_T=F(\theta_0)+F(\pi-\theta_0)=F(\theta_{0,1})+F(\theta_{0,2}),
        \label{eq:PP_F2}
    \end{equation}
    where $F_T$ is the total observed flux, and $F$ is the flux produced by each of the caps, given by Eq. \eqref{eq:flux}.
    
\subsection{Summary} 
\label{c2s4ss1}
 
    For the reader's convenience, we list below the steps to be followed to obtain the pulse profile of a neutron star with circular and uniform, single or antipodal emission caps.
    
    \begin{enumerate}
        \item Choose the metric elements $A(r)$, $B(r)$ and $C(r)$.
        \item Set $R$, $q^*$, $\epsilon$ and $h$, taking into account their respective limitations.
        \item Select the angles between the axis of rotation and the line of sight $\xi$ ($0\leq\xi\leq\pi/2$), between the rotation axis and the center of the cap $\chi$ ($0\leq\chi\leq\pi$) and half-opening of the cap $\theta_c$ ($0\leq\theta_c\leq\pi$).
        \item Choose a surface emission function $f_B[\delta(x)]$.
        \item Calculate the maximum impact parameter $x_{max}$ [Eq. \eqref{eq:b}].
        \item Discretize the values to be used for the phase $\gamma_p\in[0,2\pi]$ and the impact parameter $x\in[0,x_{max}]$ and evaluate on them $\theta_0(t)$ by Eq.~\eqref{eq:t0(t)}, $\theta(x)$ by Eq.~\eqref{eq:t(b)} and $h(\theta,\theta_c,\theta_0)$ by Eq.~\eqref{eq:h2}.
        \item For each $\gamma_p$, integrate the observed flux $F$ using Eq.~\eqref{eq:flux}.
        \item Repeat the procedure sweeping values of $\gamma_p$ between $0$ and $\pi$ (the single-cap profile is symmetric with respect to $\pi$) to obtain the single-cap pulse profile. 
        \item In the case of presence of a second cap identical and antipodal to the first one, add the flux of the latter according to Eq. \eqref{eq:PP_F2} by sweeping the values of $\gamma_p$ between $0$ and $\pi/2$ (the profile of the antipodal caps is symmetric with respect to $\pi/2$). In this case, we take $\chi_1$ such that $0\leq\chi\leq\pi/2$.
    \end{enumerate}
    
\subsection{Results}
\label{c2s4ss2}

    We implemented in Fortran 90 a program that uses Simpson's rule to solve the integral in Eq. \eqref{eq:flux}.
    
    To obtain the luminosity curves we must plot the observed flux $F$ as a function of the star period $\gamma_p$, recalling the time dependence of $\theta_0$. This procedure was performed for a variety of frequency ratios $\epsilon=\omega_e(R)/\omega(R)$, charges $q^*$ 
    and radius ratios $R/r_S$, with $r_S=2M$ being the radius of the event horizon for the Schwarzschild metric.
    The most compact radius ratio used is such that it presents multiple images of the surface, which is now completely visible. 
    The plasma density distributions used are the same as in the previous section, resulting in the refractive indices given by Eq.\eqref{eq:NS_n_h=3} and Eq.\eqref{eq:NS_n_h=0}. 
    We consider polar caps with angular half-opening $\theta_c=5^o$, single or antipodal, assuming isotropic emission ($f_B(\delta)=1$). The specific value of $I_0$ only affects the scale of the pulse profile leaving its morphology unchanged, so $I_0=1$ was taken.
     
\begin{figure}
        \centering
        \begin{tabular}{cccc}
            & $R/2M=1.675$ & $R/2M=2$ & \\
            \rotatebox{90}{\hspace{0.5cm}Flux ($\cdot100$)} &  
            {\includegraphics[width=3.1cm,height=2.2cm,trim=0 0 0 0]{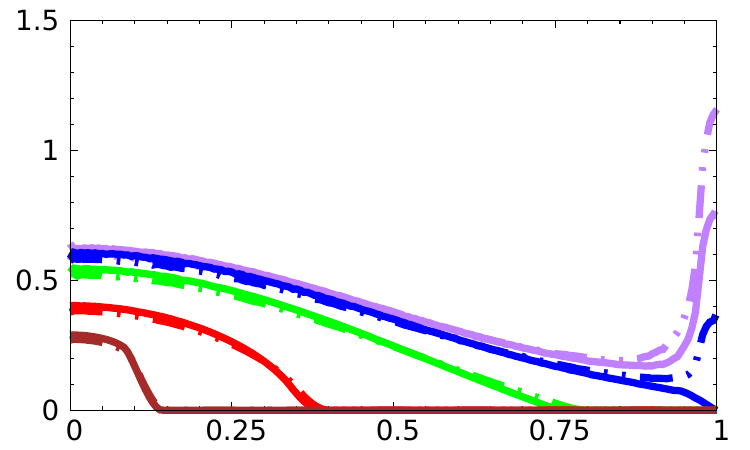}} &
            {\includegraphics[width=3.1cm,height=2.2cm,trim=0 0 0 0]{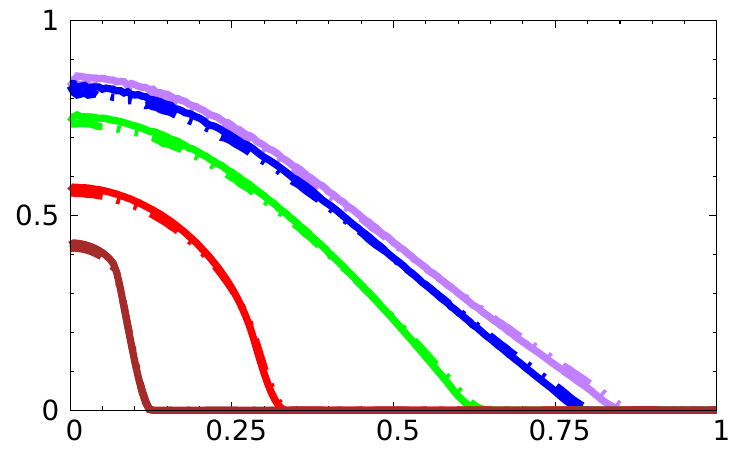}} &
            \rotatebox{90}{\hspace{0.15cm}$|q^*|=(0.25M)^2$} \\
            \rotatebox{90}{\hspace{0.5cm}Flux ($\cdot100$)} &  
            {\includegraphics[width=3.1cm,height=2.2cm,trim=0 0 0 0]{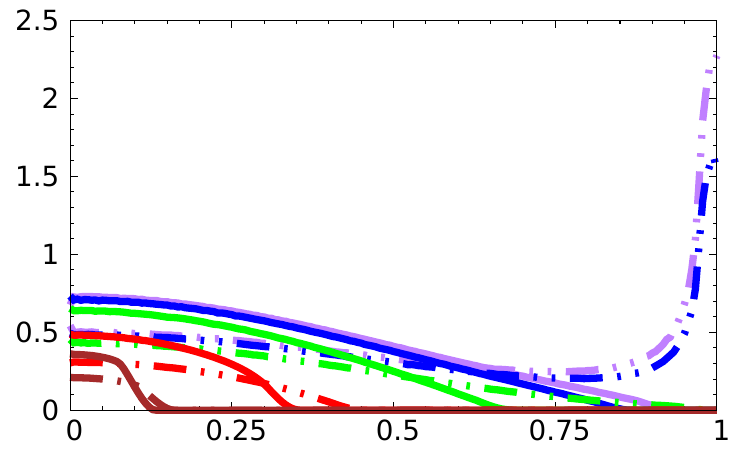}} &
            {\includegraphics[width=3.1cm,height=2.2cm,trim=0 0 0 0]{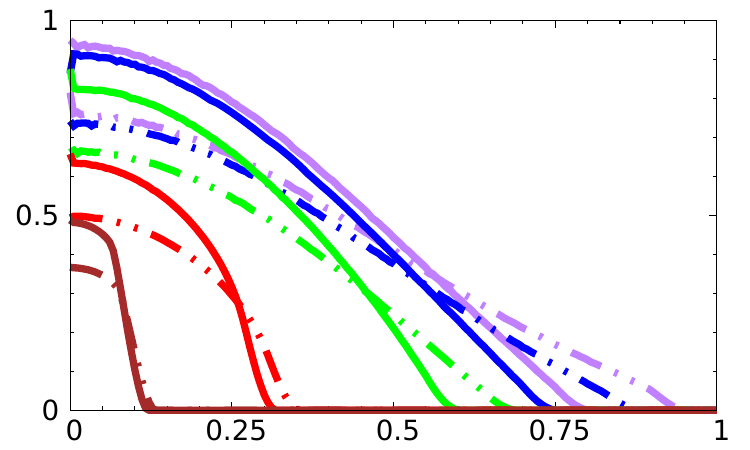}} &
            \rotatebox{90}{\hspace{0.15cm}$|q^*|=(0.75M)^2$} \\
            & {~~$\Omega t/\pi$} & {~~$\Omega t/\pi$} &
        \end{tabular}
        \caption{Pulse profile for single caps. $h=3$, $\xi=\chi=\pi/2$. The purple, blue, green, red, and brown lines correspond to $\epsilon=0.00$, $0.30$, $0.60$, $0.90$ and $0.99$ respectively. Solid lines correspond to $q^*>0$ while dashed lines correspond to $q^*<0$. In the left column we see the characteristic peak due to the cap in opposition generating multiple images. If the star is not compact enough, the flux goes to zero when the cap leaves the zone of visibility. In the top row, for small values of $q^*$ the curves obtained for $q^*>0$ and $q^*<0$ are almost identical. In the bottom row, we see that for $q^*<0$ the initial flux is smaller than for $q^*>0$ but decays more slowly.}
        \label{fig:PP_SHS_h=3}
    \end{figure}

\begin{figure}
        \centering
        \begin{tabular}{cccc}
            & $R/2M=1.675$ & $R/2M=2$ & \\
            \rotatebox{90}{\hspace{0.5cm}Flux ($\cdot100$)} &  
            {\includegraphics[width=3.1cm,height=2.2cm,trim=0 0 0 0]{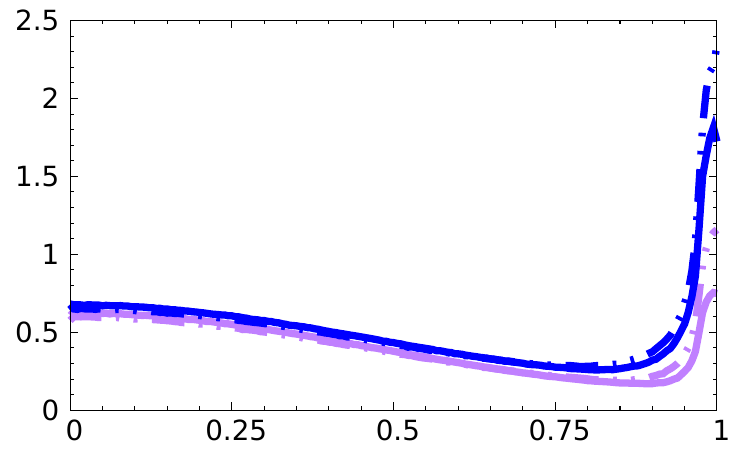}} &
            {\includegraphics[width=3.1cm,height=2.2cm,trim=0 0 0 0]{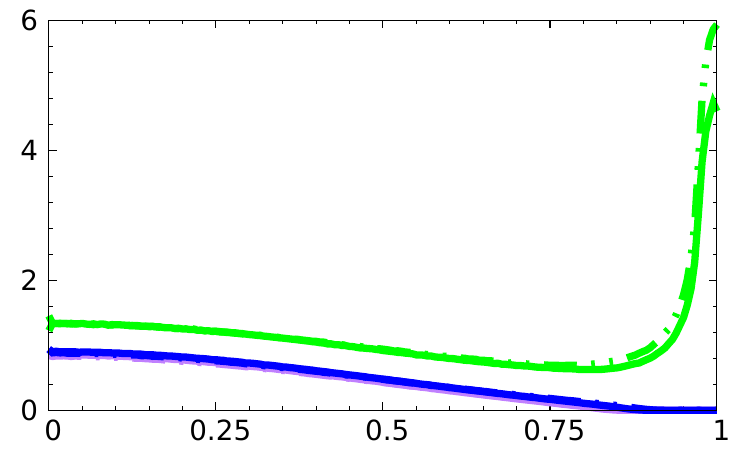}} &
            \rotatebox{90}{\hspace{0.15cm}$|q^*|=(0.25M)^2$} \\
            \rotatebox{90}{\hspace{0.5cm}Flux ($\cdot100$)} &  
            {\includegraphics[width=3.1cm,height=2.2cm,trim=0 0 0 0]{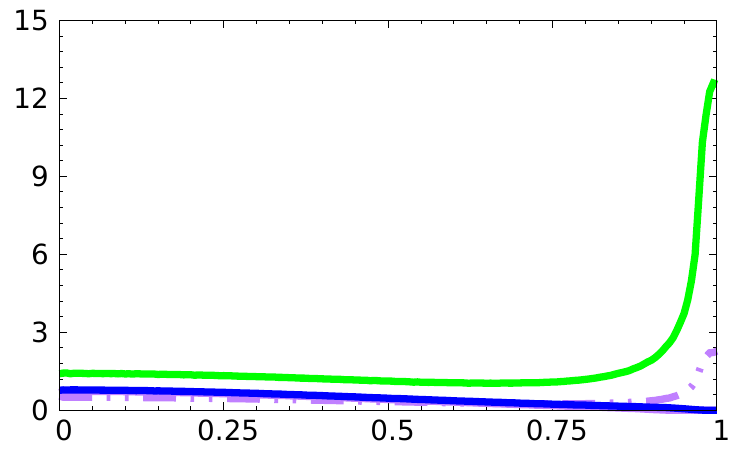}} &
            {\includegraphics[width=3.1cm,height=2.2cm,trim=0 0 0 0]{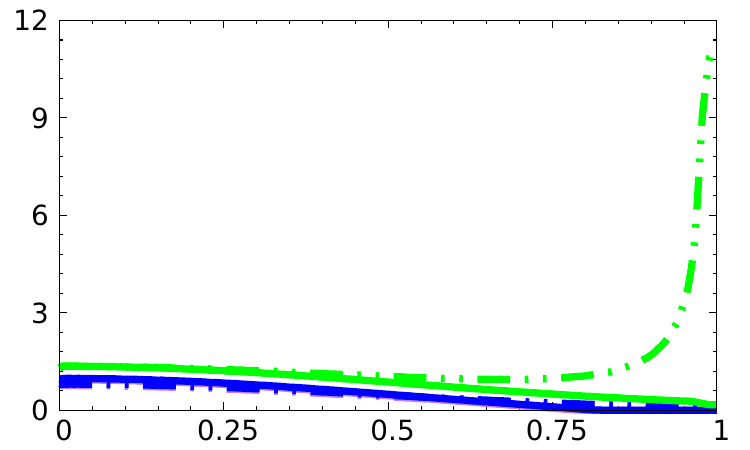}} &
            \rotatebox{90}{\hspace{0.15cm}$|q^*|=(0.75M)^2$} \\
            & {~~$\Omega t/\pi$} & {~~$\Omega t/\pi$} &
        \end{tabular}
        \caption{Same configuration as in Fig. \ref{fig:PP_SHS_h=3} but for $h=0$. We see that, for a constant plasma density, low frequencies do not satisfy the propagation conditions in the vicinity of the star.}
        \label{fig:PP_SHS_h=0}
    \end{figure}

    First, a single cap was considered in an orthogonal configuration ($\chi=\xi=\pi/2$). The results are shown in Figures \ref{fig:PP_SHS_h=3} and \ref{fig:PP_SHS_h=0}.
   
    Highly relativistic stars with radii $R/2M=1.675$ and $2$ were considered. The values $|q^*|=(0.25M)^2$ and $(0.75M)^2$ were taken. In turn, each figure corresponds to a particular plasma distribution, and compares the results obtained for different frequency ratios, with values $\epsilon=0.00$, $0.30$, $0.60$, $0.90$ and $0.99$.
        
    The first thing to note is that, for extremely compact stars and under certain conditions, the observed flux is much higher when the emitting cap is located at the back of the star ($\Omega t=\pi$) than when it is located at the front ($\Omega t=0$). This highly counterintuitive phenomenon is due to the fact that, because the star is so compact, a spot at its back produces multiple images. Thus, when the cap is in this position, its image forms a ring, not too thick but with a large angular aperture, which occupies a bigger part of the observer's sky. As the cap enters into the multiple imaging zone (i.e., for sufficiently large $\theta_0$), the luminosity is boosted resulting in the observed pulse profiles. In general, this effect increases with the compactness of the star, canceling out for stars of larger radius whose surface is no longer fully visible, so they do not form multiple images. We also see that the greater the radius of the star, the greater its relative luminosity at $\Omega t=0$. This is because, being less compact, the trajectories of the light rays are less disturbed by the gravitational field, so they are not so strongly deflected and more of them manage to reach the observer. At the same time, increasing the radius decreases the portion of the period during which the cap is visible.

    As already discussed in \cite{Sotani_2017} for the RN case, we see that, as increasing $q^*$, the fraction of the period during which the cap is visible or, alternatively, the magnitude of the peak when multiple images occur decrease, 
    {while it increases as decreasing $q^*$}. 
    This agrees with the results obtained in the previous section, where we saw that increasing $q^*$ decreases the observable surface of the star (period in which it is observable) 
    {and vice versa.} 
    At the same time, we see that the greater the value of $\epsilon$ the smaller the observed flux and the faster it decays to zero. Again, this is in agreement with what was observed in the previous section, where increasing the plasma density (or decreasing the observation frequency) reduced both the apparent size of the star and its observable surface. 
    On the other hand, we note that the flux at $\Omega t=0$ is higher for $q^*>0$, while as $\Omega t$ increases, decreases more rapidly than for $q^*<0$, which presents a higher peak at $\Omega t=\pi$ but decaying to zero later. These differences are accentuated by increasing $|q^*|$.

    \begin{figure}
        \centering
        \begin{tabular}{cccc}
            & $R/2M=1.675$ & $R/2M=2$ & \\
            \rotatebox{90}{\hspace{0.5cm}Flux ($\cdot100$)} &  
            {\includegraphics[width=3.1cm,height=2.2cm,trim=0 0 0 0]{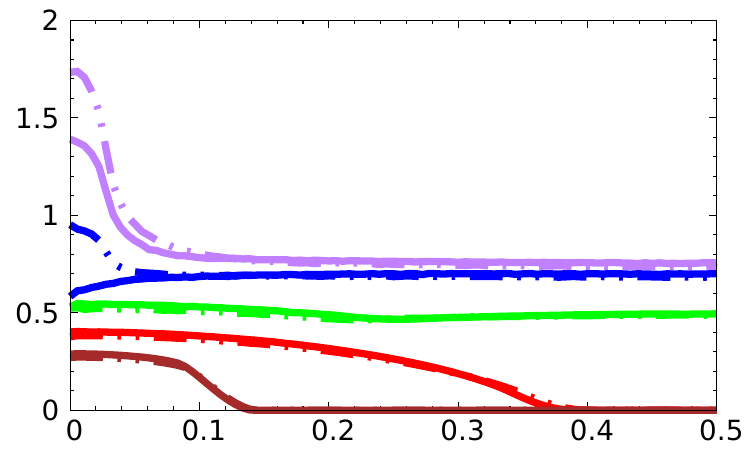}} &
            {\includegraphics[width=3.1cm,height=2.2cm,trim=0 0 0 0]{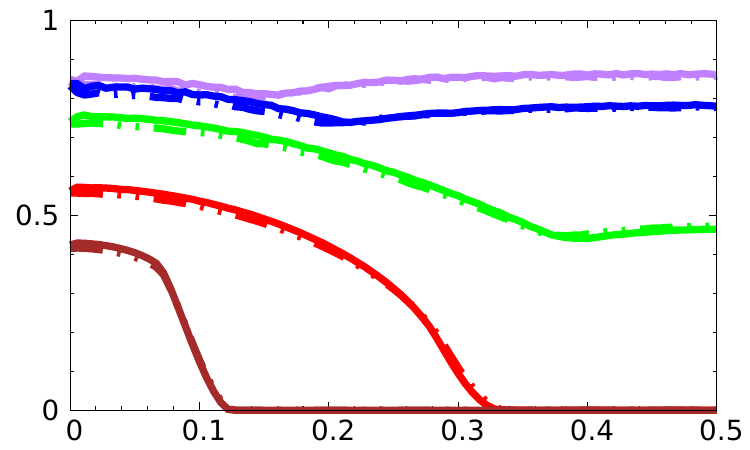}} &
            \rotatebox{90}{\hspace{0.15cm}$|q^*|=(0.25M)^2$}  \\
            \rotatebox{90}{\hspace{0.5cm}Flux ($\cdot100$)} &  
            {\includegraphics[width=3.1cm,height=2.2cm,trim=0 0 0 0]{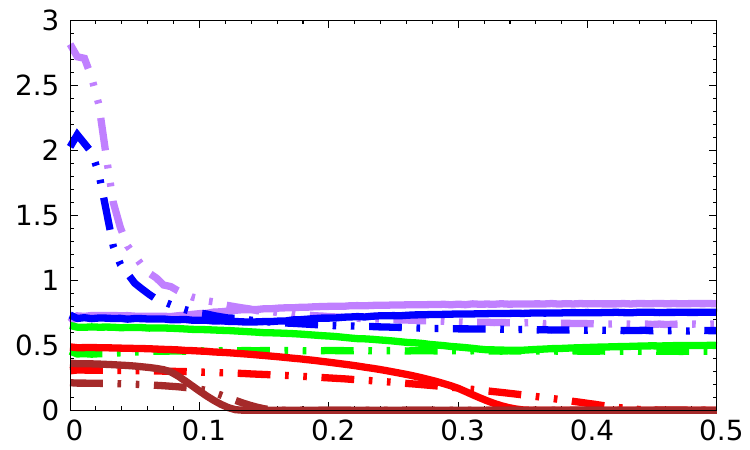}} &
            {\includegraphics[width=3.1cm,height=2.2cm,trim=0 0 0 0]{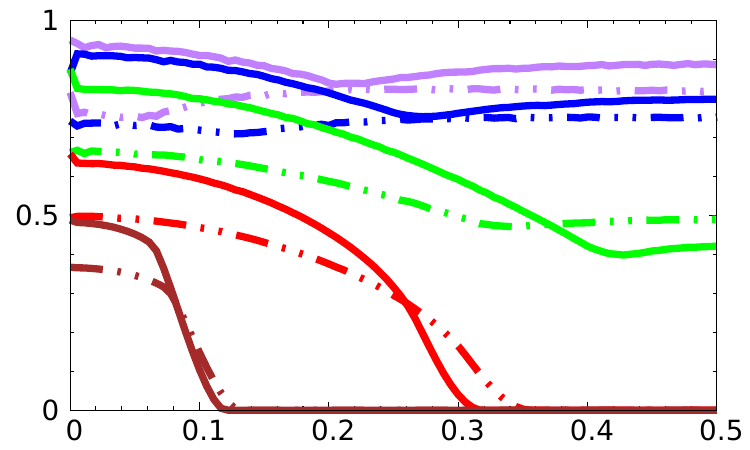}} &
            \rotatebox{90}{\hspace{0.15cm}$|q^*|=(0.75M)^2$}  \\
            & {~~$\Omega t/\pi$} & {~~$\Omega t/\pi$} &
        \end{tabular}
        \caption{Pulse profile for antipodal caps. $h=3$, $\xi=\chi=\pi/2$. Purple, blue, green, red and brown lines correspond to $\epsilon=0.00$, $0.30$, $0.60$, $0.90$ and $0.99$ respectively. Solid lines correspond to $q^*>0$ while dashed lines correspond to $q^*<0$. In the left column we see the characteristic peak due to the cap in opposition generating multiple images. When no cap produces multiple images or is out of the zone of visibility, the flux remains almost constant. In the top row, we see that for small values of $|q^*|$ the curves obtained for $q^*>0$ and $q^*<0$ are almost identical. In the bottom row, we see that the light curves associated to $q^*<0$ differ from those in $q^*>0$, 
        as observed in Fig.\eqref{fig:PP_SHS_h=3}.}
        \label{fig:PP_DHS_90_90_h=3}
    \end{figure}

    \begin{figure}
        \centering
        \begin{tabular}{cccc}
            & $R/2M=1.675$ & $R/2M=2$ & \\
            \rotatebox{90}{\hspace{0.5cm}Flux ($\cdot100$)} &  
            {\includegraphics[width=3.1cm,height=2.2cm,trim=0 0 0 0]{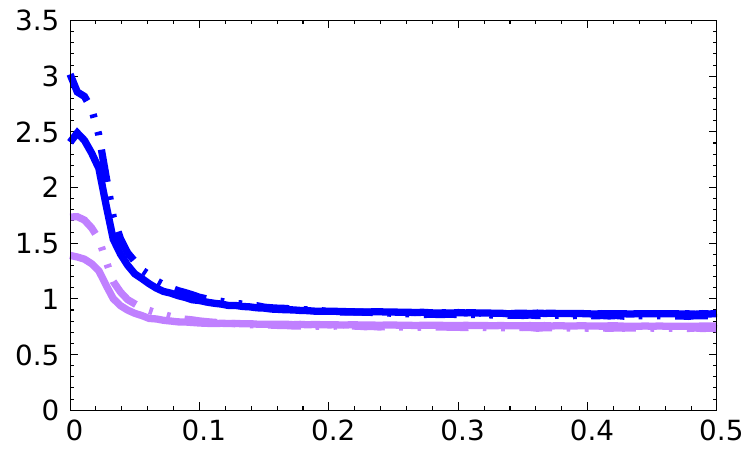}} &
            {\includegraphics[width=3.1cm,height=2.2cm,trim=0 0 0 0]{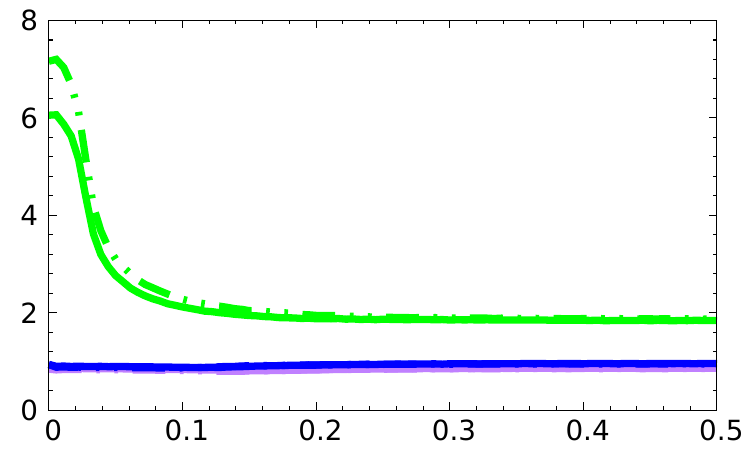}} &
            \rotatebox{90}{\hspace{0.15cm}$|q^*|=(0.25M)^2$} \\
            \rotatebox{90}{\hspace{0.5cm}Flux ($\cdot100$)} &  
            {\includegraphics[width=3.1cm,height=2.2cm,trim=0 0 0 0]{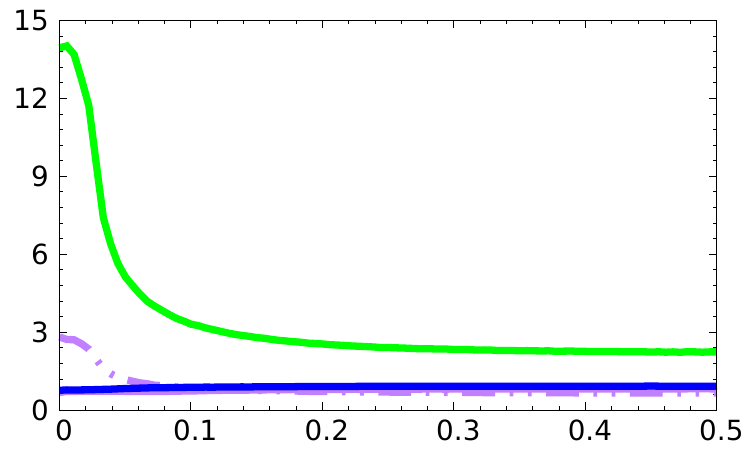}} &
            {\includegraphics[width=3.1cm,height=2.2cm,trim=0 0 0 0]{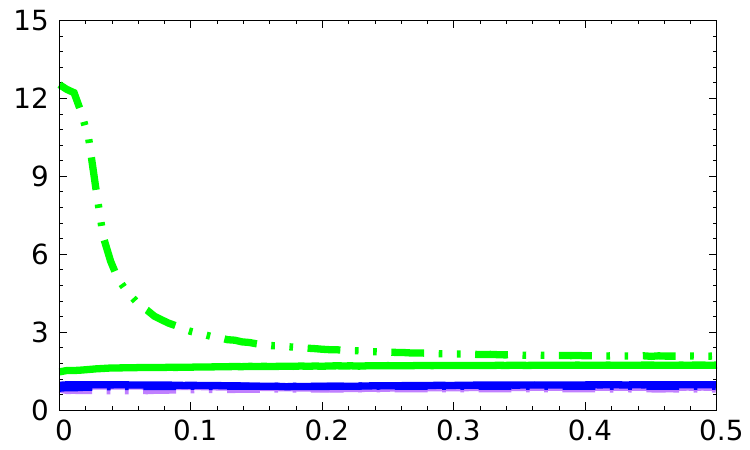}} &
            \rotatebox{90}{\hspace{0.15cm}$|q^*|=(0.75M)^2$} \\
            & {~~$\Omega t/\pi$} & {~~$\Omega t/\pi$} &
        \end{tabular}
        \caption{Same configuration as in Fig. \ref{fig:PP_DHS_90_90_h=3} but for $h=0$. We see that, for a constant plasma density, low frequencies do not satisfy the propagation conditions in the vicinity of the star.}
        \label{fig:PP_DHS_90_90_h=0}
    \end{figure}
    
    Contrary to what happens when considering plasma densities that decay asymptotically, for constant distribution, we see from Fig. \eqref{fig:PP_SHS_h=0} that higher $\epsilon$ result in higher flux. 
    However, this is a mathematical artefact due to the normalization chosen for the emitted flux in Eq. \eqref{eq:Iem}, where $I_{em}$ was set to be inversely proportional to $n_0^2$, and therefore divergent for $\epsilon$ large enough.
    Indeed, from Eq. \eqref{eq:Iobs2} we can see that, for a physically realistic finite value of $I_{em}$, $I_{obs}$ should vanish as $\epsilon$ grows.
    In addition, we see that for large enough frequency ratios that do not satisfy the propagation condition do not produce an observable pulse profile. 
    
    Let us repeat the previous procedure by adding an antipodal emission cap identical to the first one to study its impact on the pulse profile morphology. We keep the orthogonal configuration of the angles ($\chi=\xi=\pi/2$) and the same values for $R$, $\epsilon$ and $q^*$ and plasma profiles. The results are shown in Figs. \ref{fig:PP_DHS_90_90_h=3} and \ref{fig:PP_DHS_90_90_h=0}. 
    
    We see that now, given the antipodal cap, the increase in brightness due to the lensing effect also occurs at $\Omega t=0$, resulting in a flux peak at the initial instant, which will be more pronounced the higher the compactness of the star. 
    In addition, we note that at later times, when none of the caps are either producing multiple images or hidden in the star's non-visibility zone, the resulting flux remains largely constant despite the change in the position of the caps. 
    If the star is not compact enough to produce multiple images, there is a minimum with a small value in the flux when the two behaviors are spliced, but the magnitude does not vary so much from one regime to the other. 
    On the other hand, when the surface of the star is fully visible, there is no minimum when switching from one behavior to the other, and the flux intensity is much higher when multiple images are produced. Beyond the addition of a second cap, the qualitative description of the results is otherwise identical to that for the single caps.
    
    \begin{figure}
        \centering
        \begin{tabular}{cccc}
            & $|q^*|=(0.25M)^2$ & $|q^*|=(0.75M)^2$ & \\
            \rotatebox{90}{\hspace{0.5cm}Flux ($\cdot100$)} &  
            {\includegraphics[width=3.1cm,height=2.2cm,trim=0 0 0 0]{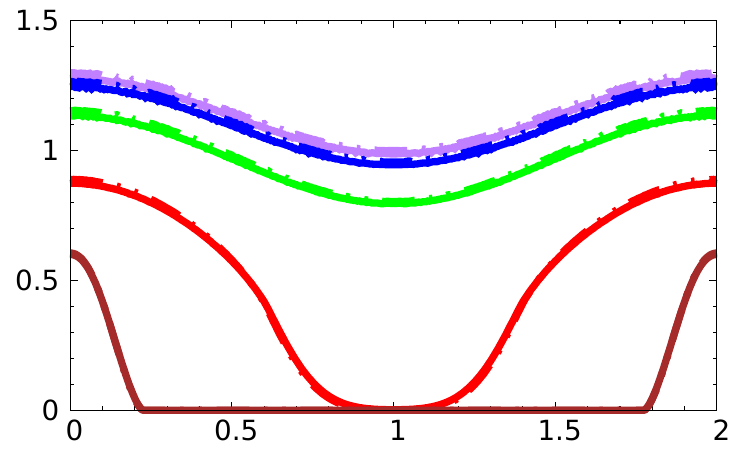}} &
            {\includegraphics[width=3.1cm,height=2.2cm,trim=0 0 0 0]{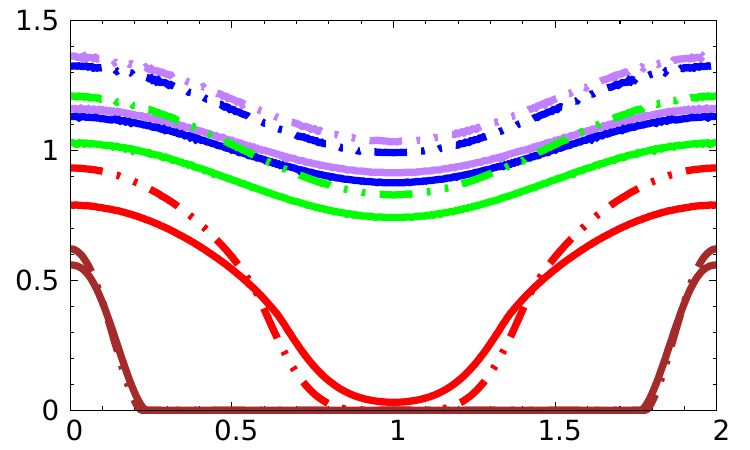}} &
            \rotatebox{90}{\hspace{0.3cm}$\chi=\frac{\pi}{9}$, $\xi=\frac{\pi}{6}$} \\
            \rotatebox{90}{\hspace{0.5cm}Flux ($\cdot100$)} &  
            {\includegraphics[width=3.1cm,height=2.2cm,trim=0 0 0 0]{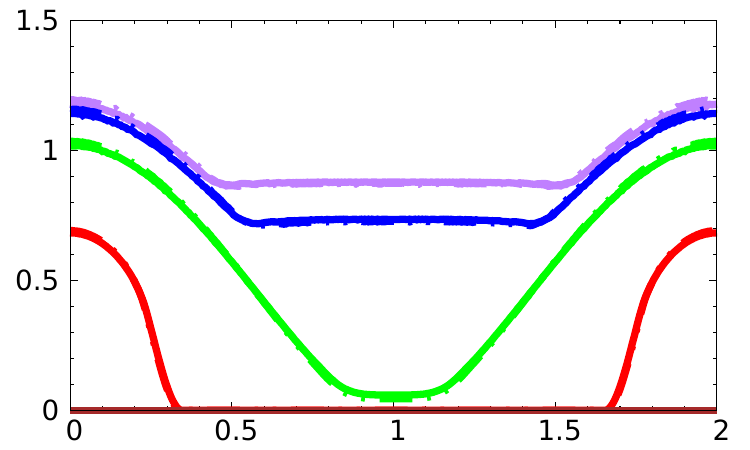}} &
            {\includegraphics[width=3.1cm,height=2.2cm,trim=0 0 0 0]{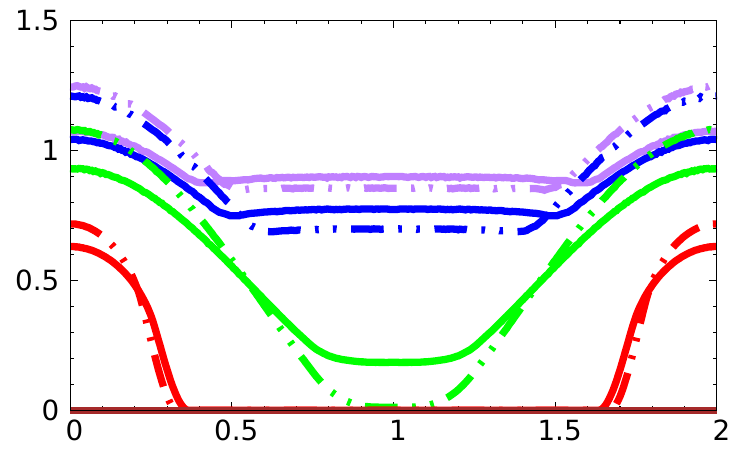}} &
            \rotatebox{90}{\hspace{0.3cm}$\chi=\frac{\pi}{6}$, $\xi=\frac{\pi}{3}$} \\
            \rotatebox{90}{\hspace{0.5cm}Flux ($\cdot100$)} &  
            {\includegraphics[width=3.1cm,height=2.2cm,trim=0 0 0 0]{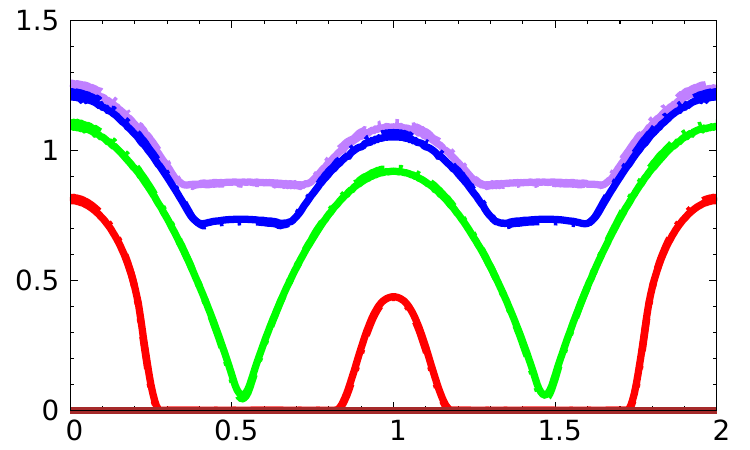}} &
            {\includegraphics[width=3.1cm,height=2.2cm,trim=0 0 0 0]{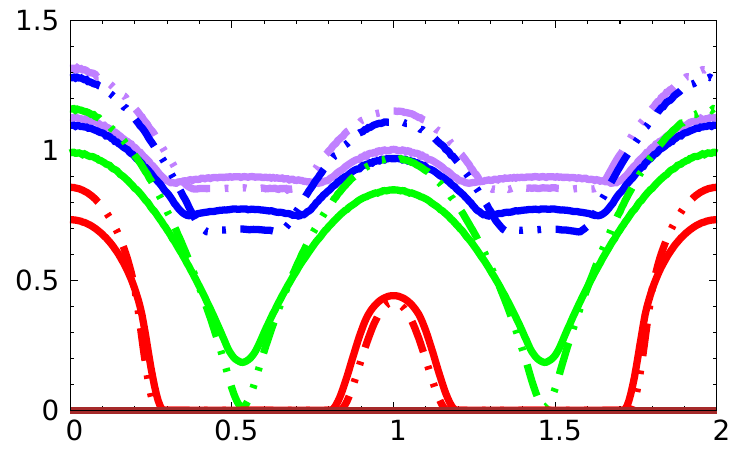}} &
            \rotatebox{90}{\hspace{0.3cm}$\chi=\frac{\pi}{3}$, $\xi=\frac{4\pi}{9}$} \\
            \rotatebox{90}{\hspace{0.5cm}Flux ($\cdot100$)} &  
            {\includegraphics[width=3.1cm,height=2.2cm,trim=0 0 0 0]{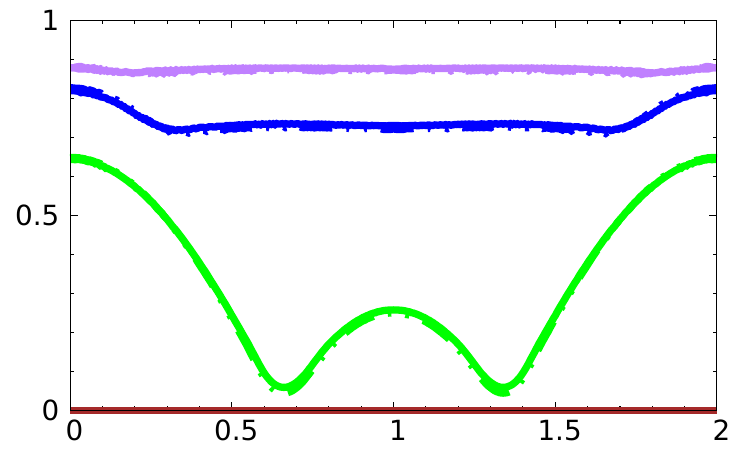}} &
            {\includegraphics[width=3.1cm,height=2.2cm,trim=0 0 0 0]{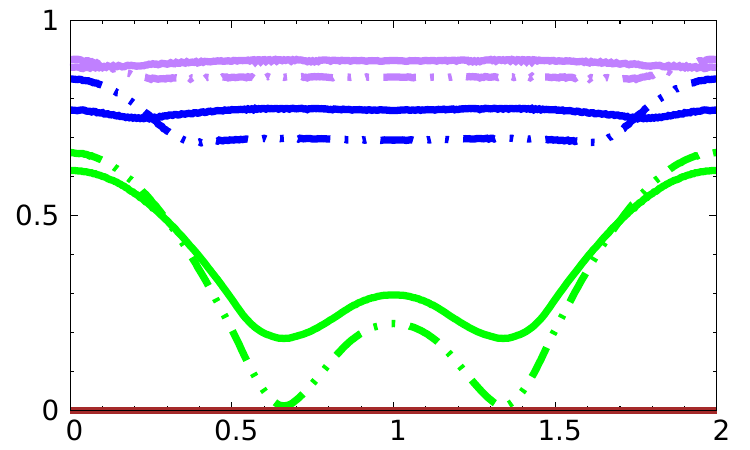}} &
            \rotatebox{90}{\hspace{0.3cm}$\chi=\frac{\pi}{9}$, $\xi=\frac{4\pi}{9}$} \\
            & {~~$\Omega t/\pi$} & {~~$\Omega t/\pi$} &
        \end{tabular}
        \caption{Pulse profile for antipodal caps. $h=3$, $R/r_h=3$. Purple, blue, green, red and brown lines correspond to $\epsilon=0.00$, $0.30$, $0.60$, $0.90$ and $0.99$ respectively. Solid lines correspond to $q^*>0$ while dashed lines to $q^*<0$. The rows, from top to bottom, correspond to Beloborodov classes I, II, III and IV. Both the $q^*$ and $\epsilon$ modify the morphology according to the metric. As $q^*$ increases, $R$ decreases.}
        \label{fig:PP_DHS_h=3}
    \end{figure}
    
    \begin{figure}
        \centering
        \begin{tabular}{cccccc}
            & $|q^*|=(0.25M)^2$ & $|q^*|=(0.75M)^2$ & \\
            \rotatebox{90}{\hspace{0.5cm}Flux ($\cdot100$)} &  
            {\includegraphics[width=3.1cm,height=2.2cm,trim=0 0 0 0]{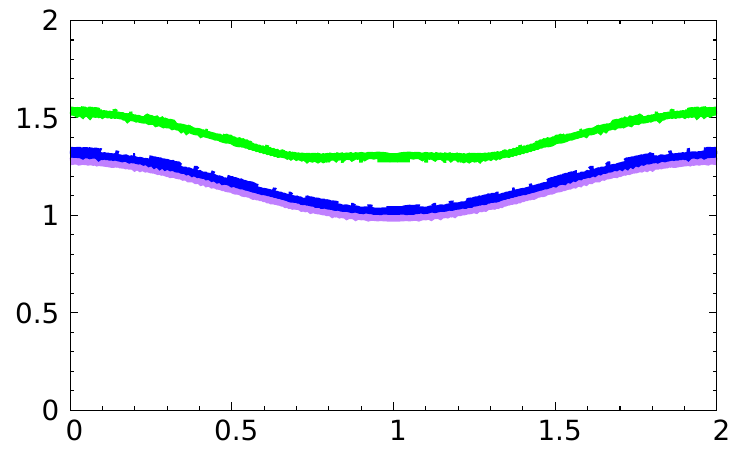}} &
            {\includegraphics[width=3.1cm,height=2.2cm,trim=0 0 0 0]{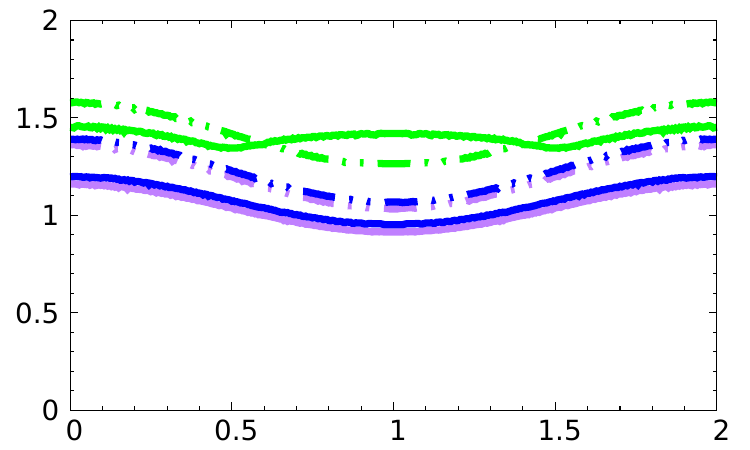}} &
            \rotatebox{90}{\hspace{0.3cm}$\chi=\frac{\pi}{9}$, $\xi=\frac{\pi}{6}$} \\
            \rotatebox{90}{\hspace{0.5cm}Flux ($\cdot100$)} &  
            {\includegraphics[width=3.1cm,height=2.2cm,trim=0 0 0 0]{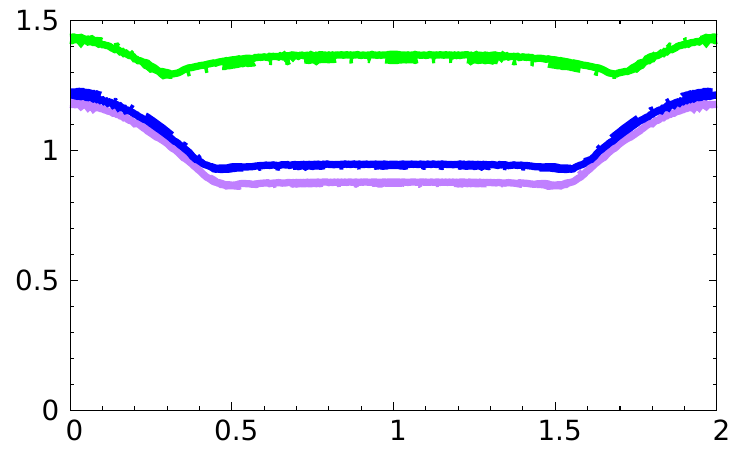}} &
            {\includegraphics[width=3.1cm,height=2.2cm,trim=0 0 0 0]{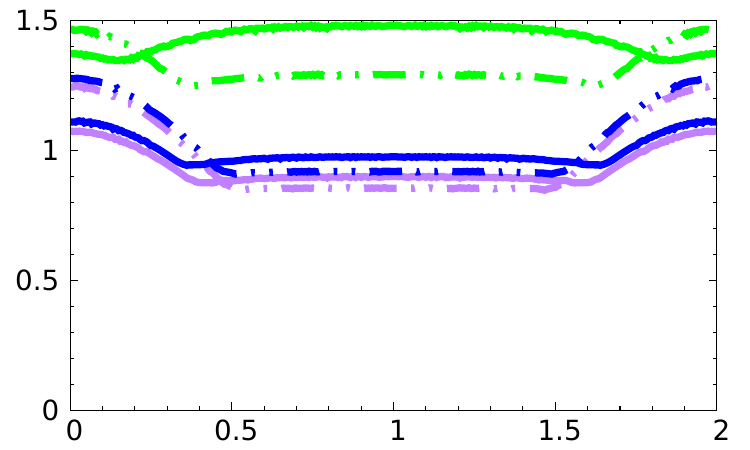}} &
            \rotatebox{90}{\hspace{0.3cm}$\chi=\frac{\pi}{6}$, $\xi=\frac{\pi}{3}$} \\
            \rotatebox{90}{\hspace{0.5cm}Flux ($\cdot100$)} &  
            {\includegraphics[width=3.1cm,height=2.2cm,trim=0 0 0 0]{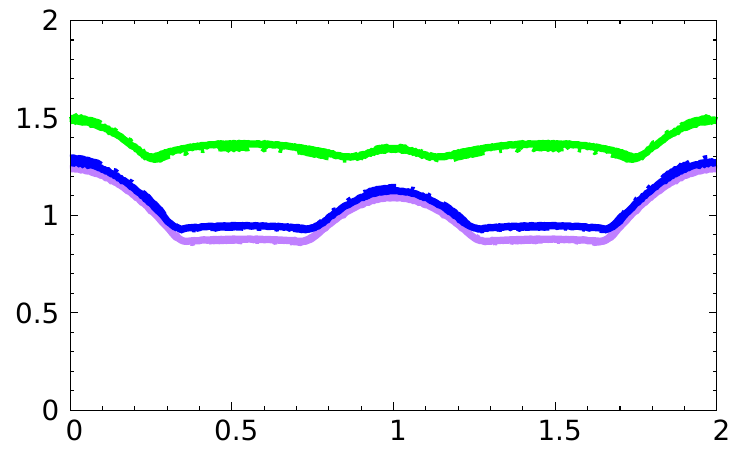}} &
            {\includegraphics[width=3.1cm,height=2.2cm,trim=0 0 0 0]{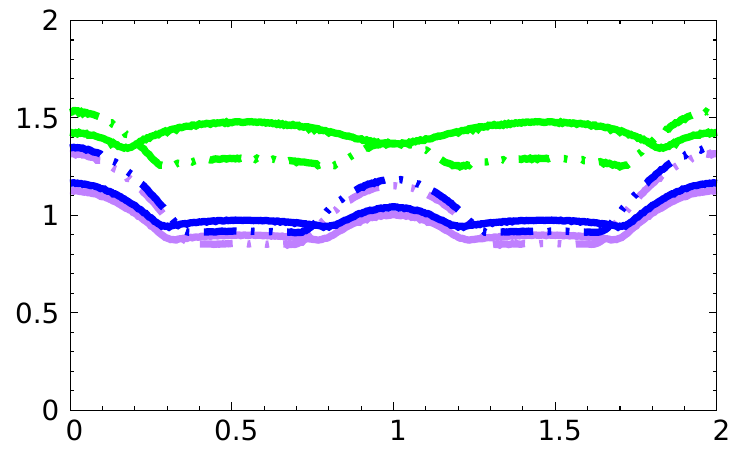}} &
            \rotatebox{90}{\hspace{0.3cm}$\chi=\frac{\pi}{3}$, $\xi=\frac{4\pi}{9}$} \\
            \rotatebox{90}{\hspace{0.5cm}Flux ($\cdot100$)} &  
            {\includegraphics[width=3.1cm,height=2.2cm,trim=0 0 0 0]{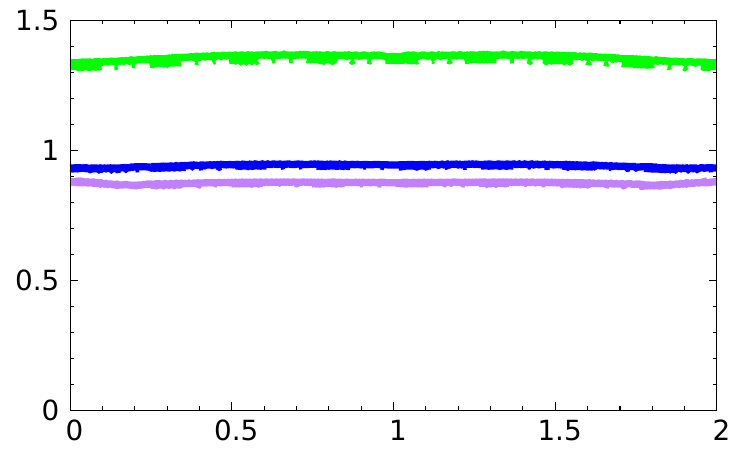}} &
            {\includegraphics[width=3.1cm,height=2.2cm,trim=0 0 0 0]{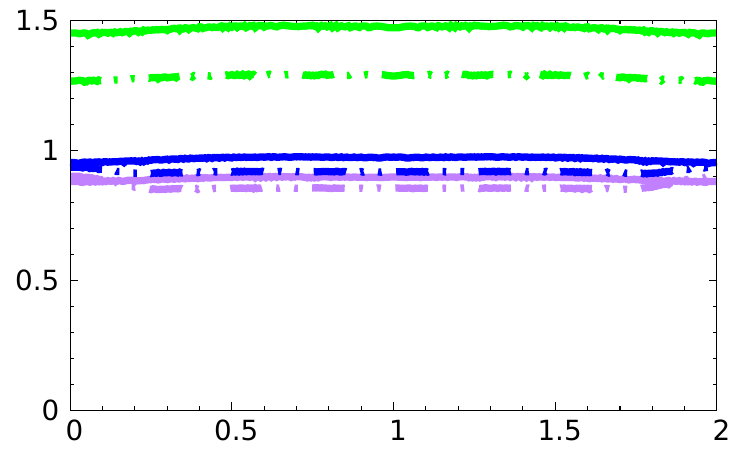}} &
            \rotatebox{90}{\hspace{0.3cm}$\chi=\frac{\pi}{9}$, $\xi=\frac{4\pi}{9}$} \\
            & {~~$\Omega t/\pi$} & {~~$\Omega t/\pi$} &
        \end{tabular}
        \caption{Same configuration as in Fig. \ref{fig:PP_DHS_h=3} but for $h=0$. We see that, for a constant plasma density, low frequencies do not satisfy the propagation conditions in the vicinity of the star.}
        \label{fig:PP_DHS_h=0}
    \end{figure}
    
    Up to this point, these results are consistent with those found in \cite{Pechenick_1983} and \cite{Dabrowski_1995}, considering the effects of compactness and charge on the pulse profile.
    
    Finally, we vary the angles $\chi$ and $\xi$ to study the frequency-dependent morphologies of the pulses for a variety of angles between the line of sight and the rotation axis and between the rotation axis and the position of the emitting cap. Employing the classification scheme proposed in \cite{Beloborodov_2002}, in Figs.  \ref{fig:PP_DHS_h=3} and \ref{fig:PP_DHS_h=0} we show examples of each pulse class for antipodal configurations. 
    For this, we use $R/r_h=3$, with $r_h$ being the radius of the event horizon, given by $r_S=2M$ for the $Sch$ metric, while for $RN$-like metrics, we have
    \begin{equation}
        r_h=\frac{r_S+\sqrt{r_S^2-4q^*}}{2}.
        \label{eq:r0}
    \end{equation}
    Note that both radius ratios, $R/2M$ and $R/r_h$, can be used to define the compactness of the star, depending on whether we want to express it in terms of its mass or the radius of its event horizon. In this work, we will use both alternatives for different purposes.
    
    In the Beloborodov's scheme, class I has a primary cap that is always visible, while the antipodal is invisible at all times. In class II, the primary cap is always visible and the antipodal is only visible for part of the time. Class III are generated when each of the caps can be visible or invisible depending on the phase and class IV is when both caps are visible at all times, producing a relatively constant profile. The angular configurations $(\chi,\xi)=(\pi/9,\pi/6)$, $(\pi/6,\pi/3)$, $(\pi/3,4\pi/9)$ and $(\pi/9,4\pi/9)$ were used, corresponding respectively to the four classes just introduced when $\epsilon=0$ and $q^*=0$. Note that, in order to represent all the classes, the star must have a low compactness, thus containing a region of its surface out of reach of the observer, where the caps are invisible. With this in mind, we use $R/r_h=3$.  
    
    Note that the considered stars do not have the same radius for $q^*<0$ and $q^*>0$. It can be seen from Eq. \eqref{eq:r0} that $r_h$ is smaller for $q^*>0$ than for $q^*<0$, so, by keeping the $R/r_h$ ratio identical, the surface radius $R$ in $q^*>0$ will be smaller.
    
    By observing all these results, it can be seen that both the metric model, the plasma frequency and the charge have influence on what will be the final classification of the pulse profile. We note that class I profiles present sinusoidal like oscillations around a positive mean value greater than the oscillation amplitude. 
    Classes II and III combine periods of sinusoidal oscillations and periods of constant flux.
    Finally, class IV profiles are almost constant, presenting slight deformations near the extreme values of the phase.

\section{Analytical Approach}
\label{c2s5}

    In a static gravitational field with spherical symmetry, the exact deflection angle $\beta=\theta-\delta$ is given by an elliptic or similar integral, corresponding to Eq. \eqref{eq:D}, which generally has no analytical solution. 
    The formalism employed in the previous sections correctly describes the physics of light rays in the vicinity of neutron stars of any size, where the deflection angle $\beta$ can become larger than $\pi/2$. 
    
    However, the integrals to be solved can be complicated and require a high computational cost. Expressions such as Eq. \eqref{eq:t(b)} or Eq. \eqref{eq:flux} may not represent a great challenge when one has a single spherical cap that emits isotropically, but the issue becomes more complicated when considering multiple sources of irregular shapes emitting anisotropically, such as the cases discussed in \cite{Pechenick_1983, Dabrowski_1995, Turolla_2013, Sotani_2020}. When working with a more realistic model, solving the flux integral to obtain the pulse profile can become a real inconvenience, and represent several hours of computation.
    
    In this section we will focus on finding analytical approximations that allow us to generalize to plasma environment situations expressions already known for the case of pure gravity and thus simplifying the calculations of luminosity curves and reduce their computational cost, without great loss of accuracy or generality, facilitating a clear understanding of the deflection effects. 
    
    Other analytical developments neglecting plasma modifications to the orbits of light rays, (similar to the ones we developed here) can be found in \cite{Silva_2019} for scalar-tensor theories of gravity, or in \cite{Sotani_2017} for a generic range of metrics. 
    
\subsection{Trajectory Equation}
\label{c2s5ss1}
    
    A simple formula that relates $\delta$ (the photon emission angle with respect to the stellar surface normal) to $\theta$ (the angle that gives the position on the stellar surface from which the emission occurs, given by Eq. \eqref{eq:D}), and which replaces the elliptic or similar integral with high accuracy, is the so called cosine relation introduced in \cite{Beloborodov_2002}\footnote{
    Beloborodov finds this relation for the particular Schwarzschild case, with $A(R)=1-2M/R$, however it is possible to verify that at least for the metrics worked here the relation is generalizable with their respective $A(R)$. Note also, that there are several analytical approaches in the literature superior to the one found by Beloborodov, such as the one shown in \cite{Sotani_2017}. However, they all agree to first order with Eq. \eqref{eq:Belo}}
    \footnote{\label{footbel}
    For $R\geq2r_S$, this Eq. estimates the deflection angle $\beta=\theta-\delta$ with high accuracy, with a maximum relative error of order $3\%$ for $R=3r_S$. Standard pulsar models, on the other hand, predict $R\geq2r_S$ with a typical $R\approx3r_S$.}
    \begin{equation}
        1-\cos\delta=(1-\cos\theta)A(R),
        \label{eq:Belo}
    \end{equation}
    being $A$ the metric element $-g_{tt}$, which must be evaluated at the radial coordinate where the angles $\delta$ and $\theta$ are being measured, in this case $R$. Eq. \eqref{eq:Belo}, however, is valid for any point along the trajectory, being $\delta$ the angle between the position vector of the ray and the tangent to its trajectory at that point.  
    
    A limitation of Beloborodov's approach is that it does not take into account the presence of plasma. For this reason, we have been forced to imitate its development, introducing now a spherically symmetric plasma distribution according to Eq. \eqref{eq:refindnum}. The influence of the plasma has been retained up to order $\epsilon^2$, obtaining a correction for the Beloborodov cosine relation that can be expressed in general form as
    {
    \begin{equation}
        (1-\cos\delta)(1-\epsilon^2P^{q^*}_h(r))=(1-\cos\theta)A(r),
        \label{eq:BeloPlas}
    \end{equation}
    where $P^{q^*}_h(r)$ is a plasma correction factor which depends on the metric (through $q^*$ and $M$) and the power $h$ in the plasma density profile,
    }
    {
    \begin{equation}
        \begin{split}
        P_h^{q^*~}(r)  &= \frac{1}{A(R)}\left(\frac{R}{r}\right)^h\left[\frac{h}{h+1}-\frac{h+1}{h+2}\frac{r_s}{r}+\frac{h+2}{h+3}\frac{q^*}{r^2}\right]. \\
        \end{split}
        \label{eq:AA_Ph}
    \end{equation}
    }
    Note that Eq. \eqref{eq:BeloPlas} tends to Eq. \eqref{eq:Belo} in the low plasma density limit {or higher observational frequencies} ($\epsilon\rightarrow0$), so that both descriptions coincide in pure gravity.  
    It can be seen that this factor vanishes ($P_h^{q^*}(r)\rightarrow0$) as we move far enough away from the star ($r\rightarrow\infty$), which corresponds to a plasma density that decays asymptotically to zero.  At the same time, as $q^*$ goes to zero, $P^{q^*}_h(r)$ tends to its Schwarzschild value $P^{q^*=0}_h(r)\equiv P_h^{Sch}(r)$ when $q^*\rightarrow0$.
    
    For the reader's convenience, we list below the steps to be followed for obtaining the plasma correction factors $P^{q*}_h$.
    
    \begin{enumerate}
        \item Express $d\phi/du$ (Eq. \eqref{eq:dpdu}) in terms of $\sqrt{1-\cos^2\delta}$.
        \item Extend the results in Taylor's series in terms of $\epsilon$, 
        keeping up to order $\epsilon^2$.
        \item Expand the above expression as a series around $\cos\delta=1$ and express the result as a polynomial.
        \item Integrate between $u=1/r$ and $u=0$ (Eq. \eqref{eq:D}), obtaining an approximation for $\theta$.
        \item Evaluate $\cos\theta$ and perform Taylor series around $\cos\delta=1$, retaining only until first order terms.
        \item Expand the result by Taylor as a power series of $\varepsilon$, retaining up to second order terms.
    \end{enumerate}
    
    Eq. \eqref{eq:BeloPlas} then gives us the relationship between the angles $\theta$ and $\delta$ for all $r$ along the ray trajectory. On the other hand, Eq. \eqref{eq:b} allows us to express $\delta$ in terms of the impact parameter $b$ (or $x$) and the radius $r$. Therefore, by combining both expressions, we obtain an equation that directly relates $\theta$ to $r$ for every point on the photon's trajectory,
    \begin{equation}
    \begin{split}
        \cos\theta_{a}= &1- \\
                        &\left(1-\sqrt{1-x^2\left(\frac{n_0}{n(r)}\right)^2\frac{A(r)}{C(r)}}\right)\frac{1-\epsilon^2P^{q*}_h(r)}{A(r)}        
    \end{split}
        \label{eq:AA_trayectoria}
    \end{equation}
    where the subscript $a$ indicates that it is an analytical approximation. This is an approximate equation for the trajectory, which will allow us to find the set of coordinates $(r,\theta)$ that describes the photon trajectory.
    While we have previously used Eq. \eqref{eq:D} and \eqref{eq:b} to describe the parameters of the photon at the instant it leaves the stellar surface, we should clarify that these are also valid for the rest of its trajectory. Thus, $\theta$ and $\theta_a$ correspond to the emission angle of the photon on the pulsar surface only when evaluated at $r=R$, while otherwise they describe the angular coordinate of the trajectory corresponding to that $r$.
    It can be seen that in \cite{Beloborodov_2002} the relation found is the inverse, i.e., Beloborodov finds an expression for $r$ as a function of $\theta$. Since the plasma corrections introduce new dependencies on $r$, and since these are, along with the metric functions, dependent on the specific spacetime model being used, we find it convenient (and much easier) to use Eq.\eqref{eq:AA_trayectoria}.
    
    It is now necessary to verify that the new expressions generalize the approximation presented by Beloborodov, improving it substantially when considering a plasma environment.
    
\begin{figure}
        \centering
        \begin{tabular}{cccc}
            & Sch & RN-like & \\
            \rotatebox{90}{\hspace{0.95cm}$\theta/\pi$} &  
            {\includegraphics[width=3.2cm,height=2.3cm,trim=0 0 0 0]{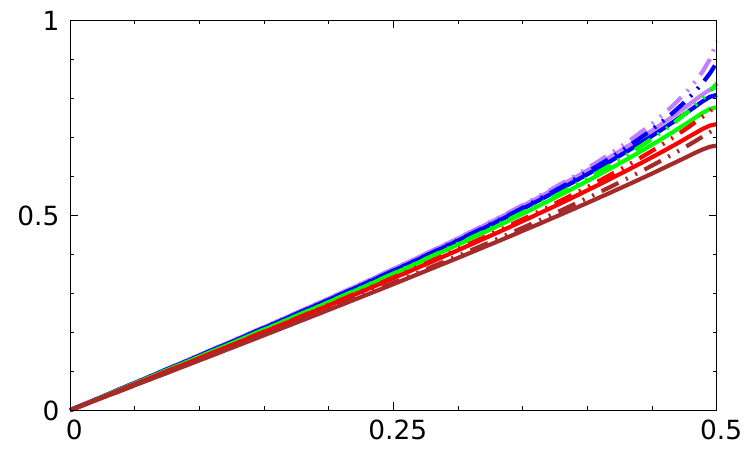}} &
            {\includegraphics[width=3.2cm,height=2.3cm,trim=0 0 0 0]{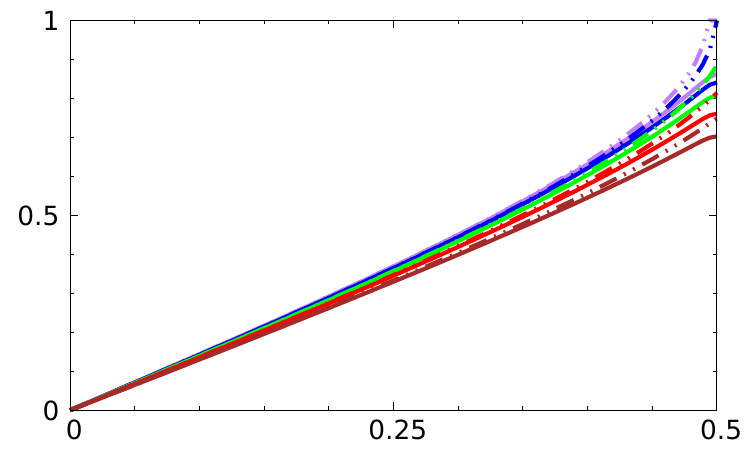}} &
            \rotatebox{90}{\hspace{0.85cm}$h=3$} \\
            \rotatebox{90}{\hspace{0.95cm}$\theta/\pi$} &  
            {\includegraphics[width=3.2cm,height=2.3cm,trim=0 0 0 0]{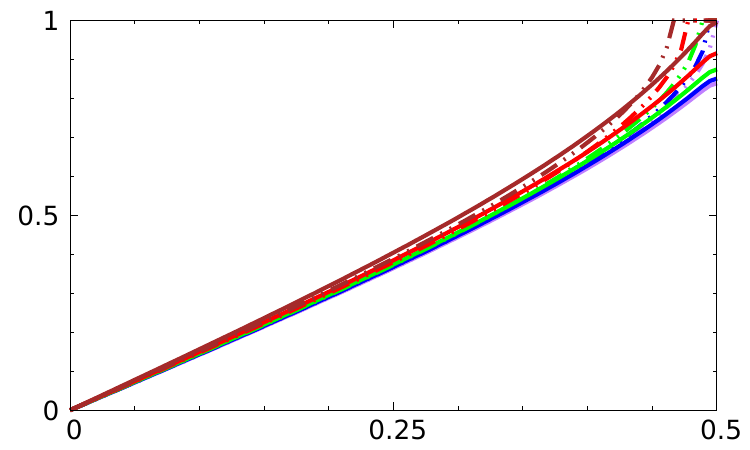}} &
            {\includegraphics[width=3.2cm,height=2.3cm,trim=0 0 0 0]{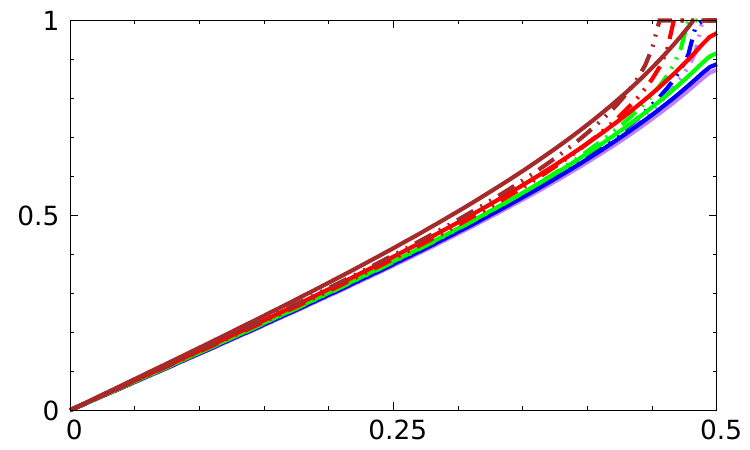}} &
            \rotatebox{90}{\hspace{0.85cm}$h=0$} \\
            & {~~$\delta/\pi$} & {~~$\delta/\pi$} &  
        \end{tabular}
        \caption{$\theta$ as a function of $\delta$. $q^*=-(0.50M)^2$, $R=4M$. The purple, blue, green, red and brown lines correspond respectively to $\epsilon=0.1$, $0.2$, $0.3$, $0.4$ and $0.5$. Continuous lines were obtained numerically [Eq. \eqref{eq:t(b)}] while dashed lines were obtained analytically from Eq. \eqref{eq:BeloPlas} (with plasma correction). Errors are minimum for small $\delta$ and grow with it, overestimating $\theta_{max}$.}
        \label{fig:AA_TR_tvd}
    \end{figure}

    With this purpose, we plot the $\theta$ and $\theta_a$ curves as a function of $\delta$, comparing the results obtained by numerical integration and by the analytical approximation, in order to get a general idea about the errors produced by our approximation.
    Since we want to know the ``standard'' errors introduced by our model, we take $R=4M$ (cf. footnote \eqref{footbel}) and $q^*=-(0.50M)^2$. On the other hand, since the development of the analytical approximation assumes small $\epsilon$, we only take values up to $\epsilon=0.5$. The results are shown in Fig. \ref{fig:AA_TR_tvd}.
    
    It can be seen that the deviations of the analytical trajectory are negligible when $\delta$ is small and that they increase as $\delta$ becomes larger, so studying the error produced for the marginal ray ($\delta=\pi/2$ and $b_{max}$) will be of great interest for our analysis. Let us note that the analytical approximation tends to overestimate the value of $\theta_{max}$, resulting in a larger visible area of the stellar surface. 
    Such a defect will have an impact on the pulse profiles, since a larger visible area implies a larger lensing contribution. Particularly for the constant plasma case ($h=0$) there is a range of $\delta$ values for which $\theta$ is underestimated by the analytical approximation. However, we see that as we get closer to $\delta=\pi/2$, $\theta$ becomes overestimated, {reaching in some cases the maximum value allowed by the approach, $\theta_a=\pi$. }
    
\begin{figure}
        \centering
        \begin{tabular}{cccc}
            & Sch & RN-like & \\
            \rotatebox{90}{\hspace{0.4cm}$\Delta\theta_{max}/\theta_{max}$} &  
            {\includegraphics[width=3.2cm,height=2.3cm,trim=0 0 0 0]{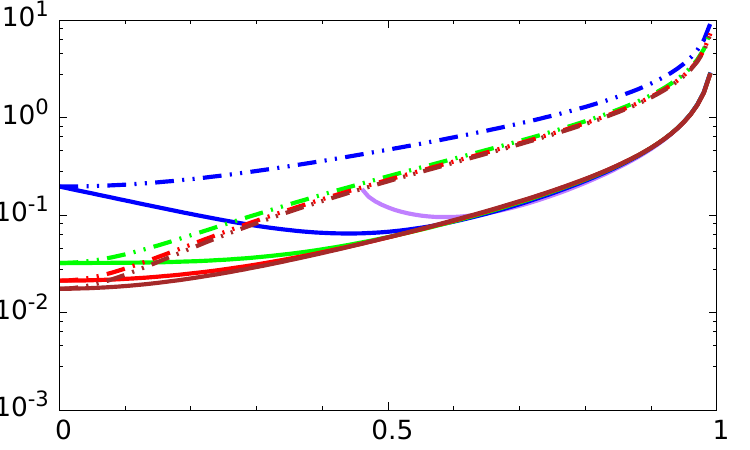}} &
            {\includegraphics[width=3.2cm,height=2.3cm,trim=0 0 0 0]{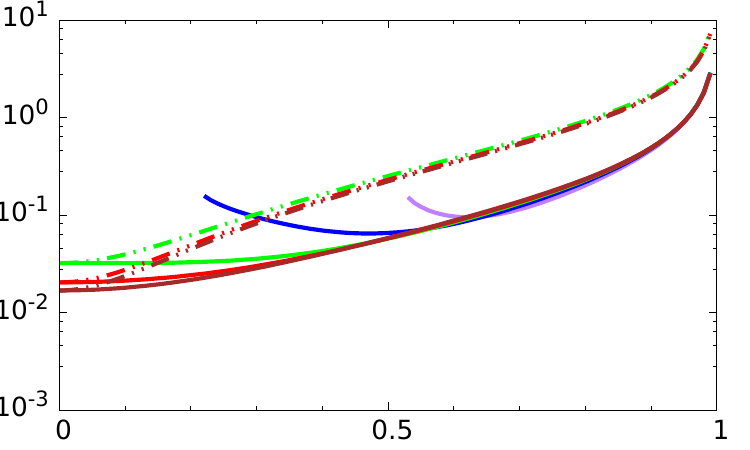}} &
            \rotatebox{90}{\hspace{0.85cm}$h=3$} \\
            \rotatebox{90}{\hspace{0.4cm}$\Delta\theta_{max}/\theta_{max}$} &   
            {\includegraphics[width=3.2cm,height=2.3cm,trim=0 0 0 0]{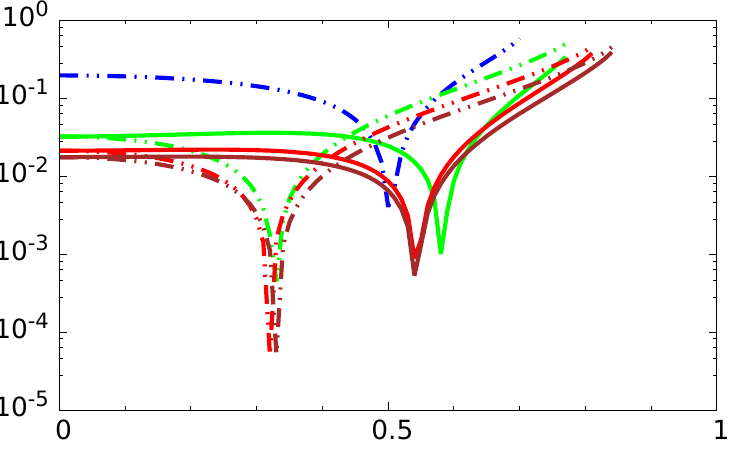}} &
            {\includegraphics[width=3.2cm,height=2.3cm,trim=0 0 0 0]{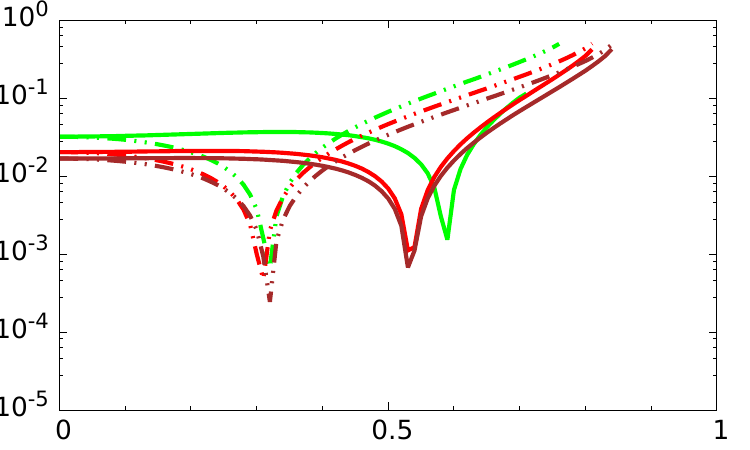}} &
            \rotatebox{90}{\hspace{0.85cm}$h=0$} \\
            & {~~$\epsilon$} & {~~$\epsilon$} &  
        \end{tabular}
        \caption{Relative error in $\theta_{max}$ ($|\theta_{max,a}-\theta_{max}|/\theta_{max}$) as a function of $\epsilon$ for $r=R$. $q^*=-(0.50M)^2$. The purple, blue, green, red and brown lines correspond respectively to $R=3.5$, $4$, $5$, $6$ and $7$. Continuous lines were obtained from Eq. \eqref{eq:BeloPlas} (with plasma correction) while dashed lines from Eq. \eqref{eq:Belo} (without plasma correction).}
        \label{fig:AA_TR_tdm}
    \end{figure}

    Let's now analyze the dependence of the error in $\theta_a$ with $\epsilon$ for $\delta=\pi/2$ and $q^*=-(0.50M)^2$, comparing the results obtained with and without the plasma correction. Also, we compare the errors in different $R$, to observe how the approach improves as the compactness decreases. The error curves obtained are shown in Fig. \ref{fig:AA_TR_tdm}. From this figure we can conclude that the analytical approximation introduced in Eq. \eqref{eq:BeloPlas} describes more accurately the photon trajectory in plasma environments than the cosine relation proposed by Beloborodov (Eq. \eqref{eq:Belo}). Except for some specific behaviors obtained for $h=0$, where anomalies generally occur, we see that the accuracy improves by up to an order of magnitude near $\epsilon=0.5$. For small $\epsilon$ values, as expected, both estimates approach until they coincide at $\epsilon=0$. On the other hand, for $\epsilon\geq0.5$ the accuracy of our method is still superior, although the error committed starts to become excessive.
    
    We see that, for small $\epsilon$, the error in $\theta_{max}$ strongly depends on the radius $R$, being considerably smaller for less compact stars. It is remarkable that, for smaller radius stars ($R\leq4M$), the error decreases with increasing $\epsilon$ until it reaches its minimum value around $\epsilon\approx0.5$, where it becomes very similar to the error obtained for larger radius stars. For larger $\epsilon$, the error exhibits a weak dependence on $R$.
    
    As always, the $h=0$ case has its peculiarities. First, we see that there are certain peaks where the error decays drastically, which is explained by the fact that at those points the plasma correction goes from underestimating to overestimating the value of $\theta$, causing the error to decay more by coincidence than by merit. We also see that the error stops being computed for $\epsilon\approx0.75$, which happens because for larger values the propagation conditions expressed in Eq. \eqref{eq:conpro1} are no longer satisfied. On the other hand, we see that in the most compact stars the error is not computed for small values of $\epsilon$, which arises because the argument within $\cos^{-1}$ in Eq. \eqref{eq:AA_trayectoria} is greater than one.
    
    \begin{figure}
        \centering
        \begin{tabular}{cccc}
            & Sch & RN-like & \\
            \rotatebox{90}{\hspace{0.8cm} $\Delta\theta/\theta$} & 
            {\includegraphics[width=3.2cm,height=2.3cm,trim=0 0 0 0]{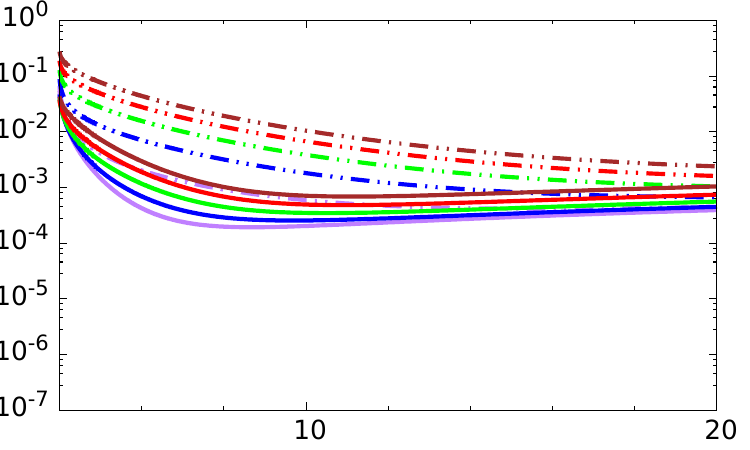}} &
            {\includegraphics[width=3.2cm,height=2.3cm,trim=0 0 0 0]{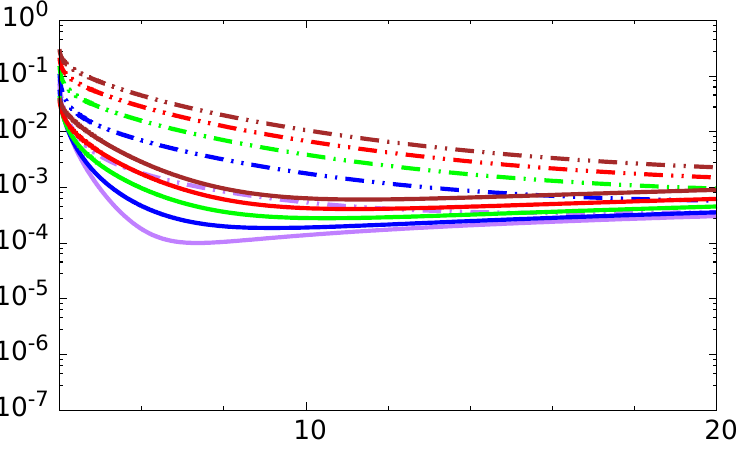}} &
            \rotatebox{90}{\hspace{0.85cm}$h=3$} \\
            \rotatebox{90}{\hspace{0.8cm} $\Delta\theta/\theta$} & 
            {\includegraphics[width=3.2cm,height=2.3cm,trim=0 0 0 0]{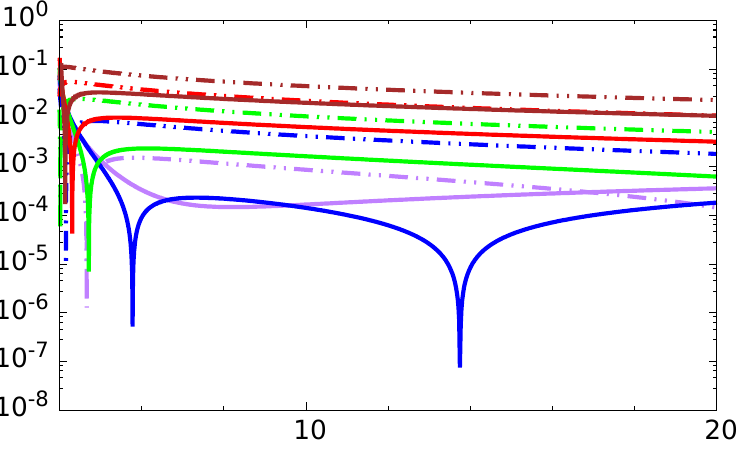}} &
            {\includegraphics[width=3.2cm,height=2.3cm,trim=0 0 0 0]{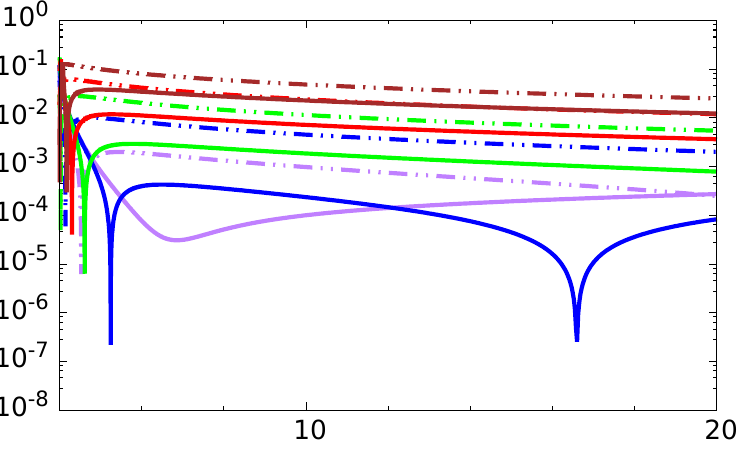}} &
            \rotatebox{90}{\hspace{0.85cm}$h=0$} \\
            & {~~$r$} & {~~$r$} &  
        \end{tabular}
        \caption{Relative error in $\theta$ ($|\theta_a-\theta|/\theta$) as a function of the radius $r$ along the marginal trajectory [$\delta(R)=\pi/2$]. $q^*=-(0.50M)^2$, $R=4M$. The purple, blue, green, red and brown lines correspond respectively to $\epsilon=0.1$, $0.2$, $0.3$, $0.4$ and $0.5$. Continuous lines were obtained from Eq. \eqref{eq:BeloPlas} (with plasma correction) while dashed lines from Eq. \eqref{eq:Belo} (without plasma correction). At all times plasma corrections increase the accuracy of the approach.}
        \label{fig:AA_TR_tvr}
    \end{figure}
    
    Finally, we study how the error in $\theta_a$ changes as we move along the trajectory. For this, we take again
    $R=4M$, $q^*=-(0.50M)^2$ and $\delta(R)=\pi/2$, comparing the results obtained with and without consideration of plasma corrections for different values of $\epsilon$. These results are shown in Fig. \ref{fig:AA_TR_tvr}. Once again, we see that the error decreases significantly when applying the plasma correction.
    Note also that the accuracy increases progressively as $r$ increases. This is because, in a sense, we are sending the rays from the observer to the star.\footnote{
    To construct the analytical photon trajectory, we first calculate the impact parameter from Eq. \eqref{eq:b} by taking $\delta=\pi/2$, being this the same as the impact parameter of the numerical trajectory. In this way, both trajectories coincide when reaching the detector at $r_O$ and separate as they approach the pulsar.
    }.
    With this in mind, it is evident that the deviation increases as $r$ decreases since we accumulate errors due to interaction with plasma. We also see that the error is smaller for lower $\epsilon$, which is based on the fact that plasma corrections in the Beloborodov formula were performed by conserving only terms of order $\epsilon^2$. 
    These results are consistent with those shown in \cite{Beloborodov_2002}.
    
\subsection{Pulse Profile}
\label{c2s5ss2}

    Making use of the analytical expressions introduced in the previous subsection, we will now focus on finding an analytical expression for the observed flux, which was previously calculated by numerical integration according to Eq. \eqref{eq:flux}. For this, we follow the paper of \cite{Turolla_2013}, introducing plasma correction and using the specific intensity as was presented in SEq. \ref{c2s4}. 
    The procedure is also similar to the one found in \cite{Silva_2019}.
    With these considerations, we can express the 
    flux differential as\footnote{
    This is the generalization of Eq. (3) of \cite{Turolla_2013}.
    }
    \begin{equation}
        dF=I_0\frac{A^{1/2}(R)}{C(R)}\frac{n^2(R)}{n_0^2}f_B(\delta)\cos\delta\frac{d\cos\delta}{d\cos\theta}dS,
        \label{eq:AA_dF}
    \end{equation}
    being $I_0$ a constant proportional to the surface intensity emitted by the cap, whose value can be taken as required. For example, in \cite{Turolla_2013} a Planckian emission at uniform $T$ temperature is considered, although other choices for $I_0$ are equally valid. The total observed flux is obtained by integrating the above expression over the entire visible area of the emitting cap, which we denote $S_V$. We remark that this approximation is valid only if $I_0=const$ over the entire cap surface $S_V$. Thus, the flux results as follows 
    \begin{equation}
        F=I_0\frac{A^{1/2}(R)}{C(R)}\frac{n^2(R)}{n_0^2}\int_{S_V} f_B(\delta)\cos\delta\frac{d\cos\delta}{d\cos\theta}dS.
        \label{eq:AA_F1}
    \end{equation}
    This expression is consistent with those that can be found in \cite{Sotani_2017} and \cite{Silva_2019} for point caps.
    
    Using the plasma corrected cosine relation (Eq. \eqref{eq:BeloPlas}) we can calculate $d\cos\delta/d\cos\theta=A(R)/(1-P^{q*}_h(R)\epsilon^2)$. Then, expressing $\cos\delta$ in terms of $\theta$, we arrive at the following expression
    \begin{equation}
    \begin{split}
        F&=I_0\frac{A^{3/2}(R)}{C(R)}\frac{n^2(R)}{n_0^2}\frac{1}{1-P^{q*}_h(R)\epsilon^2}\times\\
&\int_{S_V}f_B[\delta(\theta)]\left[
        \left(1-\frac{A(R)}{1-P^{q*}_h(R)\epsilon^2}\right)\right.\\
        &\left.+
        \left(\frac{A(R)}{1-P^{q*}_h(R)\epsilon^2}\right)\cos\theta
        \right]dS.
        \label{eq:AA_F2}
        \end{split}
    \end{equation}
    
    The flux is then expressed as the sum of two contributions, the first proportional to the surface area and the second to the projected area of the visible part of the emitting region, both modulated by the anisotropic emission function $f_B(\delta)$. Let us note that, for $A(R)\rightarrow1$ and $\epsilon\rightarrow0$, this expression is analogous to the Newtonian result, obtaining the projection of the observed surface. The problem of calculating the flux, once the geometry is fixed, is therefore reduced to determining $S_V$ and evaluating the following integrals
    \begin{equation}
        \begin{split}
        I_p &= \int_{S_V}f_B[\delta(\theta)]\cos\theta\sin\theta d\theta d\phi, \\
        I_s &= \int_{S_V}f_B[\delta(\theta)]          \sin\theta d\theta d\phi.
        \end{split}
        \label{eq:AA_IpIs}
    \end{equation}
    
    Let us consider first a simple example, a uniform circular cap of half-opening $\theta_c$ centered at $\theta_0$. For simplicity, we take $\theta_c\leq\pi/2$ and $0\leq\theta_0\leq\pi$. The integral in $\phi$ in Eq. \eqref{eq:AA_IpIs}, which yields the angular length of the linear intersection between the parallel given by $\theta$ and the emission cap, was solved earlier, resulting in the function $h(\theta,\theta_c,\theta_0)$ which is expressed in Eq. \eqref{eq:h2} \cite{Turolla_2013}. 
    Thus, it only remains to find the $\theta_{min}$ and $\theta_{max}$ limits of the interval of integration in $\theta$ for Eq. \eqref{eq:AA_IpIs}.
    
    To be visible to the observer, a point on the stellar surface must satisfy the condition $\delta\leq\pi/2$. The extreme case occurs when photons leave the star with a trajectory tangential to its surface. These trajectories correspond to the maximum observable angle of the star, $\theta_F$ (we change the notation to avoid confusion with the upper limit of integration, which we will introduce later), which can be obtained from Eq. \eqref{eq:BeloPlas} by taking $\delta=\pi/2$, resulting in
    \begin{equation}
        \theta_F=\cos^{-1}\left(1-\frac{1-P^{q*}_h(R)\epsilon^2}{A(R)}\right),
        \label{eq:AA_t_F}
    \end{equation}
    from which the following propagation conditions can be deduced
    \begin{equation}
        \frac{1-2A(R)}{P^{q*}_h(R)}\leq\epsilon^2\leq\frac{1}{P^{q*}_h(R)}.
        \label{eq:AA_condprop}
    \end{equation}
    
    Note that this restricts the compactness of the star to cases where not every point on its surface is visible to the observer, being the limiting case precisely $\theta_F=\pi$. This is in agreement with the limitations we have been imposing, since in the analytical development of the cosine relation $r_h/R$ is assumed to be small.
    
    Since any point on the surface of the star with $\theta>\theta_F$ will not be visible to the observer, and the cap extends from $\theta_0-\theta_c$ to $\theta_0+\theta_c$, it can be seen that the limits of integration will be given by~ \cite{Turolla_2013}
    \begin{equation}
        \begin{split}
        \theta_{min} &= \min[\theta_F,\max(0,\theta_0-\theta_c)], \\
        \theta_{max} &= \min(\theta_F,\theta_0+\theta_c).
        \end{split}
        \label{eq:AA_tminmax}
    \end{equation}
    
    The integrals to be solved can now be expressed as
    \begin{equation}
        \begin{split}
        I_1 &= 2\int f_B[\delta(\theta)] \cos^{-1}{\left(\frac{\cos{\theta_c}-\cos{\theta_0}\cos{\theta}}{\sin{\theta_0}\sin{\theta}}\right)}\sin\theta\cos\theta d\theta, \\
        I_2 &= 2\int f_B[\delta(\theta)] \cos^{-1}{\left(\frac{\cos{\theta_c}-\cos{\theta_0}\cos{\theta}}{\sin{\theta_0}\sin{\theta}}\right)}\sin\theta d\theta,
        \end{split}
        \label{eq:AA_I1I2}
    \end{equation}
    being
    \begin{equation}
        \begin{split}
        I_p &= I_1(\theta_{max})-I_1(\theta_{min}), \\
        I_s &= I_2(\theta_{max})-I_2(\theta_{min}).
        \end{split}
        \label{eq:AA_IpIs12}
    \end{equation}
    
    For simplicity, from now on we will consider an isotropic emission with $f_B(\delta)=1$. When the cap is completely visible and its center is over the visual ($\theta_0=0$), we have $h(\theta,\theta_c,\theta_0)=2\pi$, so that the integrals are trivially solved, resulting in
    \begin{equation}
        \begin{split}
        I_p &= \pi(\sin^2\theta_{max}-\sin^2\theta_{min})=\pi\sin^2\theta_c, \\
        I_s &= 2\pi(\cos\theta_{min}-\cos\theta_{max})=2\pi(1-\cos\theta_c).
        \end{split}
        \label{eq:AA_I_h=2pi}
    \end{equation}
    
    In the case that $\theta_0\neq0$, we must find the indefinite integrals $I_1$ and $I_2$. This calculation is developed in the appendix of ~\cite{Turolla_2013} and gives the following results\footnote{
    In Eq. \eqref{eq:AA_I1I2_} two sign typo corresponding to Eq. (20) of \cite{Turolla_2013} have been corrected.
    } 
    \begin{equation}
        \begin{split}
        I_1 =& \sin^2\theta\cos^{-1}\left[\frac{\cos\theta_c-\cos\theta_0\cos\theta}{\sin\theta_0\sin\theta}\right]\\
        &- \sin^2\theta_c\cos\theta_0\sin^{-1}\left[\frac{\cos\theta-\cos\theta_0\cos\theta_c}{\sin\theta_0\sin\theta_c}\right] \\
             &- \cos\theta_c\sqrt{-[\cos\theta-\cos(\theta_0+\theta_c)][\cos\theta-\cos(\theta_0-\theta_c)]}, \\
             & \\
        I_2 =&-2\cos\theta\cos^{-1}\left[\frac{\cos\theta_c-\cos\theta_0\cos\theta}{\sin\theta_0\sin\theta}\right]\\
              &+2\cos\theta_c\sin^{-1}\left[\frac{\cos\theta-\cos\theta_0\cos\theta_c}{\sin\theta_0\sin\theta_c}\right]\\ &+sign(\theta_0+\theta_c-\pi)\sin^{-1}\Delta_1\\ &+sign(\theta_0-\theta_c)\sin^{-1}\Delta_2,
        \end{split}
        \label{eq:AA_I1I2_}
    \end{equation}
    being
    \begin{equation}
    \begin{split}
        \Delta_1&=\frac{(\cos\theta_0\cos\theta_c+1)\cos\theta+\sin^2\theta_0-\cos^2\theta_c-\cos\theta_0\cos\theta_c}{(1+\cos\theta)|\sin\theta_0\sin\theta_c|},\\
        \Delta_2&=\frac{(\cos\theta_0\cos\theta_c-1)\cos\theta+\sin^2\theta_0-\cos^2\theta_c+\cos\theta_0\cos\theta_c}{(1-\cos\theta)|\sin\theta_0\sin\theta_c|},
    \end{split}
    \end{equation}
    where the arbitrary constant has been taken zero. It can be seen that, if the cap is completely visible, results in
    \begin{equation}
        \begin{split}
        I_p &=\pi\cos\theta_0\sin^2\theta_c, \\
        I_s &=2\pi(1-\cos\theta_c),
        \end{split}
        \label{eq:AA_I_cv}
    \end{equation}
    which hugely simplifies the calculations. Thus, Eq. \eqref{eq:AA_I1I2_} is only required when a part of the cap escapes the visible area, in which case it must be evaluated in $\theta_F$. 
    
    Introducing now the effective area
    \begin{equation}
        \begin{split}
        A_{eff}(\theta_c,\theta_0)& = C(R)\times \\
                                  &\left[
        \left(1-\frac{A(R)}{1-P^{q*}_h(R)\epsilon^2}\right)I_s+
        \left(\frac{A(R)}{1-P^{q*}_h(R)\epsilon^2}\right)I_p
        \right],
        \end{split}
        \label{eq:AA_A_eff}
    \end{equation}
    we can express the observed net flux as
    \begin{equation}
        F=I_0\frac{A^{3/2}(R)}{C(R)}\frac{n^2(R)}{n_0^2}\frac{1}{1-P^{q*}_h(R)\epsilon^2}A_{eff}(\theta_c,\theta_0).
        \label{eq:AA_F3}
    \end{equation}
    
    For more than one emitting cap, and considering that each one may have a different intensity, the total flux is obtained by simply adding the contributions of each cap,
    \begin{equation}
        F_{tot}=\frac{A^{3/2}(R)}{C(R)}\frac{n^2(R)}{n_0^2}\frac{1}{1-P^{q*}_h(R)\epsilon^2}\sum_i I_{0,i}A_{eff}(\theta_{c,i},\theta_{0,i}),
        \label{eq:AA_Ftot}
    \end{equation}
    being able to approximate in this way non-spherical and non-homogeneous caps, obtained by combination. 
    
    For the reader's convenience, we list the steps to be followed to obtain the analytical pulse profile of a neutron star.
    
    \begin{enumerate}
        \item Choose the metric elements $A(r)$, $B(r)$ and $C(r)$.
        \item Set the parameters $R$, $q^*$, $\epsilon$ and $h$, taking into account their respective constraints.
        \item Select the angle between the rotation axis and the line of sight $\xi$ ($0\leq\xi\leq\pi/2$), 
        \item Set for each cap its intensity $I_{0,i}$, the angle between the axis of rotation and the center of the cap $\chi_i$ ($0\leq\chi_i\leq\pi$) and its angular half-opening $\theta_{c,i}$ ($0\leq\theta_{c,i}\leq\pi$).
        \item Choose a surface emission function $f_B[\delta(x)]$.
        Note that if $f_B[\delta(x)]\neq1$. the Eqs. \eqref{eq:AA_I_h=2pi}, \eqref{eq:AA_I1I2_} and \eqref{eq:AA_I_cv} must be modified, and if the choice of $f_B$ leads to the integrals no longer being analytic, this procedure is no longer valid.
        \item Calculate the maximum visible angle on the surface of the star $\theta_F$ from Eq. \eqref{eq:AA_t_F}.
        \item Discretize the values of the phase $\gamma_p\in[0,2\pi]$ and evaluate on them $\theta_{0,i}(t)$ for each cap by Eq.~\eqref{eq:t0(t)}.
        \item For each cap, calculate the limits $\theta_{min,i}$ and $\theta_{max,i}$ according to Eq. \eqref{eq:AA_tminmax}.
        \item Evaluate $I_{1,i}$ and $I_{2,i}$ on the limits $\theta_{min,i}$ and $\theta_{max,i}$ to obtain the integrals $I_{p,i}$ and $I_{s,i}$ as given in Eq. \eqref{eq:AA_IpIs12}.
        \item Calculate the effective area of each cap $A_{eff,i}$ from Eq. \eqref{eq:AA_A_eff}. 
        \item Sum the contributions from each cap to obtain the total flux $F_{tot}$ at a given phase $\gamma_p$ according to Eq. \eqref{eq:AA_Ftot}.
        \item Repeat the procedure for each value of the phase $\gamma_p\in[0,2\pi]$ to be evaluated.
    \end{enumerate}
    
\begin{figure}
        \centering
        \begin{tabular}{cccc}
            & $h=3$ & $h=0$ & \\
            \rotatebox{90}{\hspace{0.5cm}Flux ($\cdot100$)} &  
            {\includegraphics[width=3.1cm,height=2.2cm,trim=0 0 0 0]{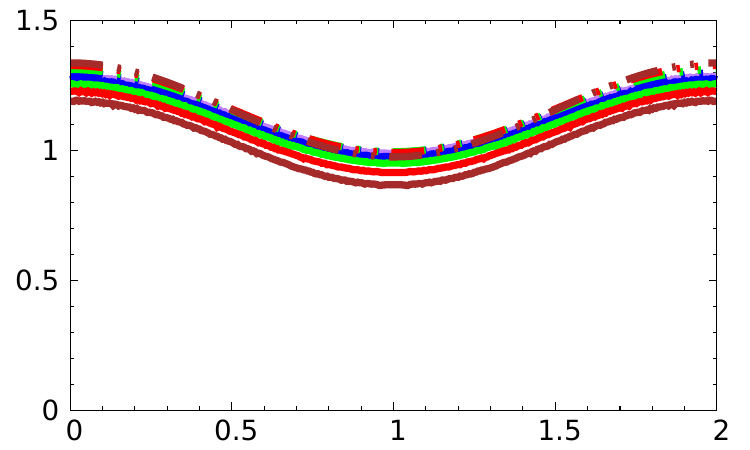}} &
            {\includegraphics[width=3.1cm,height=2.2cm,trim=0 0 0 0]{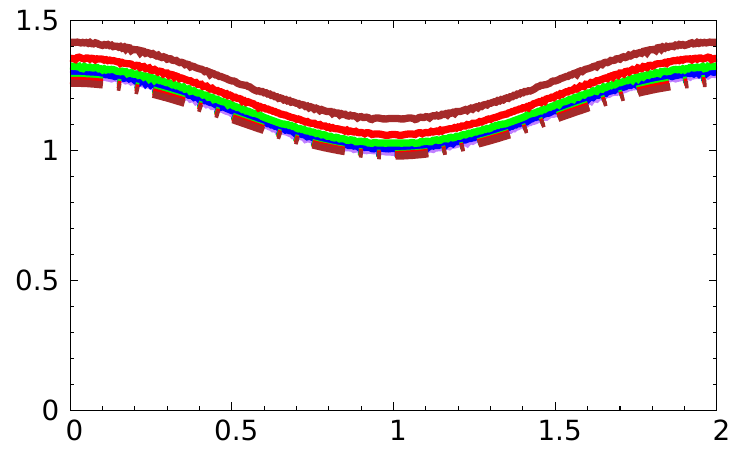}} &
            \rotatebox{90}{\hspace{0.3cm}$\chi=\frac{\pi}{9}$, $\xi=\frac{\pi}{6}$} \\
            \rotatebox{90}{\hspace{0.5cm}Flux ($\cdot100$)} &  
            {\includegraphics[width=3.1cm,height=2.2cm,trim=0 0 0 0]{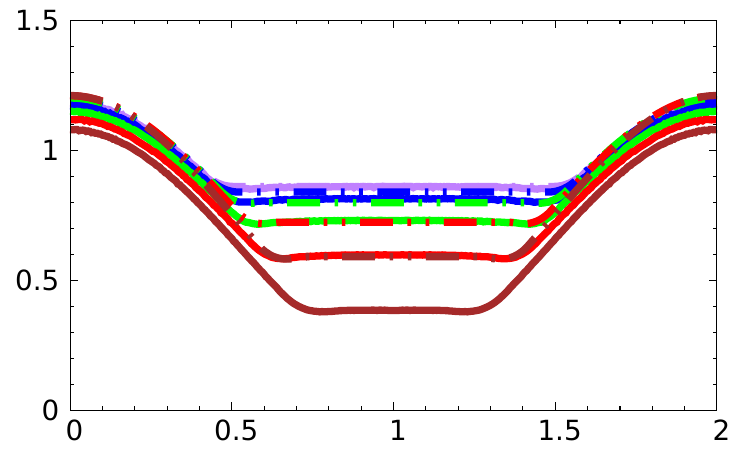}} &
            {\includegraphics[width=3.1cm,height=2.2cm,trim=0 0 0 0]{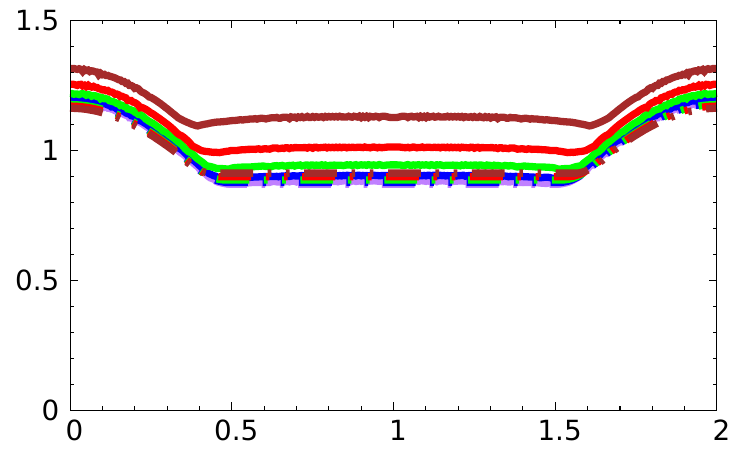}} &
            \rotatebox{90}{\hspace{0.3cm}$\chi=\frac{\pi}{6}$, $\xi=\frac{\pi}{3}$} \\
            \rotatebox{90}{\hspace{0.5cm}Flux ($\cdot100$)} &  
            {\includegraphics[width=3.1cm,height=2.2cm,trim=0 0 0 0]{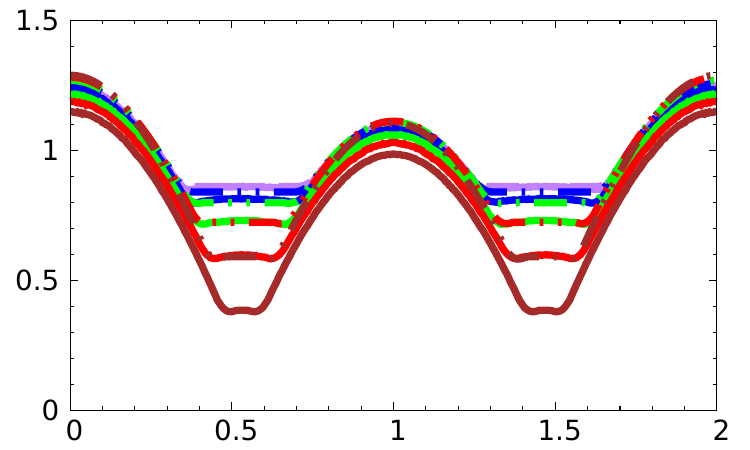}} &
            {\includegraphics[width=3.1cm,height=2.2cm,trim=0 0 0 0]{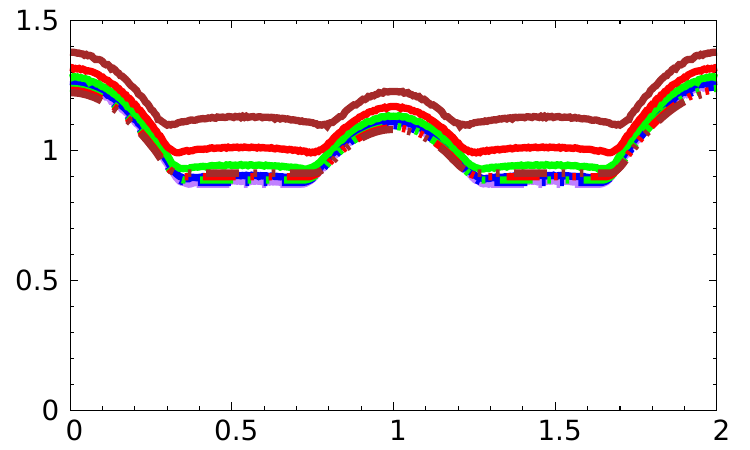}} &
            \rotatebox{90}{\hspace{0.3cm}$\chi=\frac{\pi}{3}$, $\xi=\frac{4\pi}{9}$} \\
            \rotatebox{90}{\hspace{0.5cm}Flux ($\cdot100$)} &  
            {\includegraphics[width=3.1cm,height=2.2cm,trim=0 0 0 0]{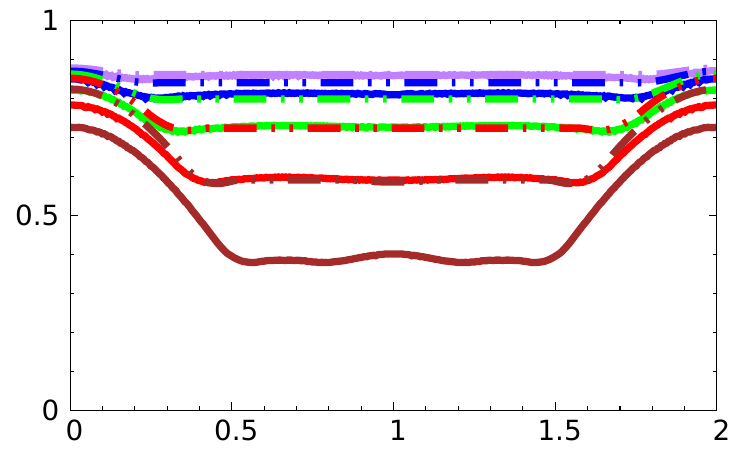}} &
            {\includegraphics[width=3.1cm,height=2.2cm,trim=0 0 0 0]{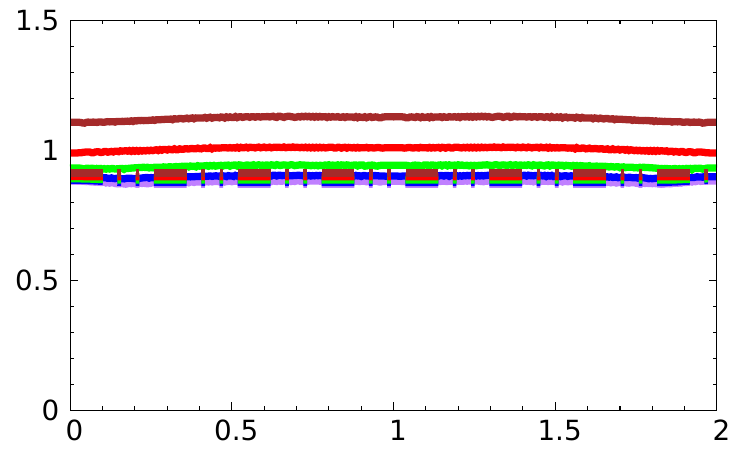}} &
            \rotatebox{90}{\hspace{0.3cm}$\chi=\frac{\pi}{9}$, $\xi=\frac{4\pi}{9}$} \\
            & {~~$\Omega t/\pi$} & {~~$\Omega t/\pi$} &
        \end{tabular}
        \caption{Analytical pulse profile for two identical, antipodal caps in Schwarzschild metric, $R/r_h=3$, $\theta_c=5^o$. Purple, blue, green, red and brown lines correspond to $\epsilon=0.1$, $0.2$, $0.3$, $0.4$ and $0.5$ respectively. Solid lines correspond to the numerical result, dashed lines to analytical. 
        The rows, from top to bottom, correspond to Beloborodov classes I, II, III and IV.
        For $h=3$, the flux is overestimated at maximums and underestimates at minimums.
        For $h=0$ is in general overestimated.}
        \label{fig:AA_PP_SH}
    \end{figure}

\begin{figure}
        \centering
        \begin{tabular}{cccc}
            & $h=3$ & $h=0$ & \\
            \rotatebox{90}{\hspace{0.5cm}Flux ($\cdot100$)} &  
            {\includegraphics[width=3.1cm,height=2.2cm,trim=0 0 0 0]{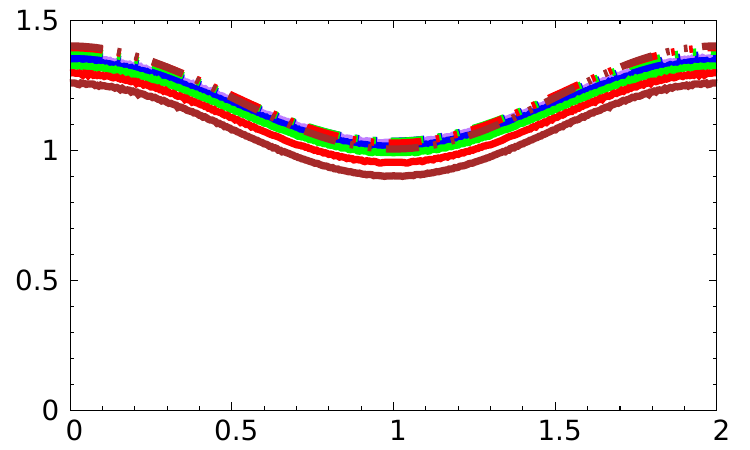}} &
            {\includegraphics[width=3.1cm,height=2.2cm,trim=0 0 0 0]{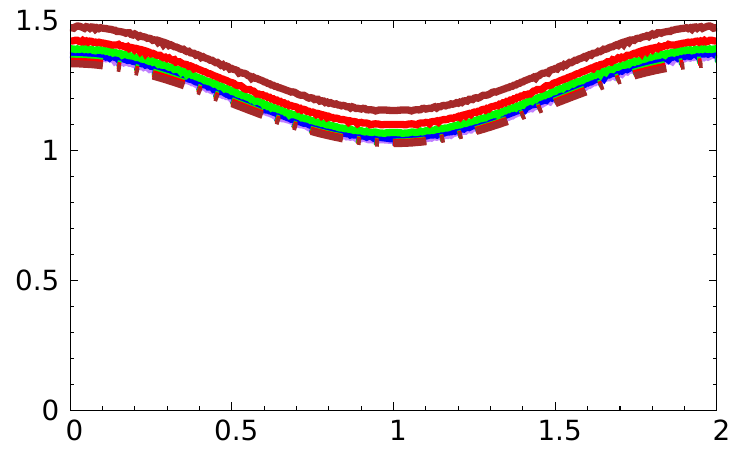}} &
            \rotatebox{90}{\hspace{0.3cm}$\chi=\frac{\pi}{9}$, $\xi=\frac{\pi}{6}$} \\
            \rotatebox{90}{\hspace{0.5cm}Flux ($\cdot100$)} &  
            {\includegraphics[width=3.1cm,height=2.2cm,trim=0 0 0 0]{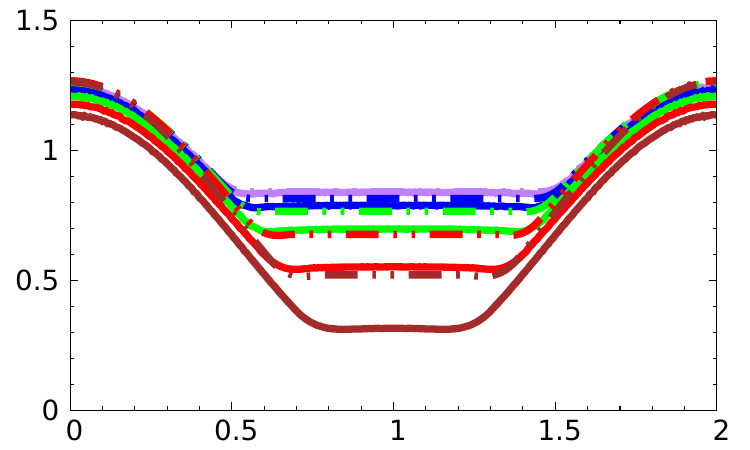}} &
            {\includegraphics[width=3.1cm,height=2.2cm,trim=0 0 0 0]{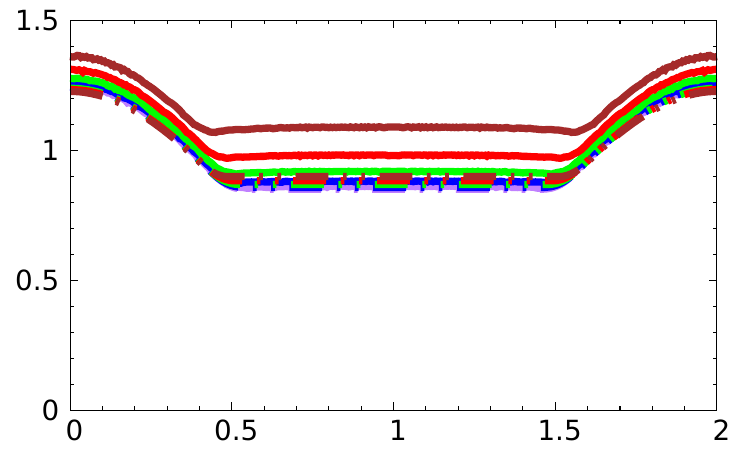}} &
            \rotatebox{90}{\hspace{0.3cm}$\chi=\frac{\pi}{6}$, $\xi=\frac{\pi}{3}$} \\
            \rotatebox{90}{\hspace{0.5cm}Flux ($\cdot100$)} &  
            {\includegraphics[width=3.1cm,height=2.2cm,trim=0 0 0 0]{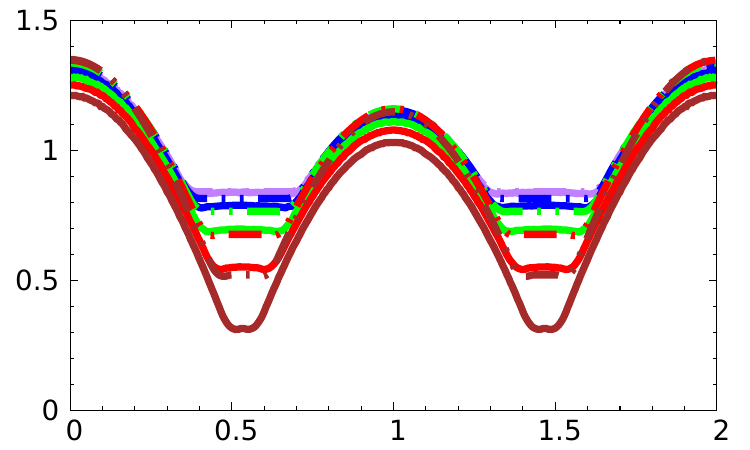}} &
            {\includegraphics[width=3.1cm,height=2.2cm,trim=0 0 0 0]{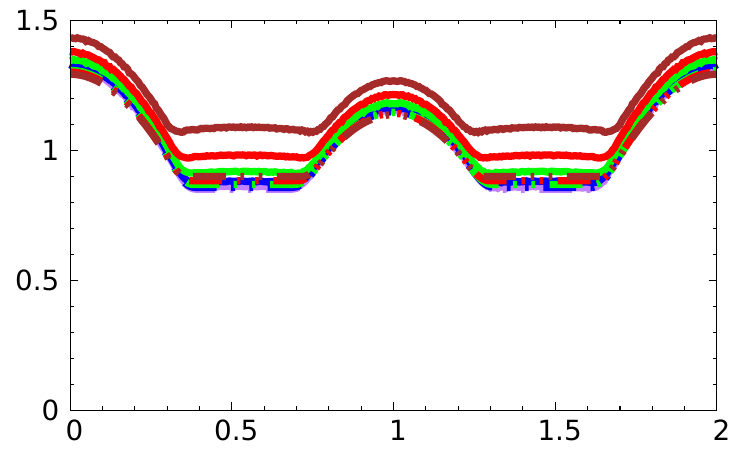}} &
            \rotatebox{90}{\hspace{0.3cm}$\chi=\frac{\pi}{3}$, $\xi=\frac{4\pi}{9}$} \\
            \rotatebox{90}{\hspace{0.5cm}Flux ($\cdot100$)} &  
            {\includegraphics[width=3.1cm,height=2.2cm,trim=0 0 0 0]{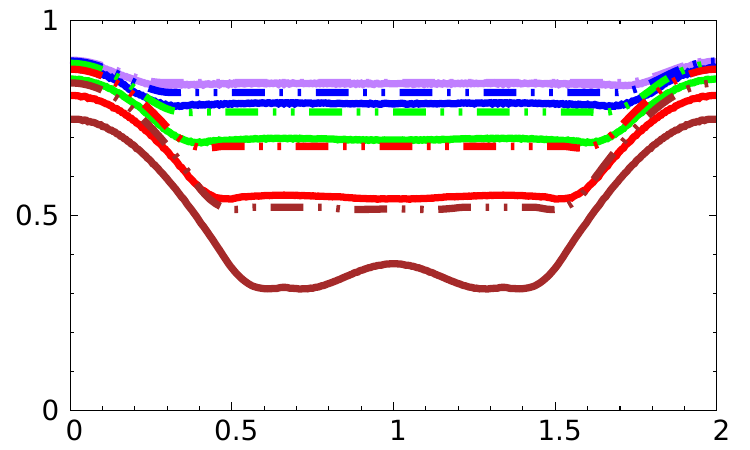}} &
            {\includegraphics[width=3.1cm,height=2.2cm,trim=0 0 0 0]{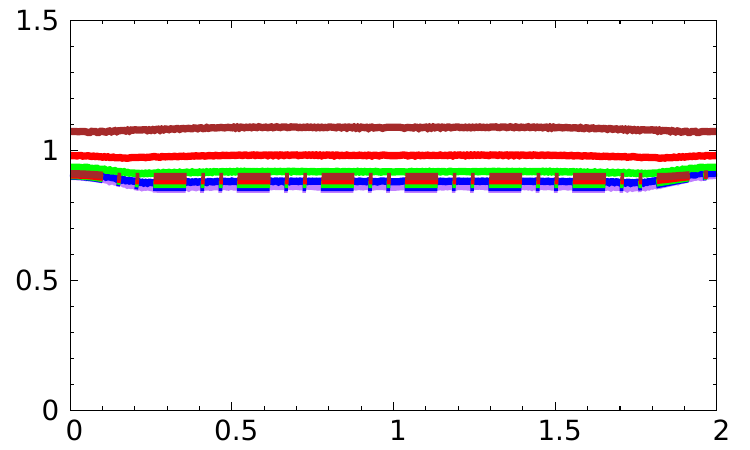}} &
            \rotatebox{90}{\hspace{0.3cm}$\chi=\frac{\pi}{9}$, $\xi=\frac{4\pi}{9}$} \\
            & {~~$\Omega t/\pi$} & {~~$\Omega t/\pi$} &
        \end{tabular}
        \caption{Same parameters and color references as in Fig. \ref{fig:AA_PP_SH} but for RN-like metric, with $q^*=-(0.75M)^2$.}
        \label{fig:AA_PP_RN}
    \end{figure}

    We now show the pulse profiles obtained analytically using the approach developed here and compare them with the corresponding numerical result. For this, we will consider pulsars of a typical size with $R/r_h=3$, charge $q^*=-(0.75M)^2$ and $\epsilon\leq0.5$, with two antipodal and identical caps of angular half-aperture $\theta_c=\pi/36$, in the angular configurations $(\chi,\xi)=(\pi/9,\pi/6)$, $(\pi/6,\pi/3)$, $(\pi/3,4\pi/9)$ and $(\pi/9,4\pi/9)$ corresponding (when $\epsilon=0$ and $q^*=0$) to classes I, II, III and IV defined in \cite{Beloborodov_2002} (these and other new classes will be discussed in Sec. \ref{c2s7}). This was performed for the Schwarzschild and RN-like
    metric models on the plasma distributions previously employed. 
    The results are shown in Figs. \ref{fig:AA_PP_SH} and \ref{fig:AA_PP_RN}.
    
    We see that the analytically obtained pulse profile significantly approximates the numerical results for frequencies with $\epsilon\leq0.3$, reaching a high accuracy for $\epsilon=0.1$, while for frequency ratios higher than $0.3$ the approximation loses accuracy, although it is still qualitatively reasonable.     

    We note that the approximation tends to overestimate the flux generated by a single cap, which corresponds to the peaks observed in Figures \ref{fig:AA_PP_SH} and \ref{fig:AA_PP_RN} at $\Omega t =0$, $\pi$ and $2\pi$, or the profiles corresponding to $(\chi,\xi)=(\pi/9,\pi/6)$, in the top row of each figure. 
    
    At the same time, this effect is magnified when both caps are visible. This is due to the fact that the approximation overestimates the maximum angle $\theta_F$, resulting in a greater lensing effect and generating an increase in the flux observed for the caps close to ``sunset''. In this regime, moreover, the approximation presents a nearly constant profile, omitting the subtle ups and downs that can be observed in the numerical flux. 
    It is in the transition between these behaviors (which is directly related to the location of $\theta_F$) where the model is less accurate.
    Thus, we conclude that the accuracy increases for plasma distributions with lower density and decreases as we consider emission areas with higher $\theta$, tending to overestimate the flux. 
    The pulse profiles analytically obtained here are consistent with those shown in \cite{Beloborodov_2002} and are morphologically very similar to those in \cite{Sotani_2017}.
    
    In summary, for class I profiles (top row of Figures \eqref{fig:AA_PP_SH}-\eqref{fig:AA_PP_RN} remembering that at the moment we are using the classification system introduced in \cite{Beloborodov_2002}, which will be revised and generalized in Sec. \eqref{c2s7})
    we see that the approximation qualitatively respects the morphology of the curves.  The class II and III are in general accurate, apart from the already mentioned irregularities that occur mainly at the peaks and at the confluence of the flux of both caps. On the other hand, the class IV are the most irregular, given that as $\epsilon$ increases, they degenerate into class II or III depending on the value of $\theta_F$, which is different depending on the used method.
    
    Regarding the plasma, we see that for $h=0$ the approximation now tends to underestimate the flux. 
   
    This could be related to the fact that, in general, a constant plasma distribution inverts the lensing effect from divergent to convergent, thus reversing the previously mentioned effects. In any case, the curves are extremely similar to the numerical curves except for a vertical translation. On the other hand, the power-law decay profiles with $h=3$ perform reasonably well, given the already mentioned peculiarities of peak accuracy and overestimation of the flux at the confluence of the two caps.

\section{Rings}
\label{c2s6}

    It has been mentioned in the literature that emission regions in neutron stars commonly have irregular shapes, sometimes resembling rings or crescent moons \cite{gralla2017inclined, lockhart2019x}. Solving these cases numerically can be relatively complicated and involve high computational cost, so the analytical approximations introduced in Sect. \ref{c2s5} represent a helpful alternative that allows us to obtain the pulse profiles in a much shorter time (about an order of magnitude, depending on the optimization of the programs) with acceptable accuracy and without loss of generality.

    In the papers by Sotani et.al., \cite{Sotani_2017, Sotani_2018, Sotani_2019, Sotani_2020} analytical approaches are also developed to deal with this type of cases, with special emphasis on ring-shaped emission caps (for more information or guidance on this topic, check the cited papers), although the influence of a plasma environment is completely neglected. 
    
    We will demonstrate the utility and power of our model by using it in the resolution of pulse profiles generated by ring-shaped caps. Suppose we have a pulsar with a single ring-shaped emitting cap, whose outer edge has an angular half-aperture $\theta_e$ while its inner edge has an angular half-aperture $\theta_i$. This can be easily obtained by calculating the flux produced by a circular cap of half-aperture $\theta_e$ and subtracting the flux produced by a circular cap of half-aperture $\theta_i$, both caps with the same position vector $\hat{C}$, 
    \begin{equation}
        \begin{split}
        F_{Ring} &=\frac{A^{3/2}(R)}{C(R)}\frac{n^2(R)}{n_0^2}\frac{1}{1-P^{q*}_h(R)\epsilon^2}I_0 \times \\
                 &\left[A_{eff}(\theta_e,\theta_0)-A_{eff}(\theta_i,\theta_0)\right],    
        \end{split}
        \label{eq:AA_Anillo}
    \end{equation}
    
\begin{figure}
        \centering
        \begin{tabular}{cccc}
            & Sch & RN-like & \\
            \rotatebox{90}{\hspace{0.5cm} F($\Omega t$)/F($0$)} &  
            {\includegraphics[width=3.2cm,height=2.3cm,,trim=0 0 0 0]{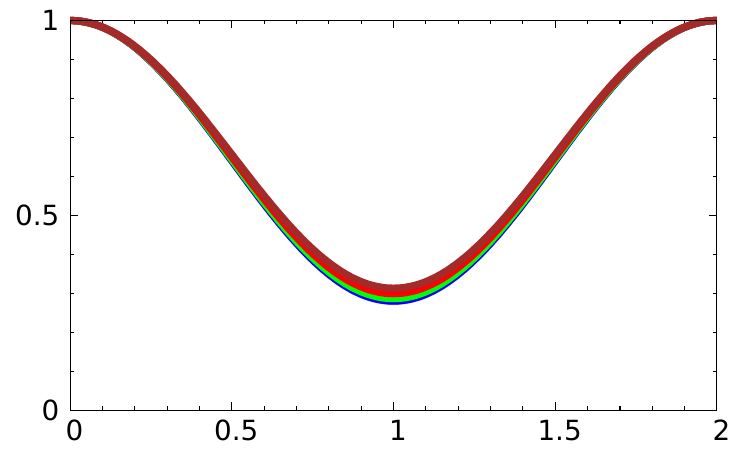}} &
            {\includegraphics[width=3.2cm,height=2.3cm,,trim=0 0 0 0]{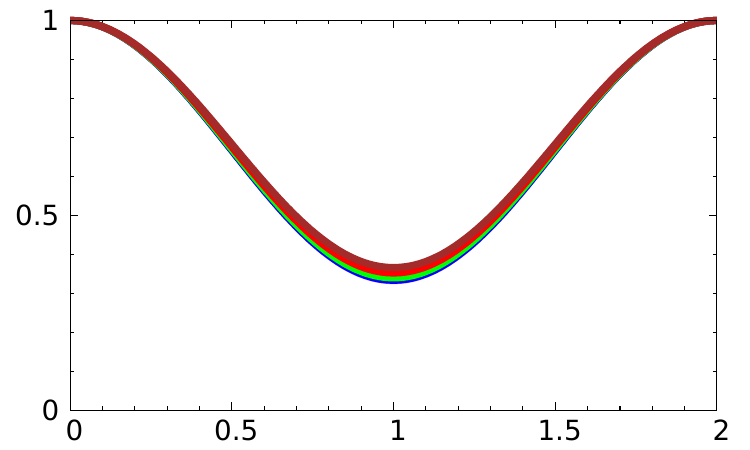}} &
            \rotatebox{90}{\hspace{0.3cm} $\chi=\xi=\pi/3$} \\
            \rotatebox{90}{\hspace{0.5cm} F($\Omega t$)/F($0$)} &  
            {\includegraphics[width=3.2cm,height=2.3cm,,trim=0 0 0 0]{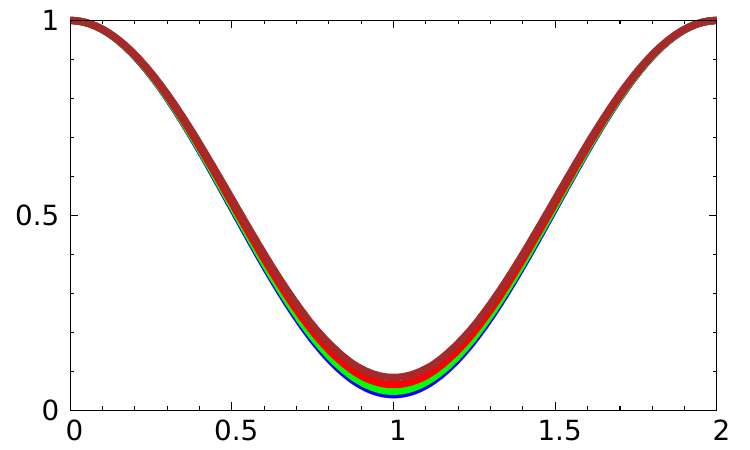}} &
            {\includegraphics[width=3.2cm,height=2.3cm,,trim=0 0 0 0]{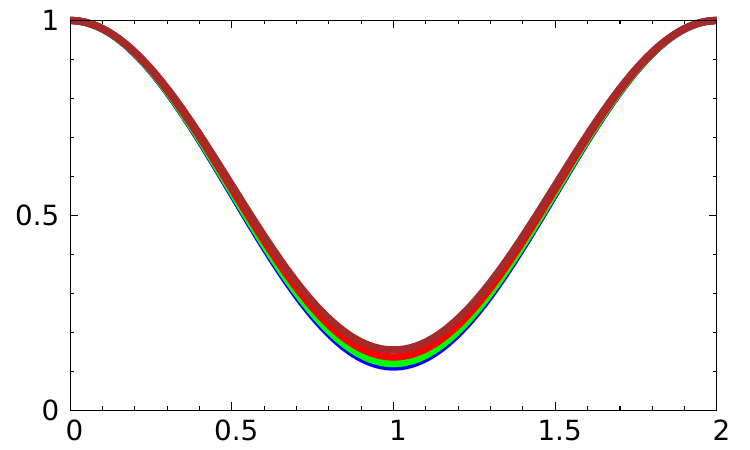}} &
            \rotatebox{90}{\hspace{0.3cm} $\chi=\xi=\pi/2$} \\
            & {~~$\Omega t/\pi$} & {~~$\Omega t/\pi$} &
        \end{tabular}
        \caption{Pulse profiles generated by ring-shaped cap emission. $R=4M$, $q^*=-(0.75M)^2$, $\epsilon=0.15$, $h=3$. The outer edge of the ring has an angular half-aperture $\theta_e=35^o$, while the purple, blue, green, red and brown curves correspond to $\theta_i=0^o$, $5^o$, $15^o$, $25^o$ and $32.5^o$ respectively.
        The width of the ring hardly modifies the pulse profile.}
        \label{fig:Anillos_35}
    \end{figure}

\begin{figure}
        \centering
        \begin{tabular}{cccc}
            & Sch & RN-like & \\
            \rotatebox{90}{\hspace{0.5cm} F($\Omega t$)/F($0$)} &  
            {\includegraphics[width=3.2cm,height=2.3cm,,trim=0 0 0 0]{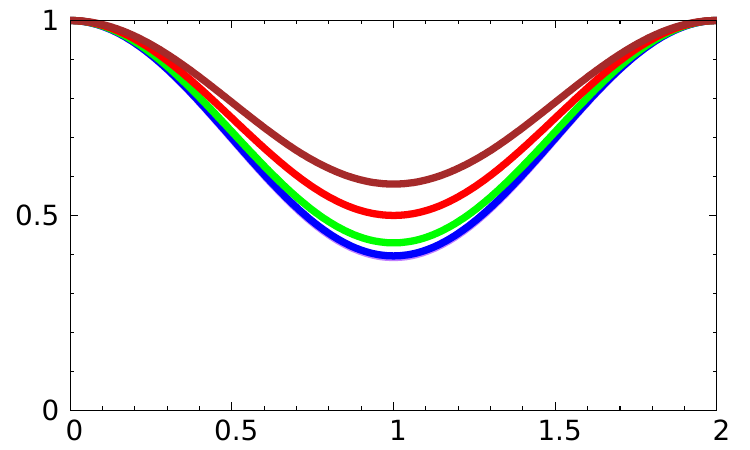}} &
            {\includegraphics[width=3.2cm,height=2.3cm,,trim=0 0 0 0]{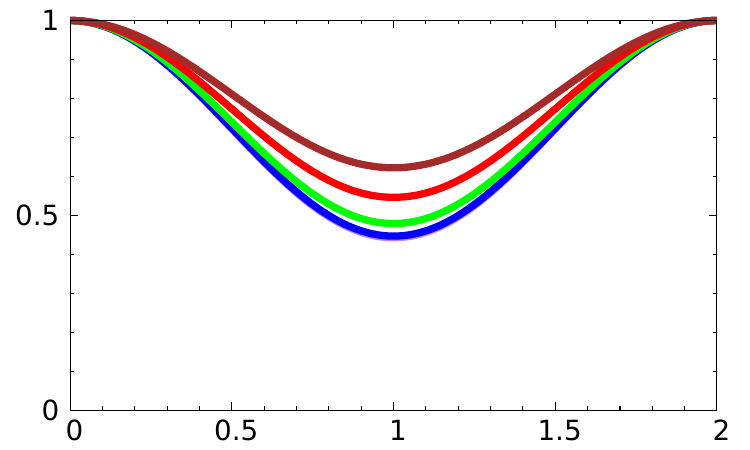}} &
            \rotatebox{90}{\hspace{0.3cm} $\chi=\xi=\pi/3$} \\
            \rotatebox{90}{\hspace{0.5cm} F($\Omega t$)/F($0$)} &  
            {\includegraphics[width=3.2cm,height=2.3cm,,trim=0 0 0 0]{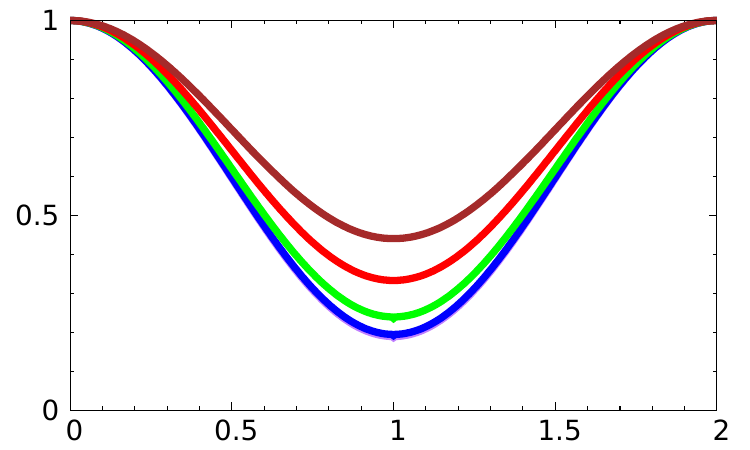}} &
            {\includegraphics[width=3.2cm,height=2.3cm,,trim=0 0 0 0]{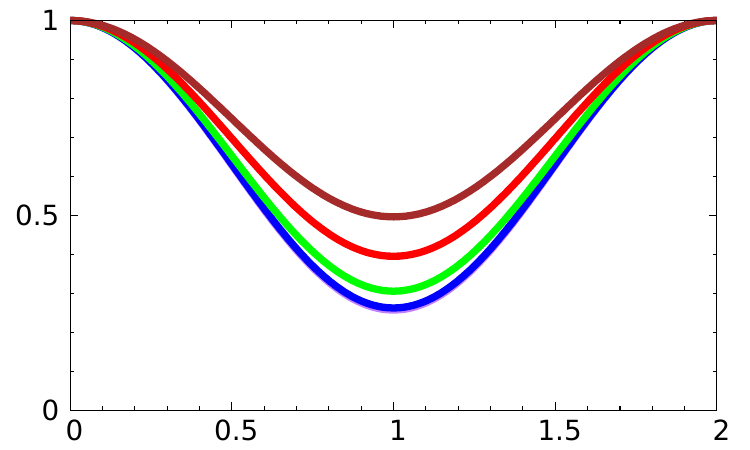}} &
            \rotatebox{90}{\hspace{0.3cm} $\chi=\xi=\pi/2$} \\
            & {~~$\Omega t/\pi$} & {~~$\Omega t/\pi$} &
        \end{tabular}
        \caption{Pulse profiles generated by ring-shaped cap emission. $R=4M$, $q^*=-(0.75M)^2$, $\epsilon=0.15$, $h=3$. The outer edge of the ring has an angular half-aperture $\theta_e=70^o$, while the purple, blue, green, red and brown curves correspond to $\theta_i=0^o$, $10^o$, $30^o$, $50^o$ and $65^o$ respectively.
        The width of the ring significantly modifies the pulse profile.}
        \label{fig:Anillos_70}
    \end{figure}

    In Figures \ref{fig:Anillos_35} and \ref{fig:Anillos_70} we show the pulse profiles generated by single ring-shaped emission caps with $\theta_e=35^o$ and $70^o$ respectively, for different values of $\theta_i$. For this, the parameters $R=4M$, $q^*=-(0.75M)^2$, $\epsilon=0.15$ and $h=3$ (see Eq. \eqref{eq:refindnum}) were used on the RN-like metrics, for the $\chi=\xi=\pi/3$ and $\chi=\xi=\pi/2$ configurations. The fluxes obtained were normalized according to their maximum value. 
    
    Although the magnitude of the brightness decreases noticeably with increasing value of $\theta_i$ (this is not visible in the plots given the normalization), the morphology of the profile minimally differs from that of the circular cap, at least qualitatively.
    The resulting flux presents the sinusoidal profile that we expect from a single cap. 
    
    By removing part of the cap (i.e., by increasing $\theta_i$) the amplitude of the oscillation decreases, tending to flatten the curve. This is because the contribution from the center of the cap that we are subtracting has less weight the farther it is from the line of sight, so the minimum value of the flux will decrease less than the maximum value (this looks inverted in graphs due to normalization, 
    where $F_{max}$ remains constant while $F_{min}$ increases).
    
    On the other hand, we see that this effect on the morphology is only noticeable in relatively large rings ($\theta_e\approx70^o$), while for small rings ($\theta_e\approx35^o$) we see nearly no difference between the different $\theta_i$. This is because, as the ring is small, it occupies an environment where the contributions of each point to the flux are similar, so removing part of the cap proportionally decreases the net flux without affecting the profile morphology. By considering small rings we are approaching the point cap approximation, so that what we are doing could be compared to subtracting two point caps of different intensity.
    
    Moreover, we see that by changing the configuration $(\xi,\chi)$ the behavior is as expected. For $(\xi,\chi)=(\pi/3,\pi/3)$, the ring does not move too far away from the front of the star, keeping the value of $\theta_0$ low, so the minimum flux is relatively high. In contrast, for the configuration $(\xi,\chi)=(\pi/2,\pi/2)$ the ring is opposite the observer at $\Omega t=\pi$, so the observed flux will be much lower, since much of it is located in the invisible region of the star, or at too high $\theta$ values.
    
    The analysis developed in this section is in full agreement with that found in \cite{Sotani_2019}, where similar results were obtained.
    We see that there is little difference between the profiles obtained by these metrics.
   
    We emphasize that this is only one of the possible applications of our analytical approach. With it, the calculation of the pulse profile produced by caps of all kinds of irregular shapes or with different surface temperatures can be simplified, making the study of the properties of these systems more accessible and facilitating their understanding. 

\section{Classification}
\label{c2s7}

    When considering two emitting caps which are no longer identical and which are also not in an antipodal configuration, the Beloborodov system \cite{Beloborodov_2002} is insufficient to classify the different observable profiles. As the caps are distinguishable, the profile is no longer symmetric with respect to which cap is which, so by inverting 
    their positions we should add a new class. On the other hand, by abandoning the antipodal configuration we give rise to profiles where, for example, neither of the caps is visible during the entire phase, or one is visible only occasionally while the other is always invisible. These phenomena can also be achieved by increasing the plasma density, resulting in new profile classes. We have then a total of nine profile classes for two distinguishable non-antipodal caps, the four classes introduced by Beloborodov, the inversions of his classes I and II, two classes with partial and null visibility caps and a null profile class. For convenience, we organize this new classification according to whether the caps have total (i.e., the cap is visible at all times for the observer), partial (it is visible at some times and not at others) or null visibility (it is not visible for the observer at any time). This classification is explained in Table \ref{tab:Clases}, where the correspondence with Beloborodov's classes is also indicated.
    
    \begin{table}
        \centering
        \begin{tabular}{|c|c|c|c|}
        \hline  
        Class & Main c. vis. & Secondary c. vis. & Beloborodov  \\ \hline     
        I     & Total                              & Total                               & IV                             \\ \hline 
        II    & Total                              & Partial                             & II                             \\ \hline 
        III   & Total                              & Null                                & I                              \\ \hline 
        IV    & Partial                            & Total                               & II                             \\ \hline 
        V     & Partial                            & Partial                             & III                            \\ \hline 
        VI    & Partial                            & Null                                &                                \\ \hline 
        VII   & Null                               & Total                               & I                              \\ \hline 
        VIII  & Null                               & Partial                             &                                \\ \hline 
        IX    & Null                               & Null                                &                                \\ \hline 
        \end{tabular}
        \caption{\label{tab:Clases} New classification system for pulsars with two distinguishable non-antipodal caps, compared to Beloborodov's.}
    \end{table}
  
\begin{figure}
        \centering
        \begin{tabular}{cccc}
            & $\Delta\chi=2\pi/3$ & $\Delta\chi=5\pi/6$ & \\
            \rotatebox{90}{\hspace{1.4cm} $\chi_p$} &  
            {\includegraphics[width=3.2cm,height=3.2cm,trim=0 0 0 0]{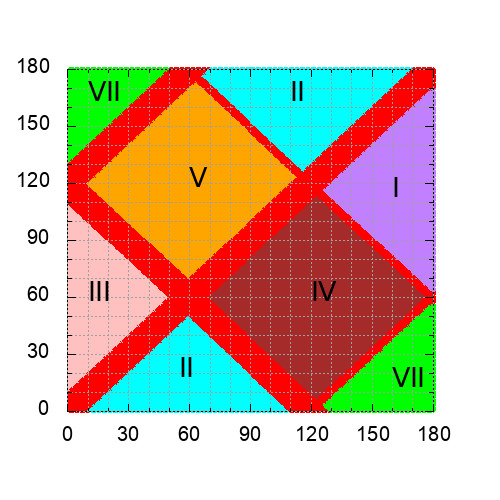}} &
            {\includegraphics[width=3.2cm,height=3.2cm,trim=0 0 0 0]{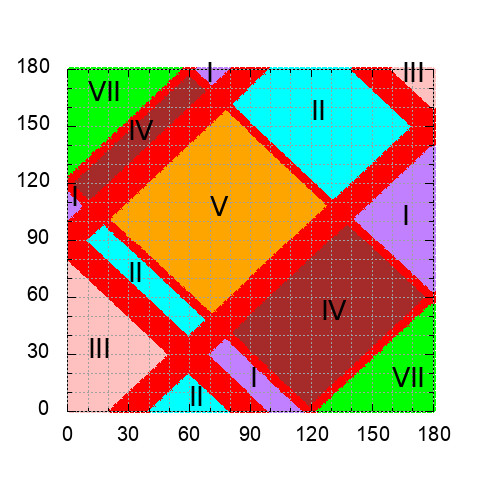}} &
            \rotatebox{90}{\hspace{1.1cm} Analytic} \\
            \rotatebox{90}{\hspace{1.4cm} $\chi_p$} &  
            {\includegraphics[width=3.2cm,height=3.2cm,trim=0 0 0 0]{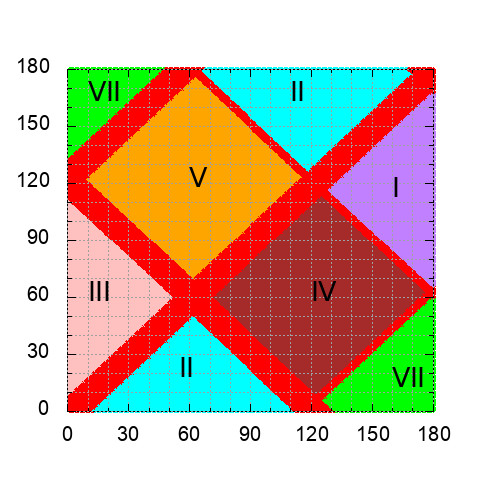}} &
            {\includegraphics[width=3.2cm,height=3.2cm,trim=0 0 0 0]{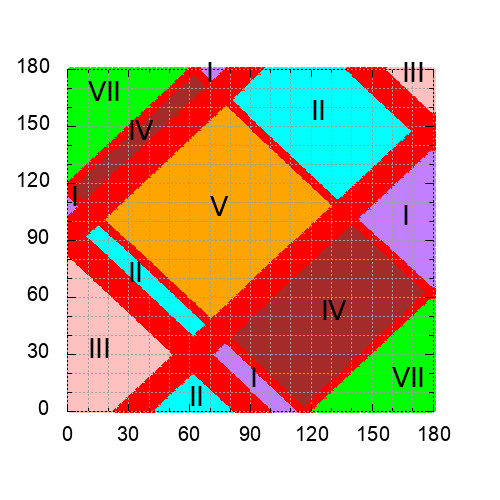}} &
            \rotatebox{90}{\hspace{1.1cm} Numeric} \\
            & {~$\xi$} & {~$\xi$} &
        \end{tabular}
        \caption{Class location maps on the $\xi-\chi_p$ plane. Schwazschild metric, $\epsilon=0.0$}
        \label{fig:Clases_SH_0.0}
    \end{figure}

\begin{figure}
        \centering
        \begin{tabular}{cccc}
            & $\Delta\chi=2\pi/3$ & $\Delta\chi=5\pi/6$ & \\
            \rotatebox{90}{\hspace{1.4cm} $\chi_p$} &  
            {\includegraphics[width=3.2cm,height=3.2cm,trim=0 0 0 0]{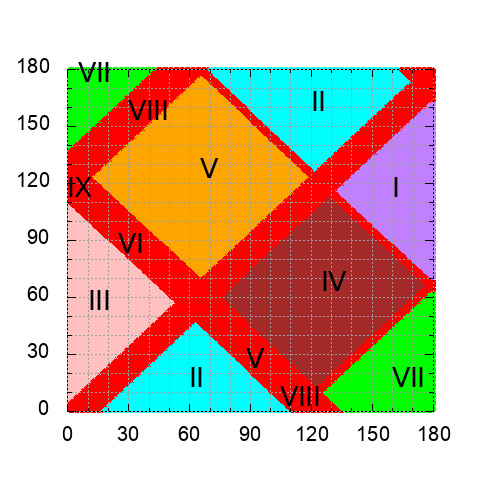}} &
            {\includegraphics[width=3.2cm,height=3.2cm,trim=0 0 0 0]{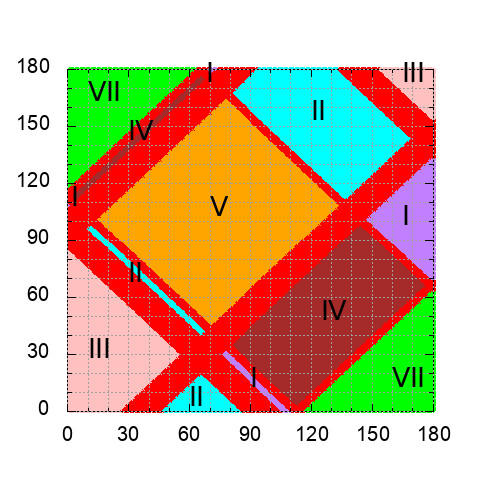}} &
            \rotatebox{90}{\hspace{1.1cm} Analytic} \\
            \rotatebox{90}{\hspace{1.4cm} $\chi_p$} &  
            {\includegraphics[width=3.2cm,height=3.2cm,trim=0 0 0 0]{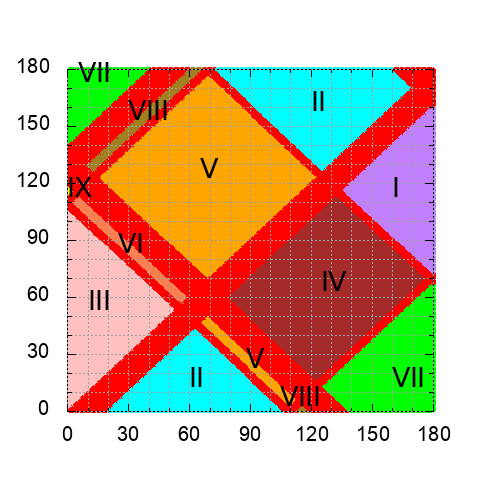}} &
            {\includegraphics[width=3.2cm,height=3.2cm,trim=0 0 0 0]{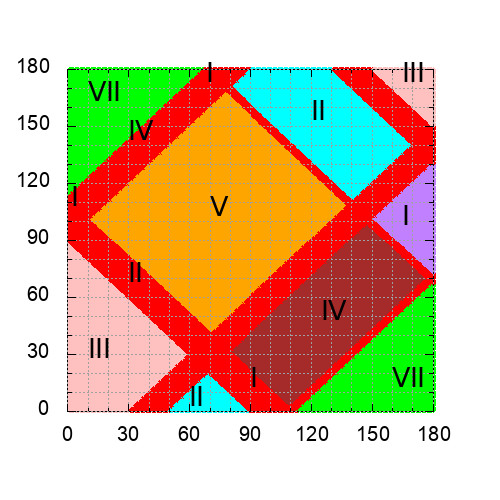}} &
            \rotatebox{90}{\hspace{1.1cm} Numeric} \\
            & {~$\xi$} & {~$\xi$} &
        \end{tabular}
        \caption{Class location maps on the $\xi-\chi_p$ plane. Schwarzschild metric, $\epsilon=0.3$}
        \label{fig:Clases_SH_0.3}
    \end{figure}

\begin{figure}
        \centering
        \begin{tabular}{cccc}
            & $\Delta\chi=2\pi/3$ & $\Delta\chi=5\pi/6$ & \\
            \rotatebox{90}{\hspace{1.4cm} $\chi_p$} &  
            {\includegraphics[width=3.2cm,height=3.2cm,trim=0 0 0 0]{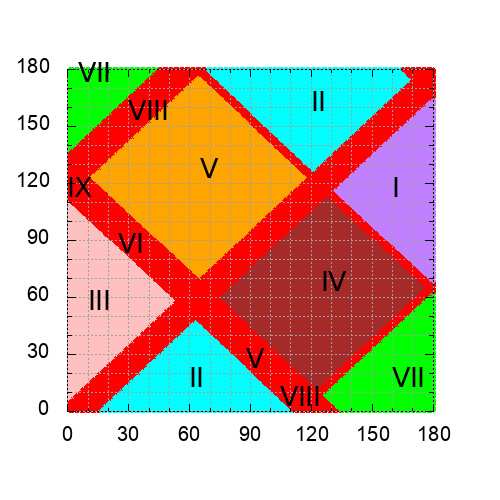}} &
            {\includegraphics[width=3.2cm,height=3.2cm,trim=0 0 0 0]{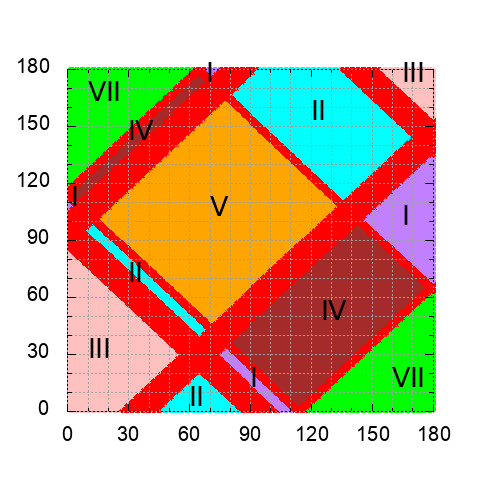}} &
            \rotatebox{90}{\hspace{1.1cm} Analytic} \\
            \rotatebox{90}{\hspace{1.4cm} $\chi_p$} &  
            {\includegraphics[width=3.2cm,height=3.2cm,trim=0 0 0 0]{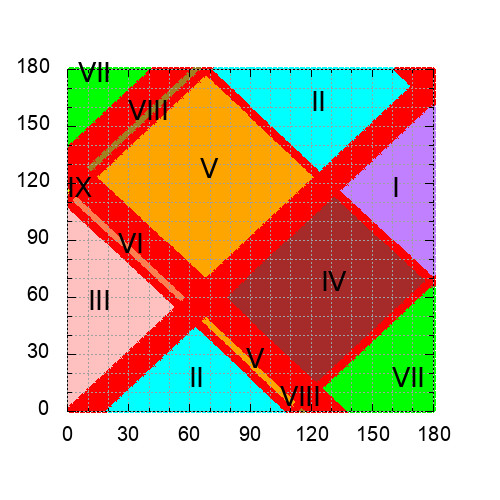}} &
            {\includegraphics[width=3.2cm,height=3.2cm,trim=0 0 0 0]{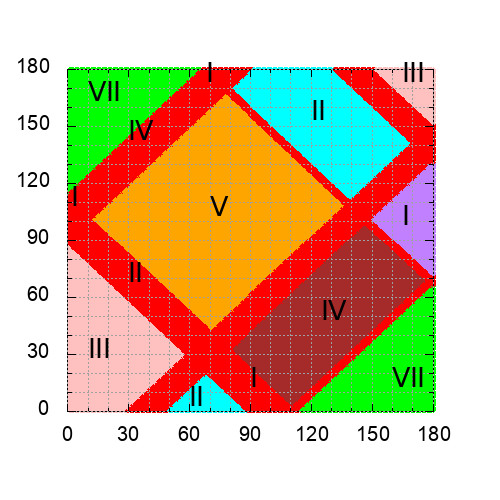}} &
            \rotatebox{90}{\hspace{1.1cm} Numeric} \\
            & {~$\xi$} & {~$\xi$} &
        \end{tabular}
        \caption{Class location maps on the $\xi-\chi_p$ plane. RN-like metric, $\epsilon=0.3$, $q^*=-(0.50M)^2$}
        \label{fig:Clases_RN}
    \end{figure}

    Following, in Figures \ref{fig:Clases_SH_0.0} to \ref{fig:Clases_RN} are shown the location maps of the classes in the $\xi-\chi_p$ plane, being $\chi_p$ the angle between the rotation axis and the center of the primary cap, for different values of $\Delta\chi=\chi_p-\chi_s$, being $\chi_s$ the angle between the rotation axis and the secondary cap. 
    {We take $R=6M$, $q^*=-(0.5M)^2$ and $\epsilon=0.3$ for Sch and RN-like metrics, and also $\epsilon=0$ for Sch.
    The primary and secondary caps have angular half-apertures $\theta_{c,p}=3^o$ and $\theta_{c,s}=10^o$ respectively.}
    
    Analyzing the class location maps, we see that the different classes are separated by straight lines marked in red on the graphs, which correspond to the values of $\xi$ and $\chi_p$ for which one of the caps, at its farthest point from the line of sight (i.e., at its maximum elongation at $\theta$), is partially in the zone of visibility and partially in the zone of invisibility, being in this way transition zones for caps from total to partial visibility, or from partial to null visibility. Thus, the thickness of these lines corresponds to the angular size of the caps {(this issue is also discussed in \cite{Sotani_2019} and \cite{Sotani_2020}, where extended caps are considered)}. Moreover, the location of these eight red lines that we see in each graph are given by the following equations
    \begin{equation}
        \begin{split}
        \chi_p &= \pm \theta_F \pm\xi, \\ 
        \chi_p &= \pm \theta_F \pm\xi - \Delta\chi, 
        \end{split}
        \label{eq:Clases_rectas}
    \end{equation}
    being the maps symmetric with respect to the straight lines $\xi=0$ and $\xi=180^o$, that is, $\text{Class}(\pi+\delta,\chi_p)=\text{Class}(\pi-\delta,\chi_p)$. On the other hand, we have that 
    $\text{Class}(\xi,\pi+\delta)=\text{Class}(\pi-\xi,\delta)$.
    Given these symmetries, we see that in each graph two slanted rectangles are formed, whose left vertex is on the $\chi_p$ axis and whose upper left and lower right edges are shorter than the lower left and upper right edges. The interior of each rectangle represents the zone of partial visibility of the corresponding cap. When we move diagonally across the edges of greater extension, we move into the zone of total visibility, while when we move across the edges of lesser extension we reach the zone of null visibility. If we move horizontally or vertically, crossing a vertex, we are again inside a rectangle of partial visibility.
    
    Given the differences between the numerical and analytical values of $\theta_F$, there is a small difference in the location of these lines which, in extreme cases, may result in the emergence of classes that should not be there, or in their omission. However, the morphology of the class maps obtained by the numerical and analytical methods are generally indistinguishable. 
    
\begin{figure}
        \centering
        \begin{tabular}{cccc}
            & Class I & Class II & Class III \\
            \rotatebox{90}{\hspace{0.3cm} Flux} &  
            {\includegraphics[width=2.0cm,height=1.4cm,trim=0 0 0 0]{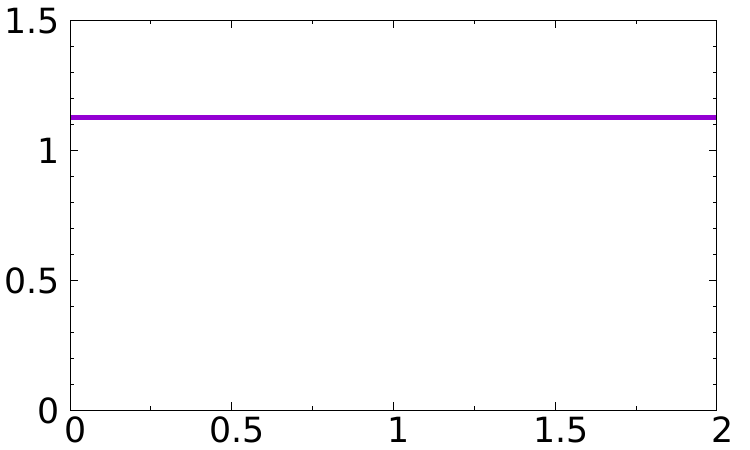}} &
            {\includegraphics[width=2.0cm,height=1.4cm,trim=0 0 0 0]{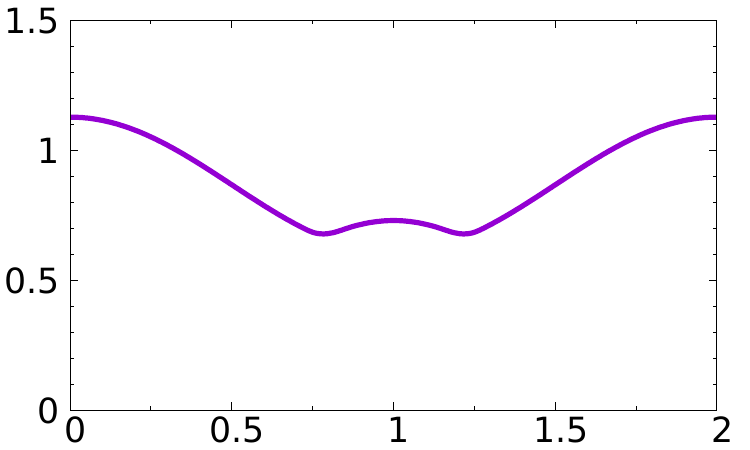}} &
            {\includegraphics[width=2.0cm,height=1.4cm,trim=0 0 0 0]{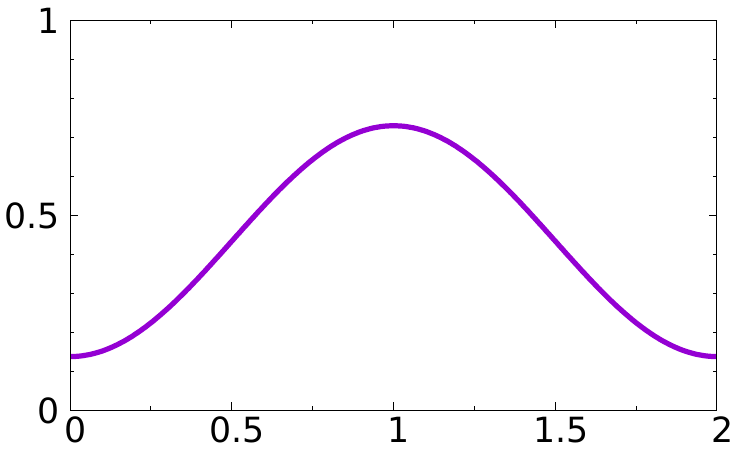}} \\
            & Class IV & Class V & Class VI \\
            \rotatebox{90}{\hspace{0.3cm} Flux} &  
            {\includegraphics[width=2.0cm,height=1.4cm,,trim=0 0 0 0]{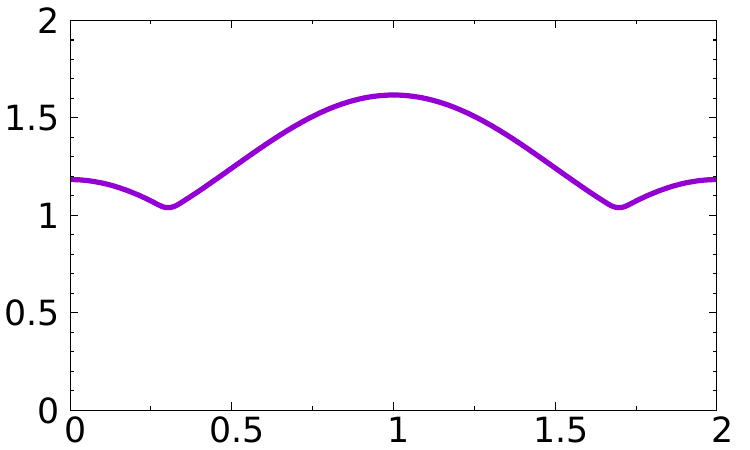}} &
            {\includegraphics[width=2.0cm,height=1.4cm,,trim=0 0 0 0]{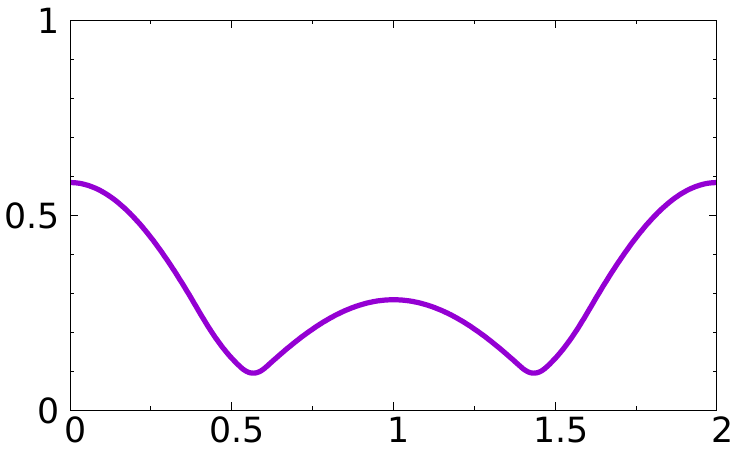}} &
            {\includegraphics[width=2.0cm,height=1.4cm,,trim=0 0 0 0]{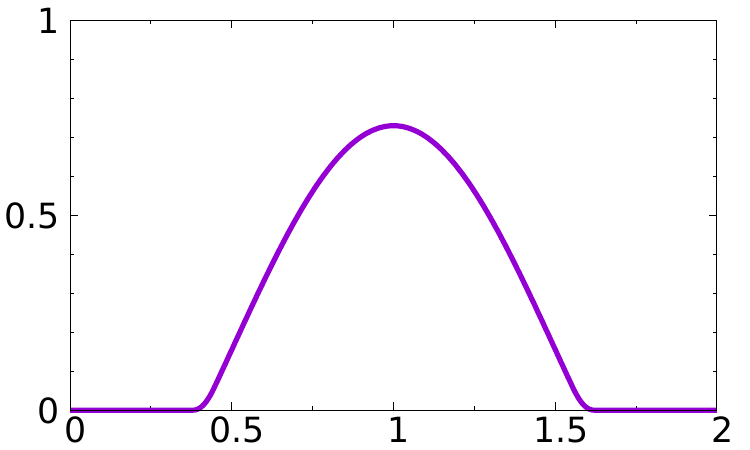}} \\
            & Class VII & Class VIII & Class IX \\
            \rotatebox{90}{\hspace{0.3cm} Flux} &  
            {\includegraphics[width=2.0cm,height=1.4cm,,trim=0 0 0 0]{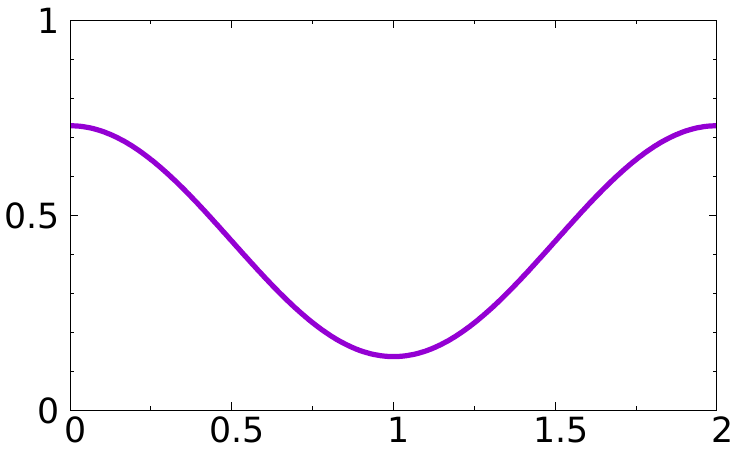}} &
            {\includegraphics[width=2.0cm,height=1.4cm,,trim=0 0 0 0]{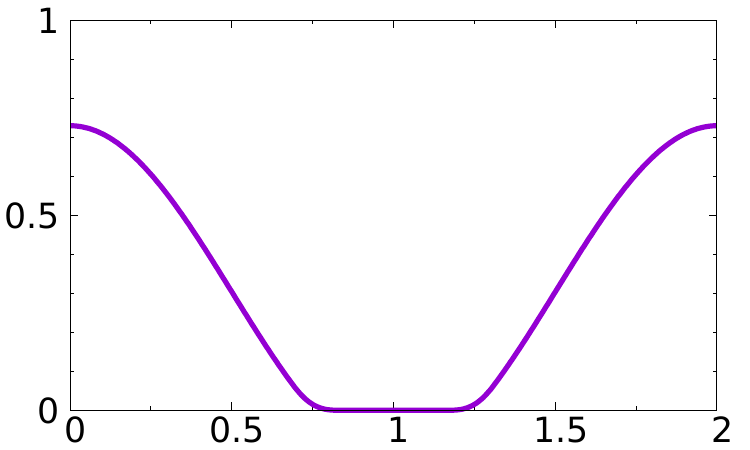}} &
            {\includegraphics[width=2.0cm,height=1.4cm,,trim=0 0 0 0]{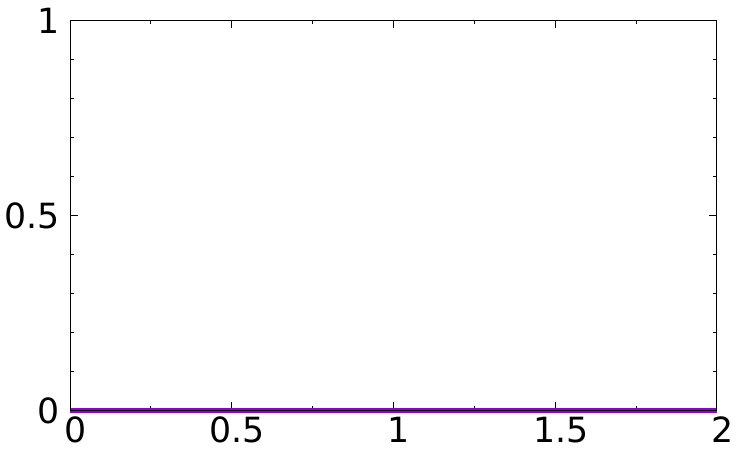}} \\
            & {~~$\Omega t/\pi$} & {~~$\Omega t/\pi$} & {~~$\Omega t/\pi$} 
        \end{tabular}
        \caption{Representative pulse profiles of each class.}
        \label{fig:Pulso_Clases}
    \end{figure}

    To complete the discussion, in Figure \ref{fig:Pulso_Clases} we show some representative pulse profiles for each classes introduced in Table \ref{tab:Clases}.
    
\begin{figure}
        \centering
        \begin{tabular}{cccc}
            & $\Delta \chi=\pi/3$ & $\Delta \chi=\pi$ & \\
            \rotatebox{90}{\hspace{0.8cm} Flux} &  
            {\includegraphics[width=3.2cm,height=2.3cm,trim=0 0 0 0]{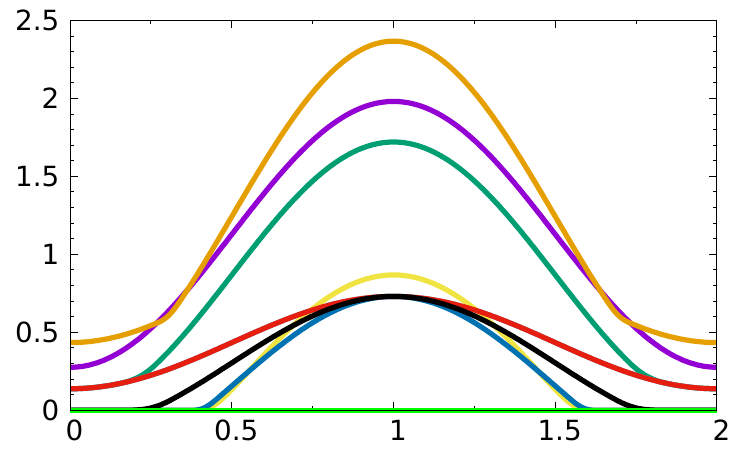}} &
            {\includegraphics[width=3.2cm,height=2.3cm,trim=0 0 0 0]{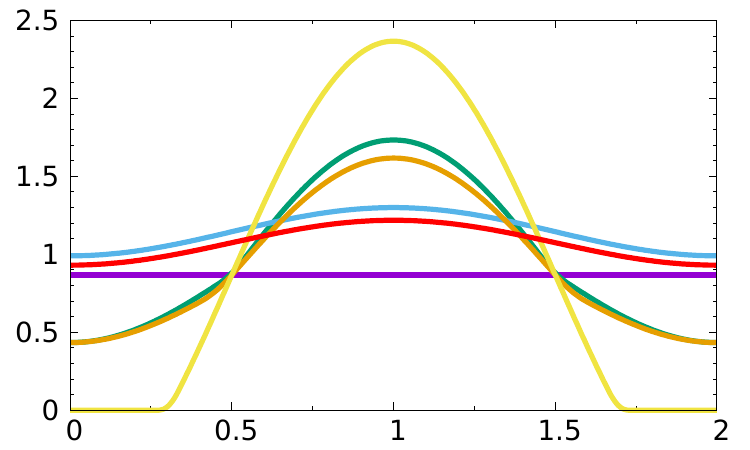}} &
            \rotatebox{90}{\hspace{0.7cm}$\Delta\gamma_p=0$}  \\
            \rotatebox{90}{\hspace{0.8cm} Flux} &  
            {\includegraphics[width=3.2cm,height=2.3cm,trim=0 0 0 0]{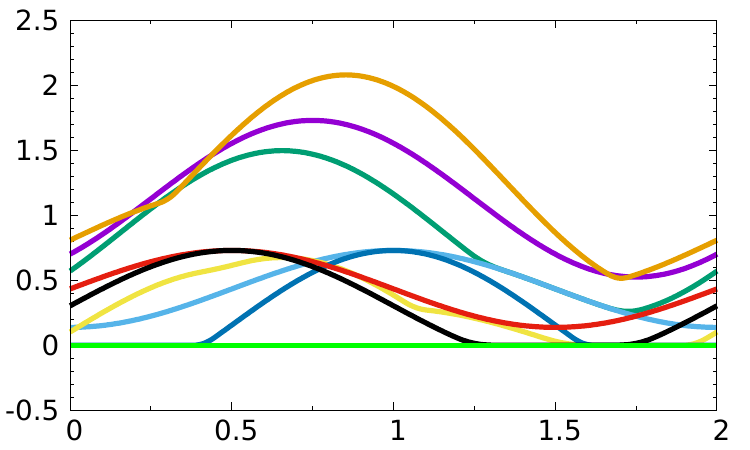}} &
            {\includegraphics[width=3.2cm,height=2.3cm,trim=0 0 0 0]{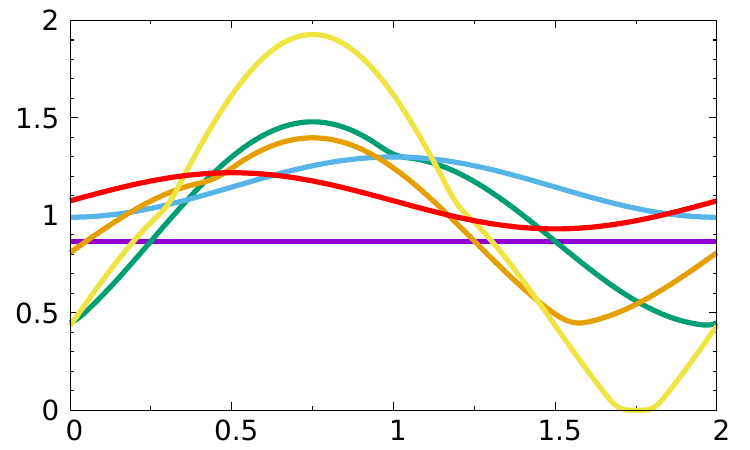}} &
            \rotatebox{90}{\hspace{0.7cm}$\Delta\gamma_p=\pi/2$}  \\
            \rotatebox{90}{\hspace{0.8cm} Flux} &  
            {\includegraphics[width=3.2cm,height=2.3cm,trim=0 0 0 0]{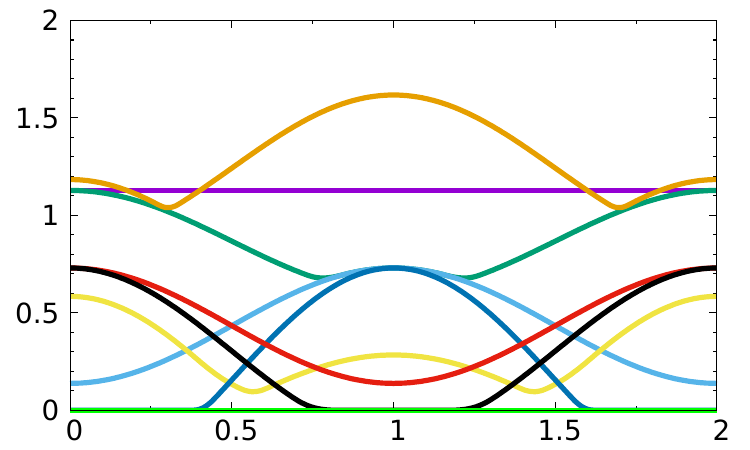}} &
            {\includegraphics[width=3.2cm,height=2.3cm,trim=0 0 0 0]{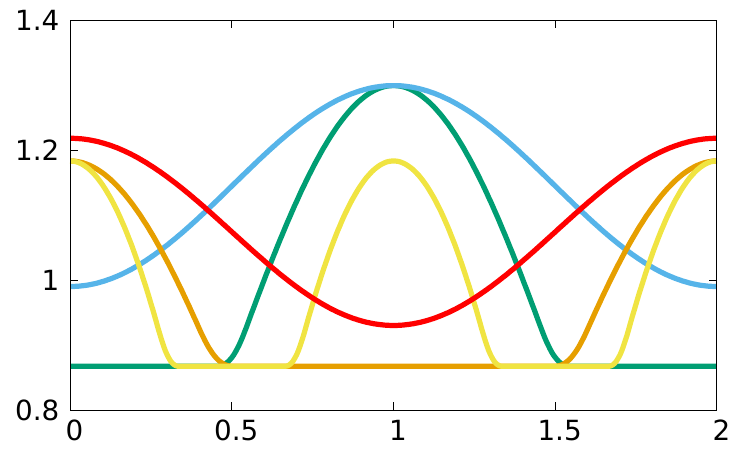}} &
            \rotatebox{90}{\hspace{0.7cm}$\Delta\gamma_p=\pi$}  \\
            & {~~$\Omega t/\pi$} & {~~$\Omega t/\pi$} &
        \end{tabular}
        \caption{Representative pulse profiles of each class. The violet, water green, light blue, orange, yellow, blue, red, black and green curves correspond respectively to classes I, II, III, IV, V, VI, VII, VIII and IX of Table \eqref{tab:Clases}.}
        \label{fig:Pulso_Varios}
    \end{figure}

    However, when we look at Figure \ref{fig:Pulso_Varios}, where pulse profiles with caps in different configurations are included, we see that, contrary to what happened in Beloborodov's classification, by allowing the caps to be not only not antipodal but also not on the same meridian (i.e., $\Delta\gamma_p=\gamma_{p,p}-\gamma_{p,s}\neq0$) we have that pulses of the same class present distinctly different profiles in terms of their morphology. 
    
    To solve this problem, we have elaborated the following subclassifications based on the scheme presented in Table \ref{tab:Clases}, taking into account only some of the most evident morphological differences between profiles of the same class.
    
    Class I:
    \begin{itemize}
        \item a: The caps are in phase opposition, so that the decay in brightness of one counteracts the increase in the other, resulting in a constant profile. 
        \item b: The caps are in phase conjunction or out of phase, so that their luminosity is not counterbalanced, resulting in a non-constant profile.
    \end{itemize}
    
    Class II and IV:
    \begin{itemize}
        \item a: The caps are out of phase, so that the profile has two distinguishable peaks.
        \item b: The caps are out of phase, so that the profile has a single peak and periods of constant flux.
        \item c:  The caps are in conjunction, so that the profile presents a single peak with two characteristic slopes. 
    \end{itemize}
    
    Class V:
    \begin{itemize}
        \item a: The caps do not share a period of visibility, resulting in two peaks separated by two periods of zero flux.
        \item b: The caps share a visibility period interval and have intervals where they are the only one visible, resulting in two overlapping peaks and a single zero flux period.
        \item c: The visibility period of one of the caps is completely contained within the visibility period of the second, resulting in a single peak and a single period of zero flux.
    \end{itemize}
    
\begin{figure}
        \centering
        \begin{tabular}{cccc}
            & Class I a & Class I b &  \\
            \rotatebox{90}{\hspace{0.3cm} Flux} &  
            {\includegraphics[width=2.0cm,height=1.4cm,,trim=0 0 0 0]{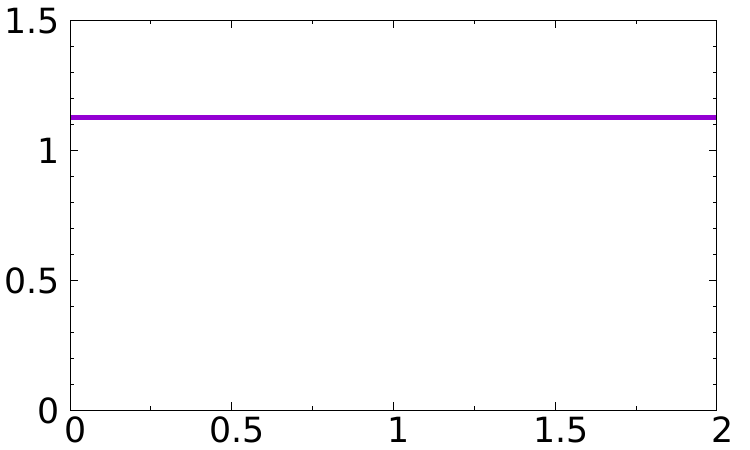}} &
            {\includegraphics[width=2.0cm,height=1.4cm,,trim=0 0 0 0]{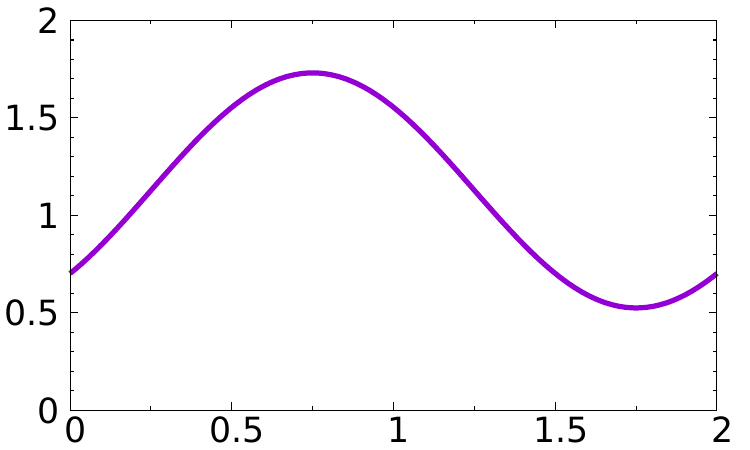}} &
             \\
            & Class II and IV a & Class II and IV b & Class II and IV c \\
            \rotatebox{90}{\hspace{0.3cm} Flux} &  
            {\includegraphics[width=2.0cm,height=1.4cm,,trim=0 0 0 0]{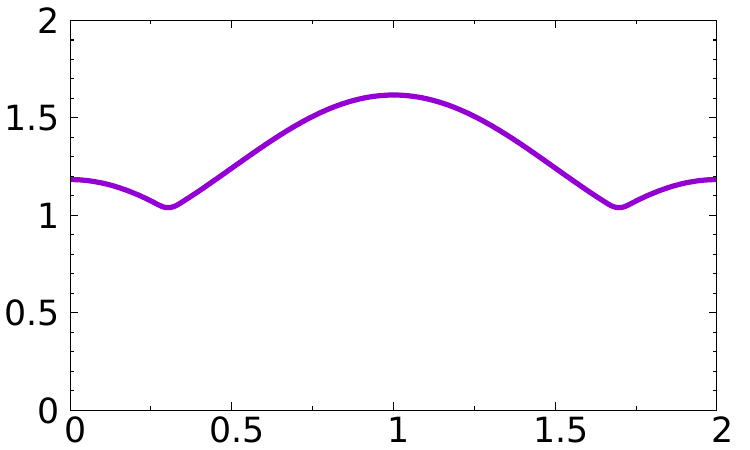}} &
            {\includegraphics[width=2.0cm,height=1.4cm,,trim=0 0 0 0]{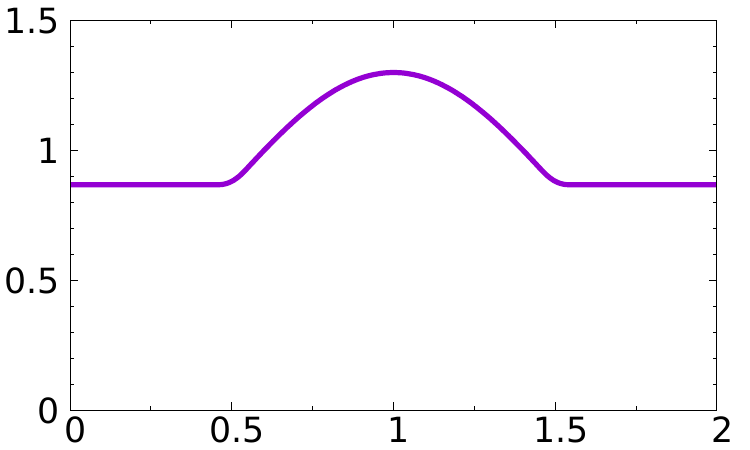}} &
            {\includegraphics[width=2.0cm,height=1.4cm,,trim=0 0 0 0]{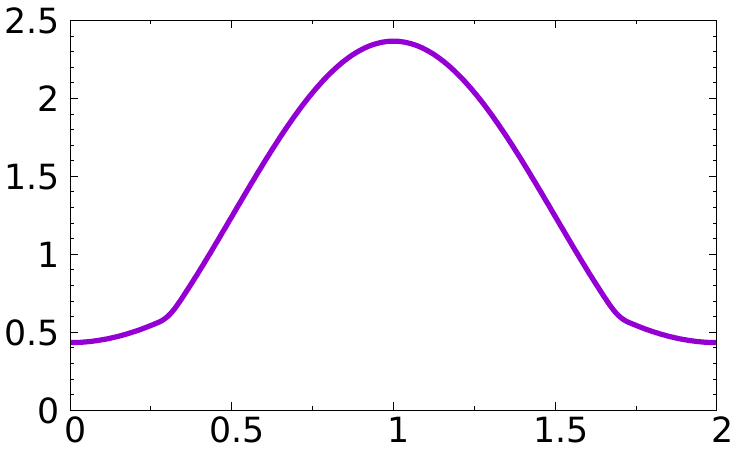}} \\
            & Class V a & Class V b & Class V c \\
            \rotatebox{90}{\hspace{0.3cm} Flux} &  
            {\includegraphics[width=2.0cm,height=1.4cm,,trim=0 0 0 0]{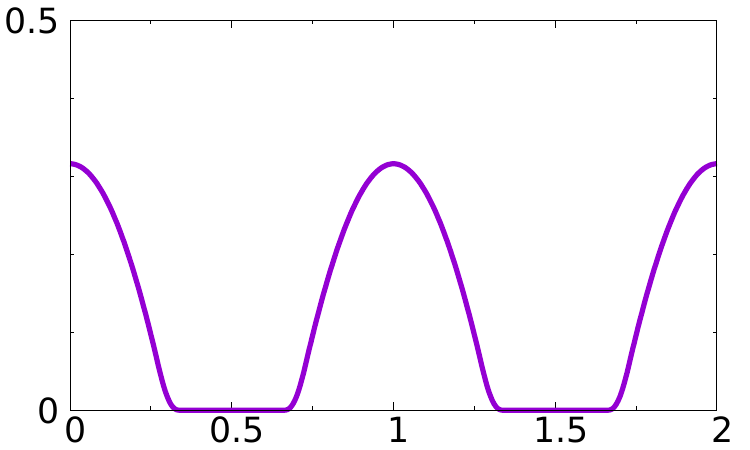}} &
            {\includegraphics[width=2.0cm,height=1.4cm,,trim=0 0 0 0]{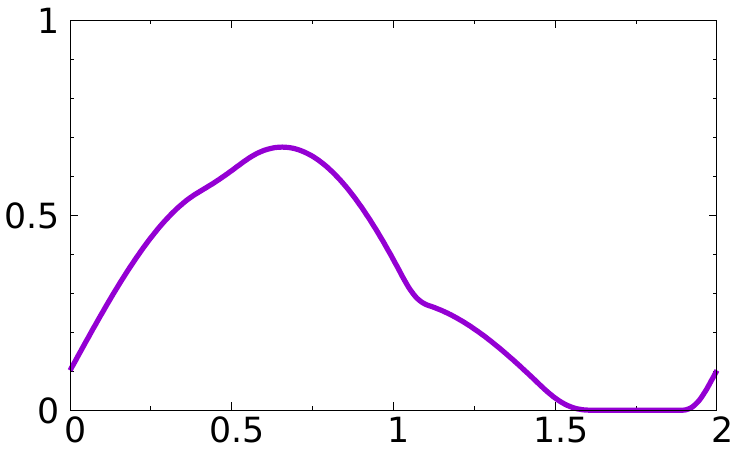}} &
            {\includegraphics[width=2.0cm,height=1.4cm,,trim=0 0 0 0]{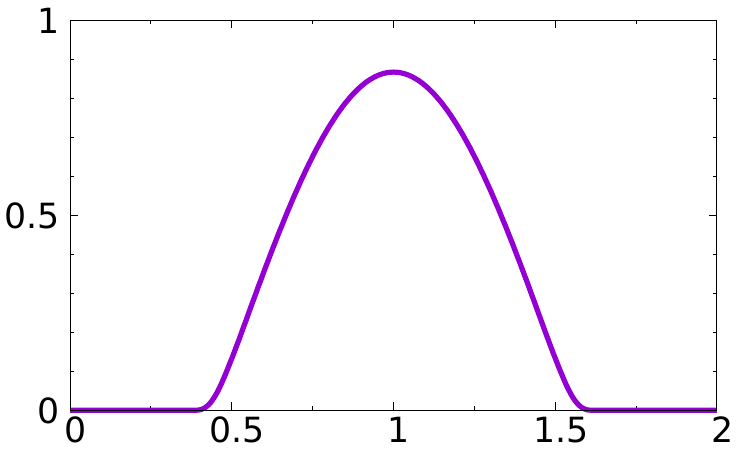}} \\
            & {~~$\Omega t/\pi$} & {~~$\Omega t/\pi$} & {~~$\Omega t/\pi$} 
        \end{tabular}
        \caption{Representative pulse profiles of each subclass.}
        \label{fig:Pulso_subclases}
    \end{figure}

    To close this section, in Figure \ref{fig:Pulso_subclases} we show some characteristic pulse profiles for the subclasses just mentioned.

\section{Conclusions}
\label{c2s8}    

    A correction to the cosine relation presented in \cite{Beloborodov_2002} [Eq. \eqref{eq:Belo}] was found to now include the refraction effects produced by a plasma environment whose density is given by a power law, arriving at Eq. \eqref{eq:BeloPlas}, whose exact form depends on the plasma profile $h$, the metric model and the quotient between frequencies $\epsilon$. This formulation was applied to the study of pulsars, allowing a much simpler treatment of the problems seen above by replacing the complicated numerical integrations \cite{Turolla_2013}, giving simple expressions for the deflection angle $\beta$ and for $\theta$ in terms of $r$ and the impact parameter $x$. The study of the errors in $\theta$ showed that our approximation is effectively an accurate improvement of the Beloborodov's formula when considering plasma environments, presenting different levels of accuracy depending on the compactness of the star, the density and distribution of the plasma or the metric model. In general, the errors are acceptable for stars of low compactness ($R\geq6$) with low frequency ratios ($\epsilon\leq0.3$). 
    We emphasize that the analytical approximations obtained can be used to considerably simplify the calculation of pulse profiles produced by
    {circular and homogeneous caps}, and even combinations of these, in environments where the plasma has a significant impact on light deflection, under realistic compactness and electric charge conditions, thus facilitating the study and understanding of these systems. In our case, we gave a demonstration of its use to solve the profile produced by ring-shaped caps \cite{Sotani_2019, Sotani_2020}, but this is only one of the wide range of models that can be treated with this approach.
    
    Given these new facilities, we were able to consider neutron stars with two non-antipodal, non-identical caps, which gave rise to new pulse profiles that did not fit Beloborodov's classification scheme. For this reason we developed our own classification system, taking into account the new peculiarities of the model. The main effect was the enlargement of the non-specular area of the class map in the $\chi-\xi$ plane, i.e., the area that cannot be obtained as a reflection or translation of the others. At the same time, due to the flexibility in the model configuration, many profiles of the same class have distinctly different morphologies, so we have extended the classification into different subclasses. 
    
    We also mention that one or more of the assumptions made in our model are likely to be violated in nature, so it is important to know their range of validity. It should be noted that the formalism presented here can only be applied to slowly rotating neutron stars. Otherwise, metrics such as Kerr's or similar ones must be considered. 
    Moreover, the caps of a realistic pulsar have no need to emit isotropically according to Lambert's law ($f_B(\delta)=1$), either because the star is covered by an atmosphere, or because the emissivity is strongly constrained to energies below the electron plasma frequency. Realistic emission models predict an angular dependence in the emitted intensity. An interesting possibility is that the emission originates above the neutron star surface, generating emission with $\delta>\pi/2$ \cite{Pechenick_1983}.
    A more realistic model would have to take into account the effect of the strong magnetic fields in the vicinity of the star on the propagation of light rays~\cite{BE_1980,2003BroBla,Broderick:2003fc}.  A study of this nature has been made in the past to study light rays in plasma environment around black holes \cite{2003BroBla}. 
    
    Even though our formulas and plots are expressed in terms of dimensionless coefficients to maintain as much universality as possible, regardless of the specific details of the model, in order to have an rough estimation of the order of magnitude in the involved parameters, let us consider a NS rotating at the angular velocity $\Omega$ and with an exterior dipole magnetic field $\boldsymbol{B}$. We also assume that the particle number density $N(R)$ of the magnetospheric plasma at the surface of the star is proportional to the Goldriech-Julian density $N_{GJ}(R)$, i.e., 
    \begin{equation}
        N(R)=\kappa N_{GJ}(R)=2\kappa\epsilon_0\frac{\Omega B}{e},
    \end{equation}
    with $\kappa$ the so called multiplication parameter, taking values in the range $10^3-10^5$ \cite{Beskin:2000si,gurevich_beskin_istomin_1993} and $B$ the characteristic value of magnetic field strength  at the surface $R$ \cite{Petri:2016tqe},
    \begin{equation}
        B=\frac{1}{R^3}\sqrt{\frac{3\mu_0c^3}{32\pi^3}I P\dot{P}},
    \end{equation}
    with $I=2/5MR^2$ the Newtonian moment of inertia of the star (assumed as a rigid sphere).
    Moreover, due to the relativistic motion of electrons and positrons along the magnetic field lines, the angular plasma frequency is decreased by a factor $\bar{\gamma}^{-3/2}$\cite{2001Ap&SS.278...77I},
    where $\bar{\gamma}$ is the average of the  Lorentz factor $\gamma=E/(m_ec^2)$ of the mentioned particles, which takes a minimum value $\bar\gamma\approx 10^2$,
    (\cite{2001Ap&SS.278...77I,gurevich_beskin_istomin_1993,PhysRevE.57.3399}).
         
    Note also that the frequency $\nu_\infty$ of the electromagnetic wave as measured for an asymptotic observer is related to $\epsilon$ by: 
    \begin{equation}
    \nu_\infty=\frac{\nu_e(R)}{\epsilon}K,
    \end{equation}
    with $K=\sqrt{1-2\frac{M}{R}+\frac{q*}{R^2}}$ the gravitational redshift factor.
    In that setting, a NS of mass $M=1.8M_\odot$ with a rotation period $P=1s$ varying as $\dot{P}=10^{-15}$ has the associated parameters shown in Table \eqref{tab:parameters}. Corrections to the Goldriech-Julian density when one consider a RN-like metric are given by \cite{Morozova:2010gg} (see their Eq.(9)). It can be checked that the correction terms are of order $\mathcal{O}(1)$, therefore the estimated values for the parameters in RN-like metrics remain of the same order of magnitude as those presented in Table\eqref{tab:parameters}.
    
    \begin{table}
    {
        \centering
        \begin{tabular}{|c|c|c|}
        \hline  
        Quantity & $R/2M=1.675$ & $R/2M=2$   \\ \hline  
        $M$ ($M_\odot$)    & 1.8                              & 1.8     \\ \hline 
        $R$ (Km)    & 9.04                              & 10.8       \\ \hline 
        $\Omega/(2\pi)$ (Hz)    & 1                            & 1            \\ \hline 
        $\dot{P}$ (s/s)  & $10^{-15}$                               & $10^{-15}$     \\ \hline 
        $\bar\gamma$  & $10^2$                              & $10^2$         \\ \hline 
        $\kappa$  & $10^5$                              & $10^5$         \\ \hline
         $N(R)$ ($m^{-3}$) & $1.02\times 10^{22}$                              &  $7.24\times 10^{21}$                                  \\ \hline 
        $B(R)$ (T)    & $1.48\times 10^8$                            &  $1.04\times 10^8$     \\ \hline 
        $\omega_e(R)/\bar\gamma^{3/2}$ (rad/s)    & $5.72\times 10^9$                            &  $4.79\times 10^9$     \\ \hline 
        $\nu_e(R)=\frac{\omega_e}{2\pi\bar\gamma^{3/2}}$  (MHz)   & 911      & 763  \\ \hline 
        \end{tabular}
        \caption{\label{tab:parameters} Characteristic values of the parameters involved in the description of the plasma magnetosphere for a assumed mass of 1.8 solar masses of the NS.}
        }
    \end{table}
        
    On the other hand, a strong magnetic field leads to an anisotropic emissivity with a preferential direction along the field \cite{Beloborodov_2002}. In such a situation, one surely should abandon semi-analytic models and perform a full numerical integration of the ray path equations taking into account a dielectric permittivity tensor.
    
    Despite all these restrictions on the range of validity and being limited to extreme cases, the present work constitutes a simple model that allows us to describe and approximate to a large extent the photon dynamics around compact objects embedded in plasma media, achieving a better understanding and comprehension of the different morphologies in the light curves and their dependence on different physical and geometric parameters involved.
    
   Finally, we would like to emphasize that the generalized Beloborodov formula that we have obtained can also be applied to the study of polarization of light rays in plasma environments, as well as to determine luminosity profiles associated with accretion disks around black holes and other possible compact objects. It is worth mentioning that in the last years alternative formulas to Beloborodov's (many of them empirical) that improve it have also been proposed in the literature~ \cite{Mutka:2002ji,Frolov:2004cu,Amore:2006pi,Semerak:2014kra,DeFalco:2016yox,LaPlaca:2019rjz,Poutanen:2019tcd}. It would be desirable to be able to obtain its analogs in situations where a plasma medium (not necessarily a cold plasma) is present.
    
\section*{Acknowledgements}

We acknowledge support from CONICET, SeCyT-UNC and
Foncyt.

\appendix

\section{ Other plasma distributions}
\label{AppendixA}

    In this appendix we show the graphs corresponding to $h=2$ and to a plasma density with exponential decay, according to the expressions
    \begin{equation}
        n_2^2(r)=1-\frac{A(r)}{A(R)}\left(\frac{R}{r}\right)^2\epsilon^2
        \label{eq:NS_n_h=2}
    \end{equation}
    and
    \begin{equation}
        n_e^2(r)=1-\frac{A(r)}{A(R)}e^{R-r}\epsilon^2.
        \label{eq:NS_n_exp}
    \end{equation}
    
    In order to use the analytical approximation in the plasma profile with exponential decay, we will assume a density of the form of Eq. \eqref{eq:Nr} with an $h$ such that both the refractive index $n$ and its first derivative with respect to $r$ match those of the exponential profile over the stellar surface. It can be seen that this is achieved by taking $h=2R$.
    
    Since the analytical approximation was developed to fit a plasma profile with a density given by a power law, it is to be expected that it does not correctly approximate the trajectories and pulse profiles obtained from a plasma profile with exponential decay, as can be seen in the tables shown in this appendix. However, we include it as a practical example.

    \begin{figure}
        \centering
        \begin{tabular}{cccc}
            & $R/2M=1.675$ & $R/2M=2$ & \\
            \rotatebox{90}{\hspace{0.5cm}Flux ($\cdot100$)} &  
            {\includegraphics[width=3.1cm,height=2.2cm,trim=0 0 0 0]{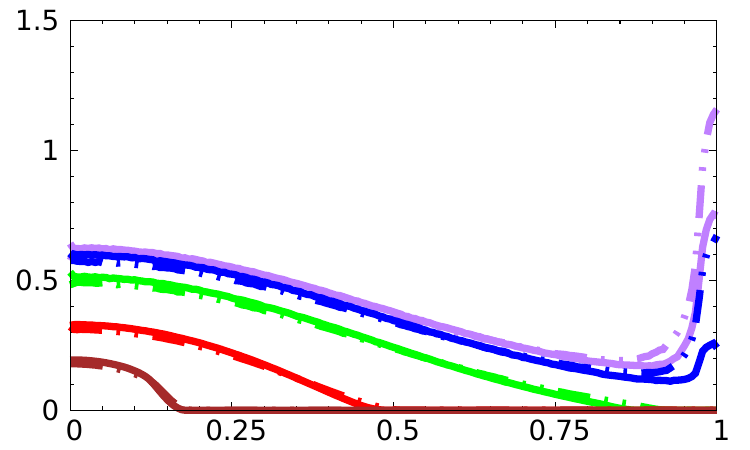}} &
            {\includegraphics[width=3.1cm,height=2.2cm,trim=0 0 0 0]{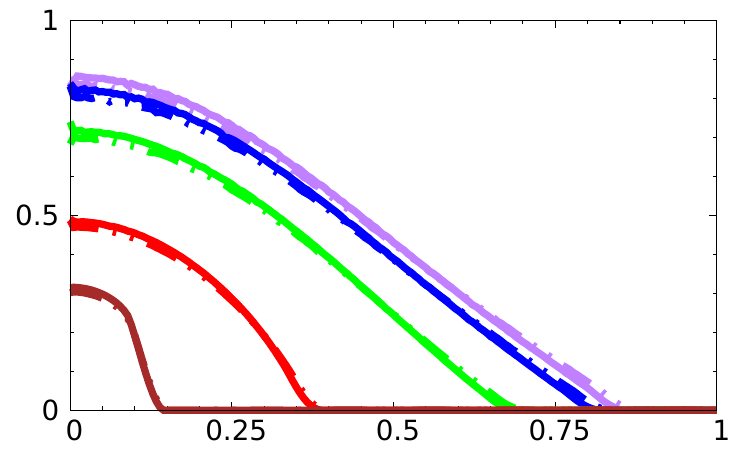}} &
            \rotatebox{90}{\hspace{0.15cm}$|q^*|=(0.25M)^2$} \\
            \rotatebox{90}{\hspace{0.5cm}Flux ($\cdot100$)} &  
            {\includegraphics[width=3.1cm,height=2.2cm,trim=0 0 0 0]{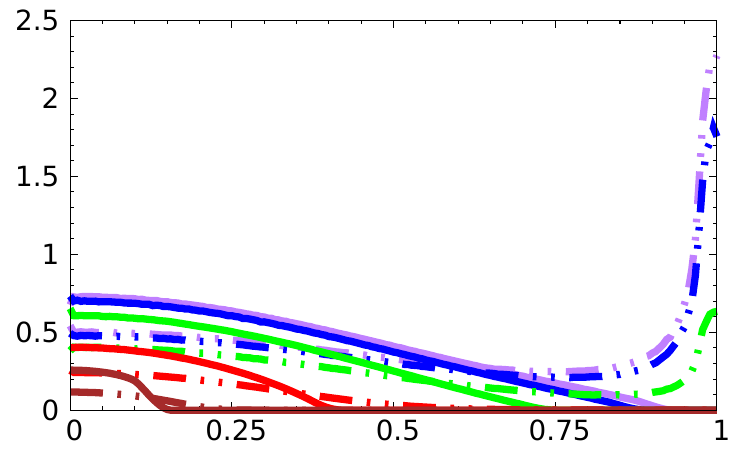}} &
            {\includegraphics[width=3.1cm,height=2.2cm,trim=0 0 0 0]{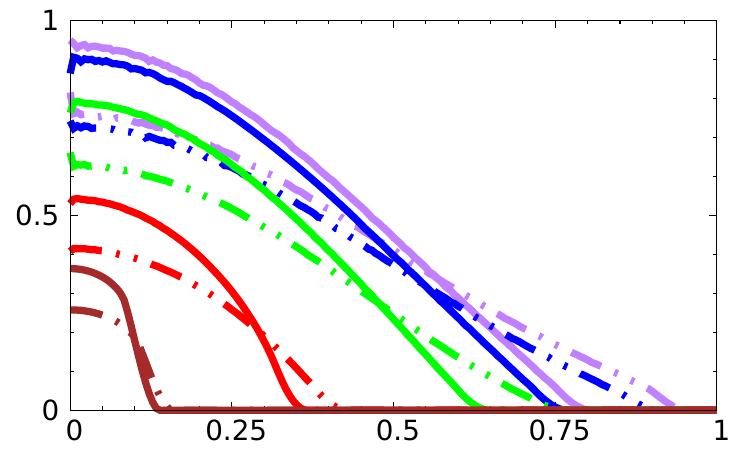}} &
            \rotatebox{90}{\hspace{0.15cm}$|q^*|=(0.75M)^2$} \\
            & {~~$\Omega t/\pi$} & {~~$\Omega t/\pi$} &
        \end{tabular}
        \caption{Pulse profile for single caps. $h=2$, $\xi=\chi=\pi/2$. The purple, blue, green, green, red, and brown lines correspond to $\epsilon=0.00$, $0.30$, $0.60$, $0.90$ and $0.99$ respectively. The solid lines correspond to $q^*>0$ while the dashed lines correspond to $q^*<0$.}
        \label{fig:PP_SHS_h=2}
    \end{figure}
    
    \begin{figure}
        \centering
        \begin{tabular}{cccc}
            & $R/2M=1.675$ & $R/2M=2$ & \\
            \rotatebox{90}{\hspace{0.5cm}Flux ($\cdot100$)} &  
            {\includegraphics[width=3.1cm,height=2.2cm,trim=0 0 0 0]{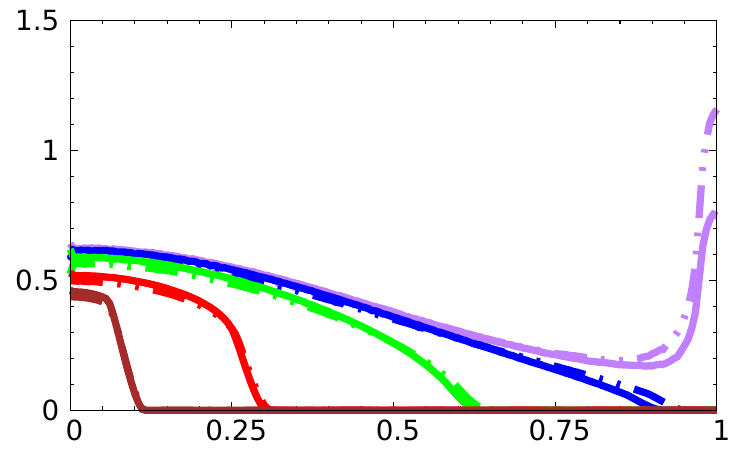}} &
            {\includegraphics[width=3.1cm,height=2.2cm,trim=0 0 0 0]{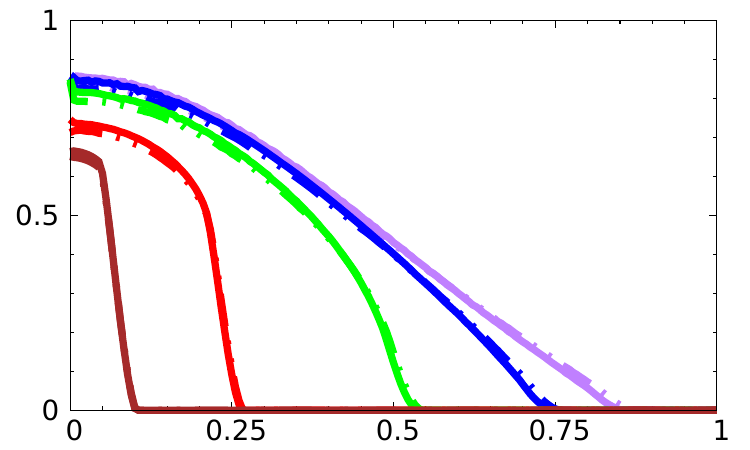}} &
            \rotatebox{90}{\hspace{0.15cm}$|q^*|=(0.25M)^2$} \\
            \rotatebox{90}{\hspace{0.5cm}Flux ($\cdot100$)} &  
            {\includegraphics[width=3.1cm,height=2.2cm,trim=0 0 0 0]{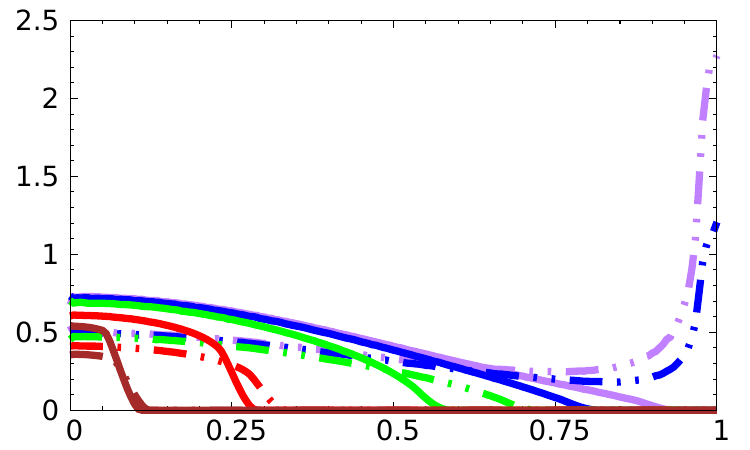}} &
            {\includegraphics[width=3.1cm,height=2.2cm,trim=0 0 0 0]{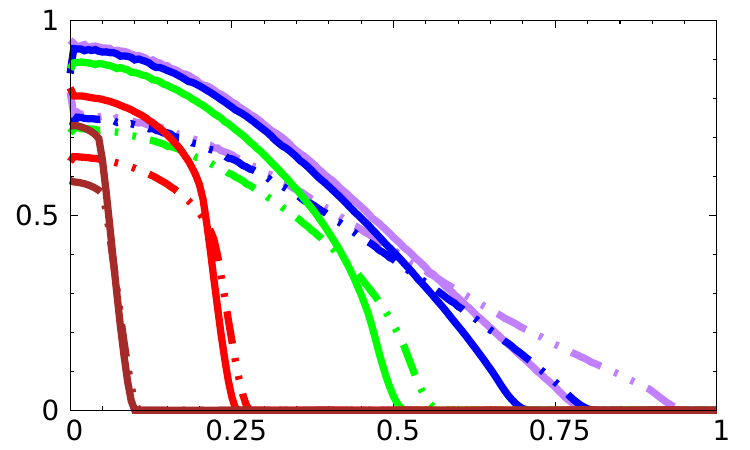}} &
            \rotatebox{90}{\hspace{0.15cm}$|q^*|=(0.75M)^2$} \\
            & {~~$\Omega t/\pi$} & {~~$\Omega t/\pi$} &
        \end{tabular}
        \caption{Pulse profile for single caps. Exponential decay, $\xi=\chi=\pi/2$. The purple, blue, green, green, red, and brown lines correspond to $\epsilon=0.00$, $0.30$, $0.60$, $0.90$ and $0.99$ respectively. The solid lines correspond to $q^*>0$ while the dashed lines correspond to $q^*<0$.}
        \label{fig:PP_SHS_exp}
    \end{figure}
    
    
    \begin{figure}
        \centering
        \begin{tabular}{cccc}
            & $R/2M=1.675$ & $R/2M=2$ & \\
            \rotatebox{90}{\hspace{0.5cm}Flux ($\cdot100$)} &  
            {\includegraphics[width=3.1cm,height=2.2cm,trim=0 0 0 0]{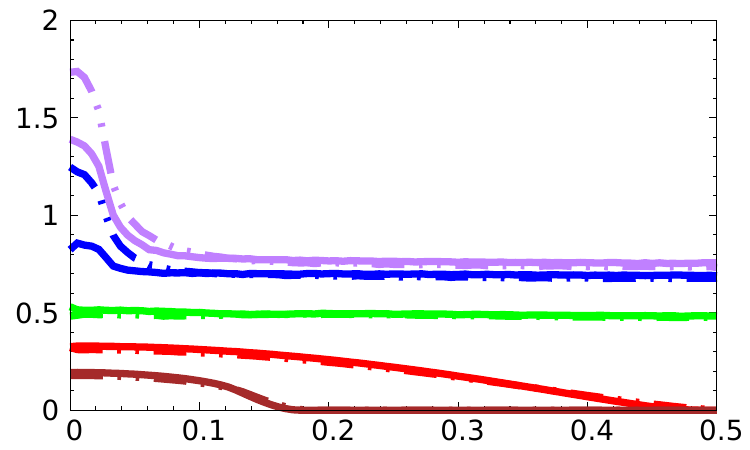}} &
            {\includegraphics[width=3.1cm,height=2.2cm,trim=0 0 0 0]{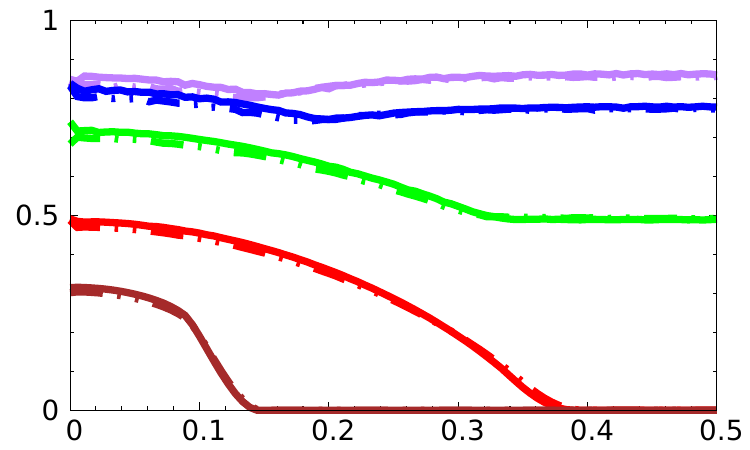}} &
            \rotatebox{90}{\hspace{0.15cm}$|q^*|=(0.25M)^2$} \\
            \rotatebox{90}{\hspace{0.5cm}Flux ($\cdot100$)} &  
            {\includegraphics[width=3.1cm,height=2.2cm,trim=0 0 0 0]{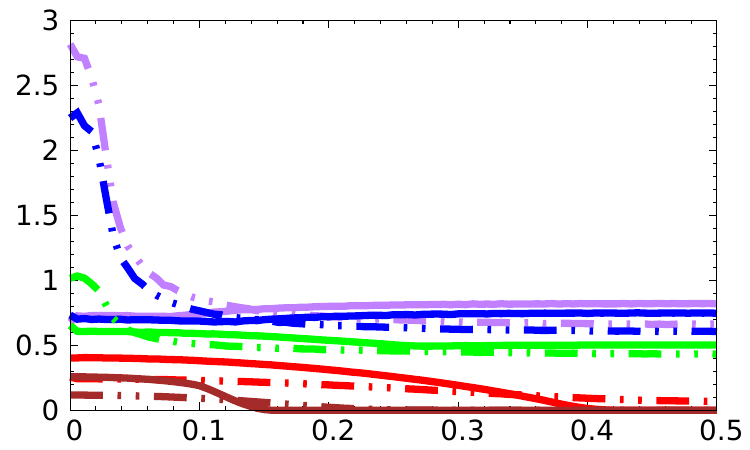}} &
            {\includegraphics[width=3.1cm,height=2.2cm,trim=0 0 0 0]{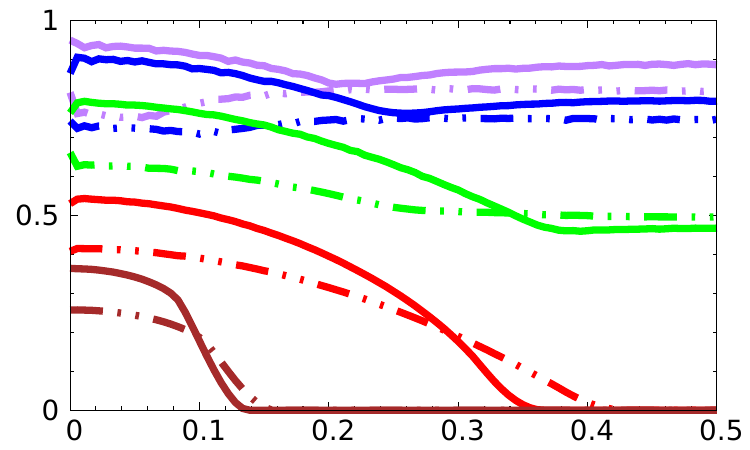}} &
            \rotatebox{90}{\hspace{0.15cm}$|q^*|=(0.75M)^2$} \\
            & {~~$\Omega t/\pi$} & {~~$\Omega t/\pi$} &
        \end{tabular}
        \caption{Pulse profile for antipodal caps. $h=2$, $\xi=\chi=\pi/2$. Purple, blue, green, green, red and brown lines correspond to $\epsilon=0.00$, $0.30$, $0.60$, $0.90$ and $0.99$ respectively. The solid lines correspond to $q^*>0$ while the dashed lines correspond to $q^*<0$.}
        \label{fig:PP_DHS_90_90_h=2}
    \end{figure}
    
    \begin{figure}
        \centering
        \begin{tabular}{cccc}
            & $R/2M=1.675$ & $R/2M=2$ & \\
            \rotatebox{90}{\hspace{0.5cm}Flux ($\cdot100$)} &  
            {\includegraphics[width=3.1cm,height=2.2cm,trim=0 0 0 0]{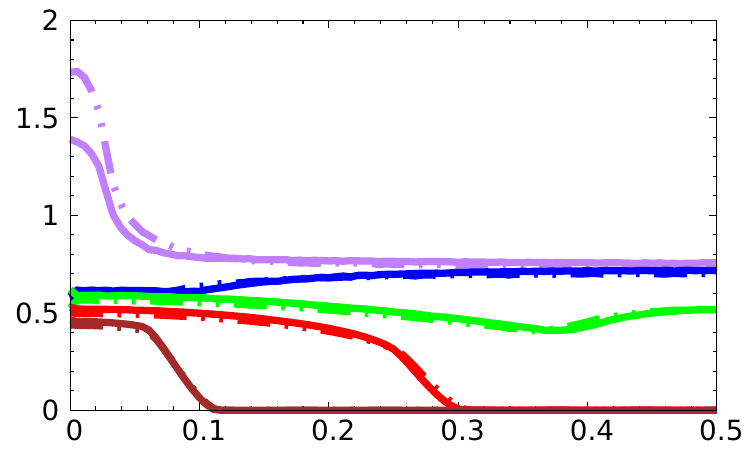}} &
            {\includegraphics[width=3.1cm,height=2.2cm,trim=0 0 0 0]{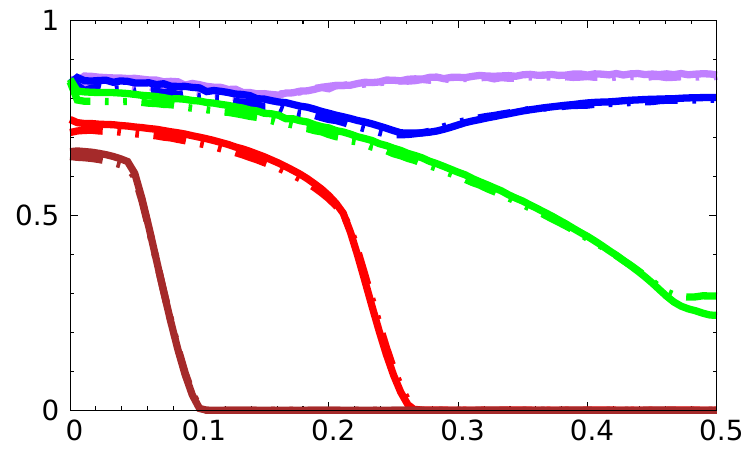}} &
            \rotatebox{90}{\hspace{0.15cm}$|q^*|=(0.25M)^2$} \\
            \rotatebox{90}{\hspace{0.5cm}Flux ($\cdot100$)} &  
            {\includegraphics[width=3.1cm,height=2.2cm,trim=0 0 0 0]{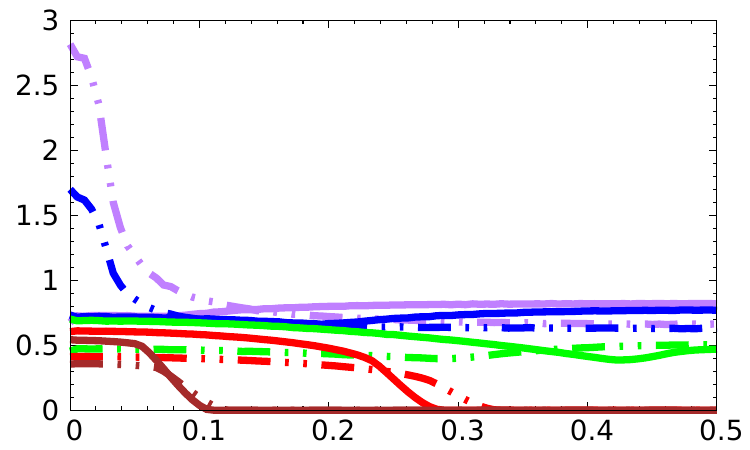}} &
            {\includegraphics[width=3.1cm,height=2.2cm,trim=0 0 0 0]{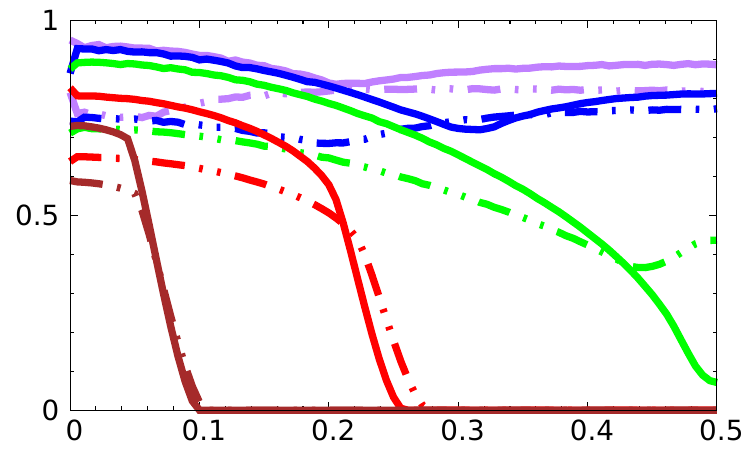}} &
            \rotatebox{90}{\hspace{0.15cm}$|q^*|=(0.75M)^2$} \\
            & {~~$\Omega t/\pi$} & {~~$\Omega t/\pi$} &
        \end{tabular}
        \caption{Pulse profile for antipodal caps. Exponential decay, $\xi=\chi=\pi/2$. Purple, blue, green, green, red and brown lines correspond to $\epsilon=0.00$, $0.30$, $0.60$, $0.90$ and $0.99$ respectively. The solid lines correspond to $q^*>0$ while the dashed lines correspond to $q^*<0$.}
        \label{fig:PP_DHS_90_90_exp}
    \end{figure}
    
    
    \begin{figure}
        \centering
        \begin{tabular}{cccc}
            & $|q|^*=(0.25M)^2$ & $|q|^*=(0.75M)^2$ & \\
            \rotatebox{90}{\hspace{0.5cm}Flux ($\cdot100$)} &  
            {\includegraphics[width=3.1cm,height=2.2cm,trim=0 0 0 0]{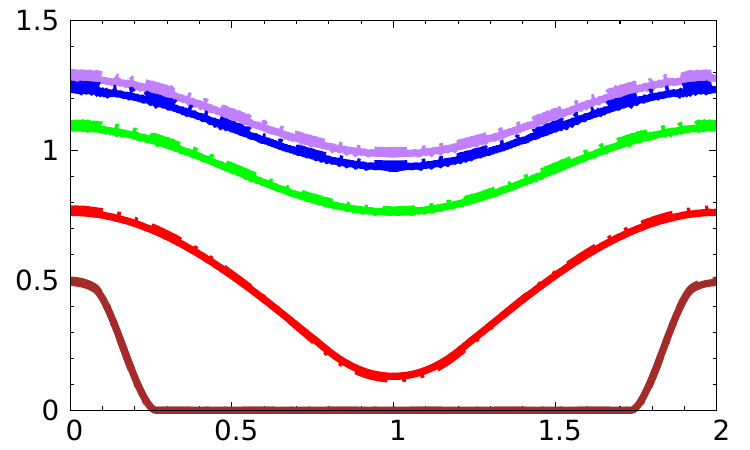}} &
            {\includegraphics[width=3.1cm,height=2.2cm,trim=0 0 0 0]{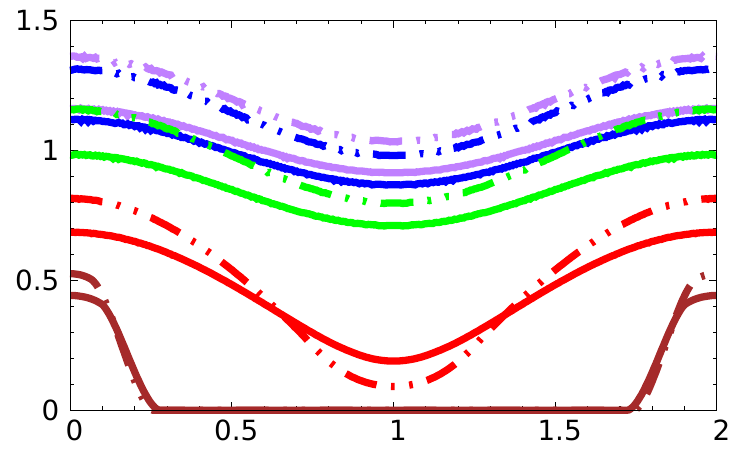}} &
            \rotatebox{90}{\hspace{0.3cm}$\chi=\frac{\pi}{9}$, $\xi=\frac{\pi}{6}$} \\
            \rotatebox{90}{\hspace{0.5cm}Flux ($\cdot100$)} &  
            {\includegraphics[width=3.1cm,height=2.2cm,trim=0 0 0 0]{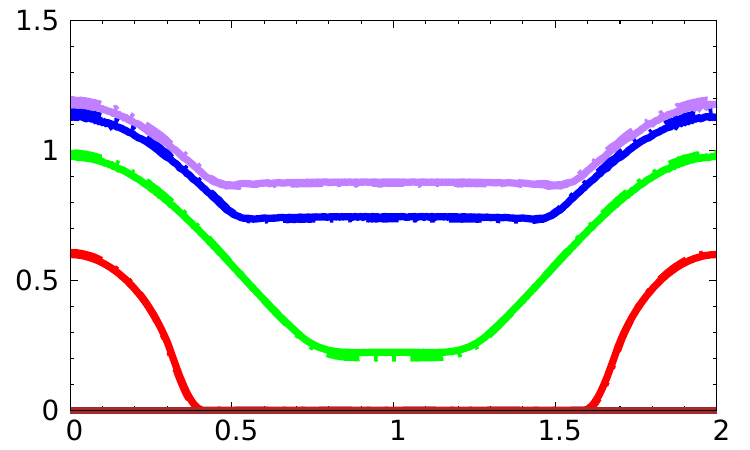}} &
            {\includegraphics[width=3.1cm,height=2.2cm,trim=0 0 0 0]{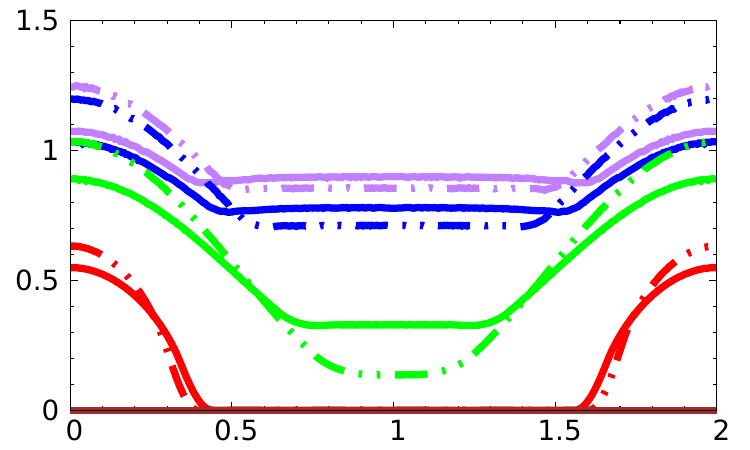}} &
            \rotatebox{90}{\hspace{0.3cm}$\chi=\frac{\pi}{6}$, $\xi=\frac{\pi}{3}$} \\
            \rotatebox{90}{\hspace{0.5cm}Flux ($\cdot100$)} &  
            {\includegraphics[width=3.1cm,height=2.2cm,trim=0 0 0 0]{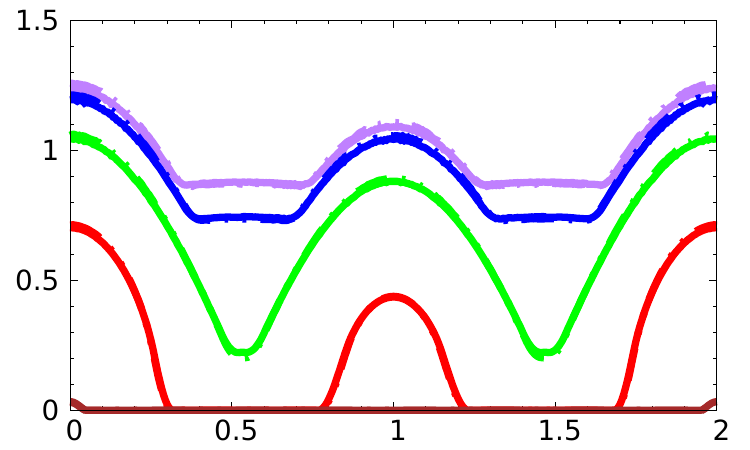}} &
            {\includegraphics[width=3.1cm,height=2.2cm,trim=0 0 0 0]{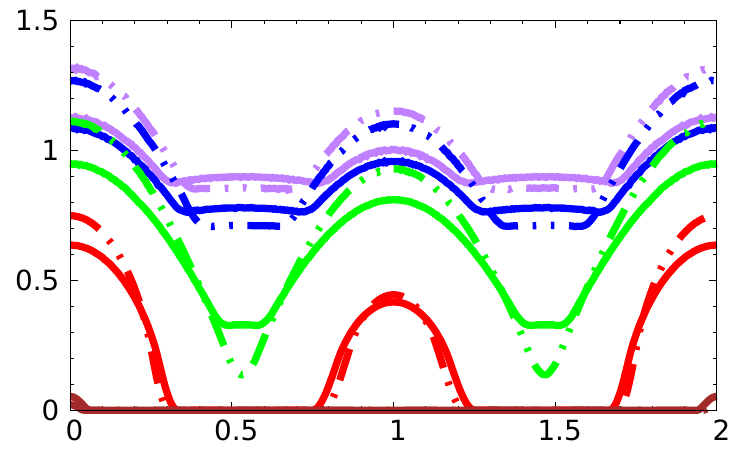}} &
            \rotatebox{90}{\hspace{0.3cm}$\chi=\frac{\pi}{3}$, $\xi=\frac{4\pi}{9}$} \\
            \rotatebox{90}{\hspace{0.5cm}Flux ($\cdot100$)} &  
            {\includegraphics[width=3.1cm,height=2.2cm,trim=0 0 0 0]{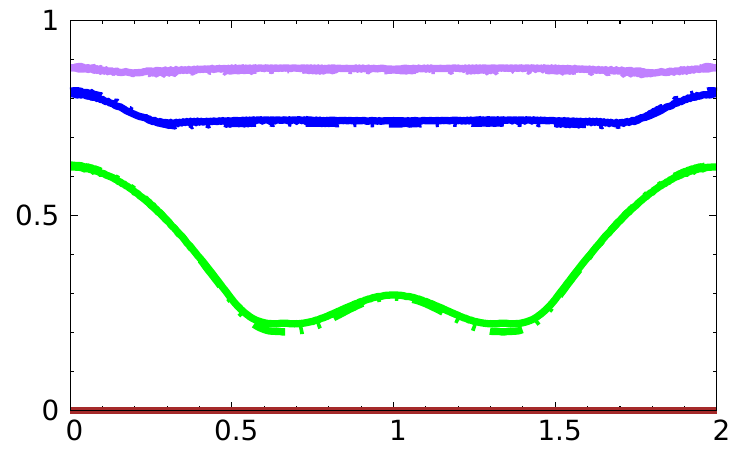}} &
            {\includegraphics[width=3.1cm,height=2.2cm,trim=0 0 0 0]{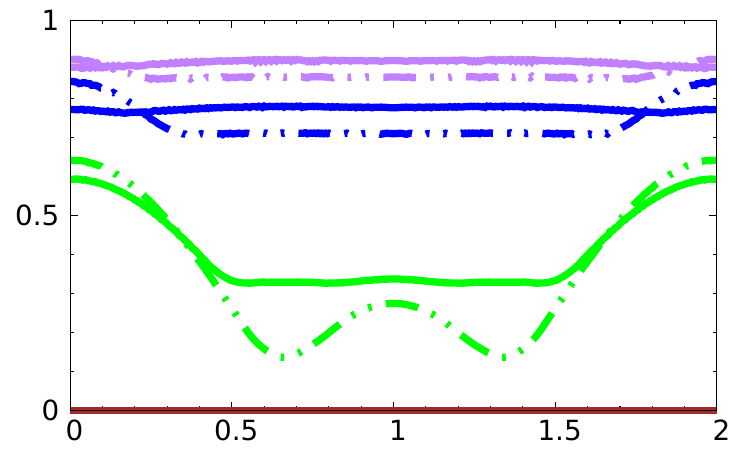}} &
            \rotatebox{90}{\hspace{0.3cm}$\chi=\frac{\pi}{9}$, $\xi=\frac{4\pi}{9}$} \\
            & {~~$\Omega t/\pi$} & {~~$\Omega t/\pi$} &
        \end{tabular}
        \caption{Pulse profile for antipodal caps. $h=2$, $R/r_h=3$. Purple, blue, green, green, red and brown lines correspond to $\epsilon=0.00$, $0.30$, $0.60$, $0.90$ and $0.99$ respectively. The solid lines correspond to $q^*>0$ while the dashed lines correspond to $q^*<0$.}
        \label{fig:PP_DHS_h=2}
    \end{figure}
    
    \begin{figure}
        \centering
        \begin{tabular}{cccc}
            & $|q|^*=(0.25M)^2$ & $|q|^*=(0.75M)^2$ &  \\
            \rotatebox{90}{\hspace{0.5cm}Flux ($\cdot100$)} &  
            {\includegraphics[width=3.1cm,height=2.2cm,trim=0 0 0 0]{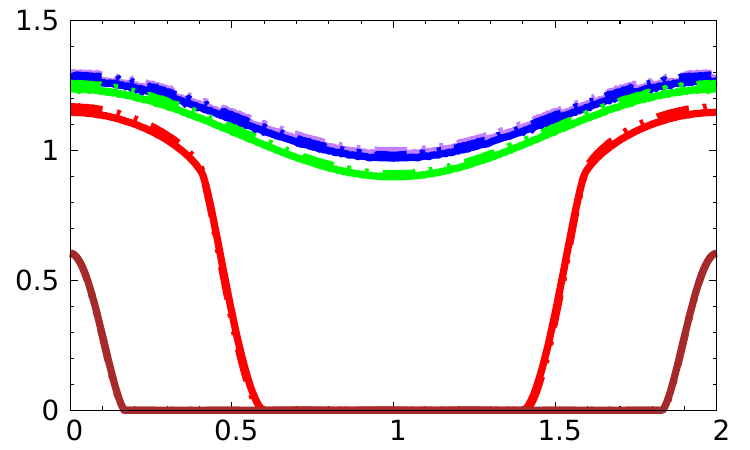}} &
            {\includegraphics[width=3.1cm,height=2.2cm,trim=0 0 0 0]{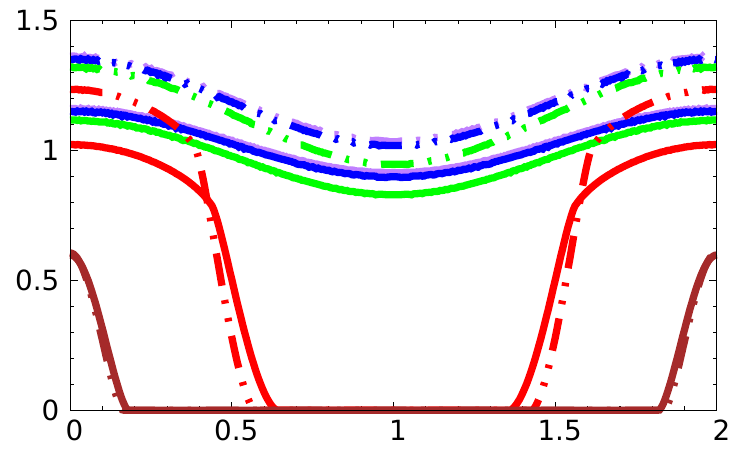}} &
            \rotatebox{90}{\hspace{0.3cm}$\chi=\frac{\pi}{9}$, $\xi=\frac{\pi}{6}$} \\
            \rotatebox{90}{\hspace{0.5cm}Flux ($\cdot100$)} &  
            {\includegraphics[width=3.1cm,height=2.2cm,trim=0 0 0 0]{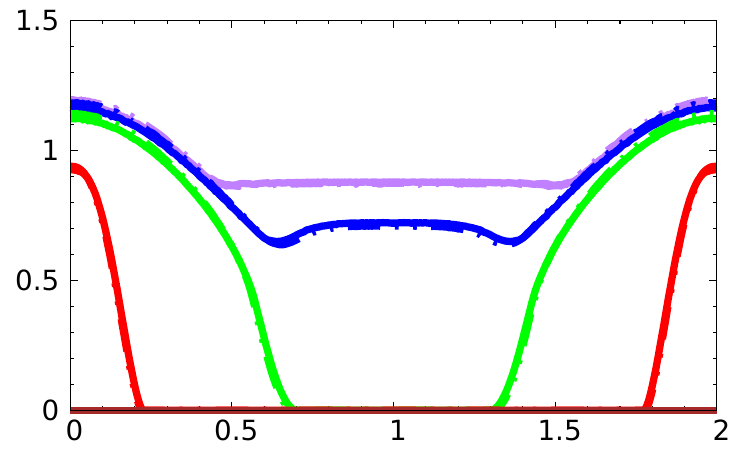}} &
            {\includegraphics[width=3.1cm,height=2.2cm,trim=0 0 0 0]{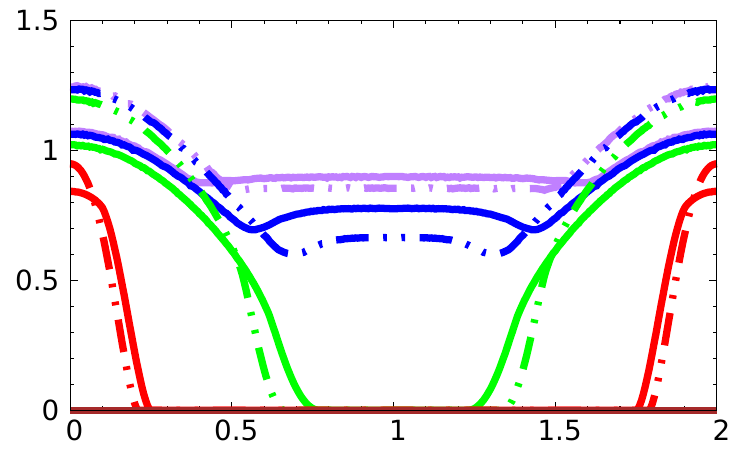}} &
            \rotatebox{90}{\hspace{0.3cm}$\chi=\frac{\pi}{6}$, $\xi=\frac{\pi}{3}$} \\
            \rotatebox{90}{\hspace{0.5cm}Flux ($\cdot100$)} &  
            {\includegraphics[width=3.1cm,height=2.2cm,trim=0 0 0 0]{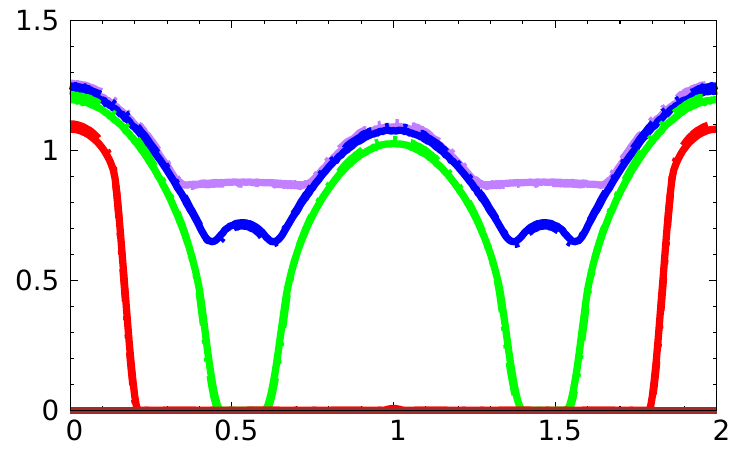}} &
            {\includegraphics[width=3.1cm,height=2.2cm,trim=0 0 0 0]{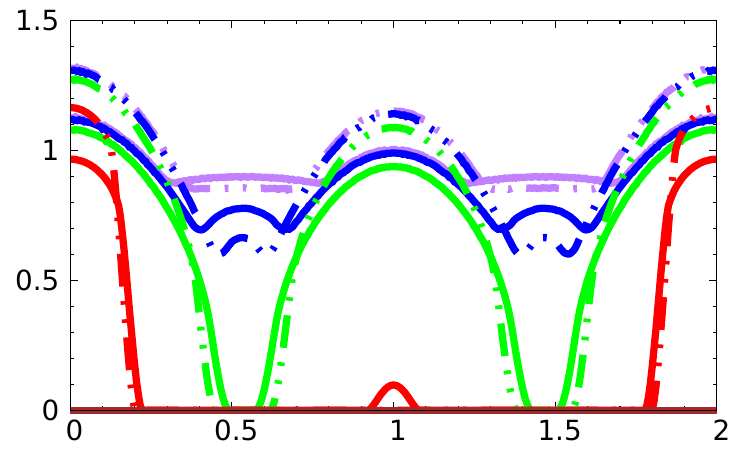}} &
            \rotatebox{90}{\hspace{0.3cm}$\chi=\frac{\pi}{3}$, $\xi=\frac{4\pi}{9}$} \\
            \rotatebox{90}{\hspace{0.5cm}Flux ($\cdot100$)} &  
            {\includegraphics[width=3.1cm,height=2.2cm,trim=0 0 0 0]{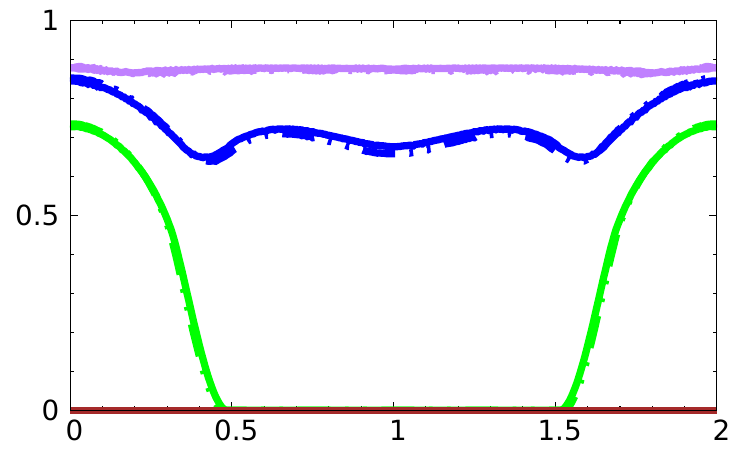}} &
            {\includegraphics[width=3.1cm,height=2.2cm,trim=0 0 0 0]{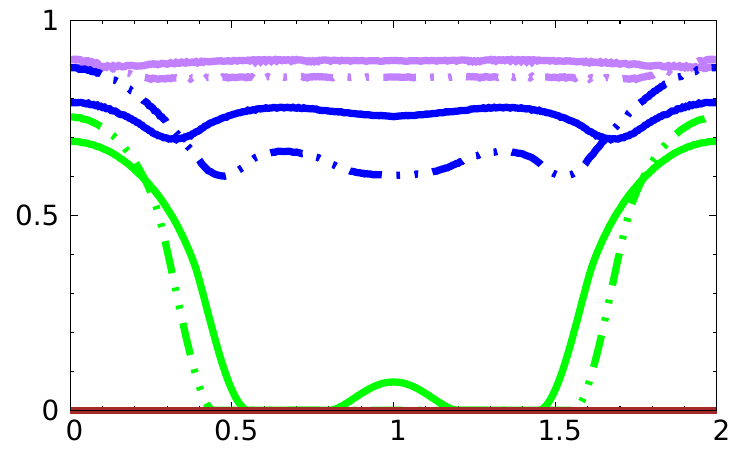}} &
            \rotatebox{90}{\hspace{0.3cm}$\chi=\frac{\pi}{9}$, $\xi=\frac{4\pi}{9}$} \\
            & {~~$\Omega t/\pi$} & {~~$\Omega t/\pi$} &
        \end{tabular}
        \caption{Pulse profile for antipodal caps. Exponential decay, $R/r_h=3$. Purple, blue, green, green, red and brown lines correspond to $\epsilon=0.00$, $0.30$, $0.60$, $0.90$ and $0.99$ respectively. The solid lines correspond to $q^*>0$ while the dashed lines correspond to $q^*<0$.}
        \label{fig:PP_DHS_exp}
    \end{figure}

    \begin{figure}
        \centering
        \begin{tabular}{cccc}
            & Sch & RN-like & \\
            \rotatebox{90}{\hspace{0.95cm}$\theta/\pi$} &  
            {\includegraphics[width=3.2cm,height=2.3cm,trim=0 0 0 0]{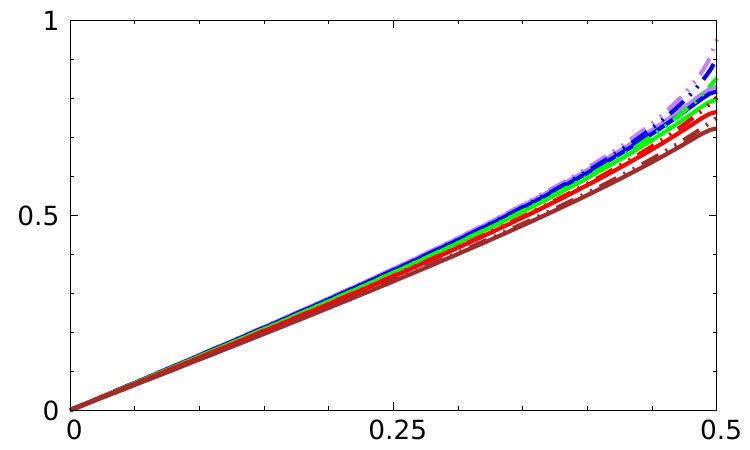}} &
            {\includegraphics[width=3.2cm,height=2.3cm,trim=0 0 0 0]{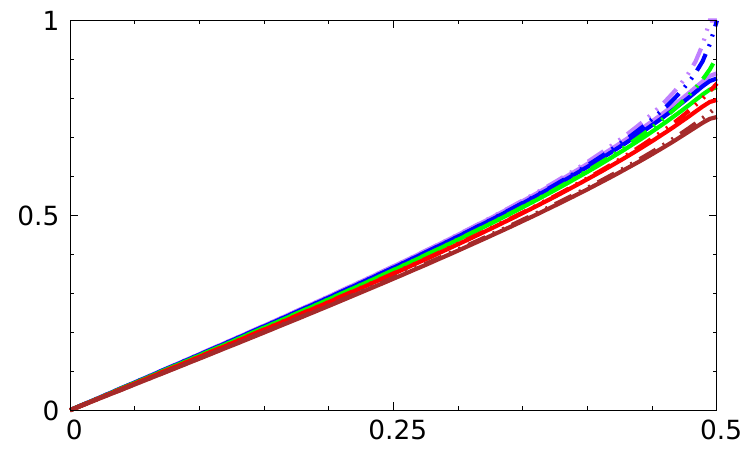}} &
            \rotatebox{90}{\hspace{0.85cm}$h=2$} \\
            \rotatebox{90}{\hspace{0.95cm}$\theta/\pi$} &  
            {\includegraphics[width=3.2cm,height=2.3cm,trim=0 0 0 0]{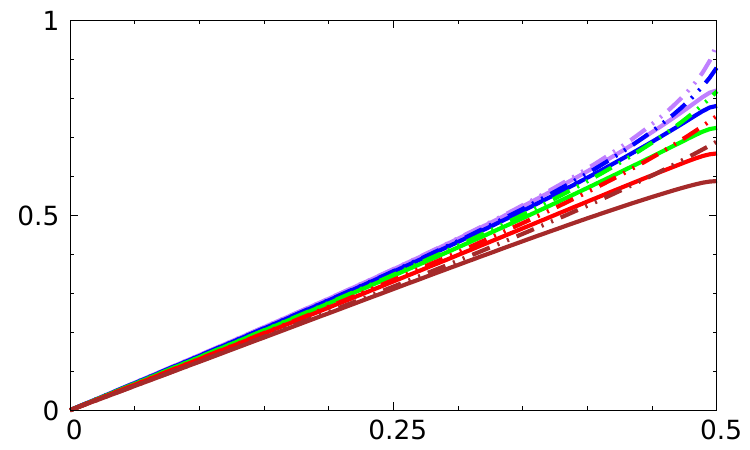}} &
            {\includegraphics[width=3.2cm,height=2.3cm,trim=0 0 0 0]{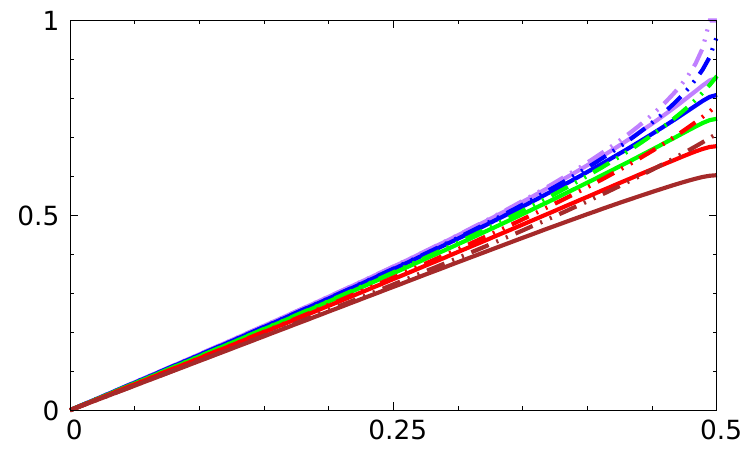}} &
            \rotatebox{90}{\hspace{1.2cm}Exp} \\
            & {~~$\delta/\pi$} & {~~$\delta/\pi$} &  
        \end{tabular}
        \caption{$\theta$ as a function of $\delta$. $q^*=-(0.50M)^2$, $R=4M$. The purple, blue, green, red and brown lines correspond respectively to $\epsilon=0.1$, $0.2$, $0.3$, $0.4$ and $0.5$. The continuous lines were obtained by numerical integration from Eq. \eqref{eq:t(b)} while the dashed lines were obtained analytically from Eq. \eqref{eq:BeloPlas} (with plasma correction).}
        \label{fig:Ap_AA_TR_tvd}
    \end{figure}
    
    \begin{figure}
        \centering
        \begin{tabular}{cccc}
            & Sch & RN-like &\\
            \rotatebox{90}{\hspace{0.4cm}$\Delta\theta_{max}/\theta_{max}$} &  
            {\includegraphics[width=3.2cm,height=2.3cm,trim=0 0 0 0]{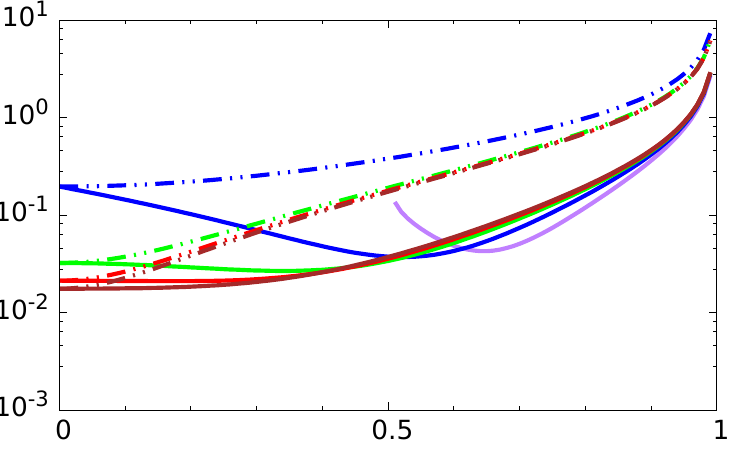}} &
            {\includegraphics[width=3.2cm,height=2.3cm,trim=0 0 0 0]{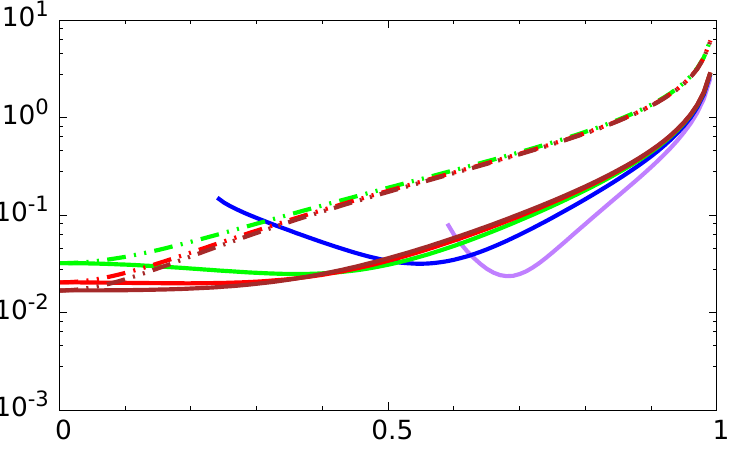}} &
            \rotatebox{90}{\hspace{0.85cm}$h=2$} \\
            \rotatebox{90}{\hspace{0.4cm}$\Delta\theta_{max}/\theta_{max}$} &  
            {\includegraphics[width=3.2cm,height=2.3cm,trim=0 0 0 0]{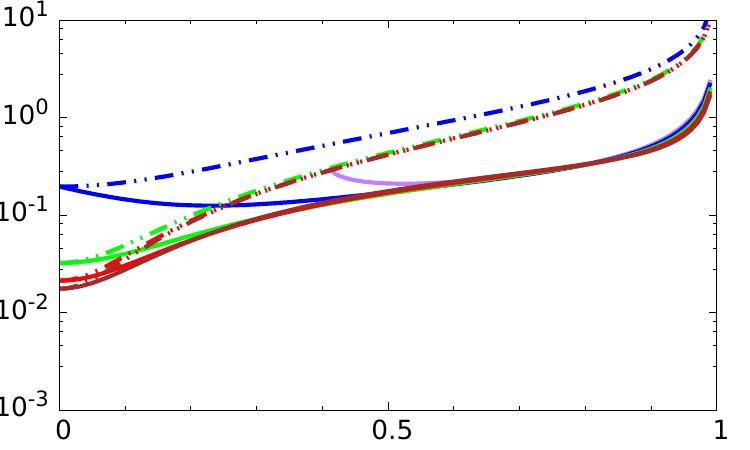}} &
            {\includegraphics[width=3.2cm,height=2.3cm,trim=0 0 0 0]{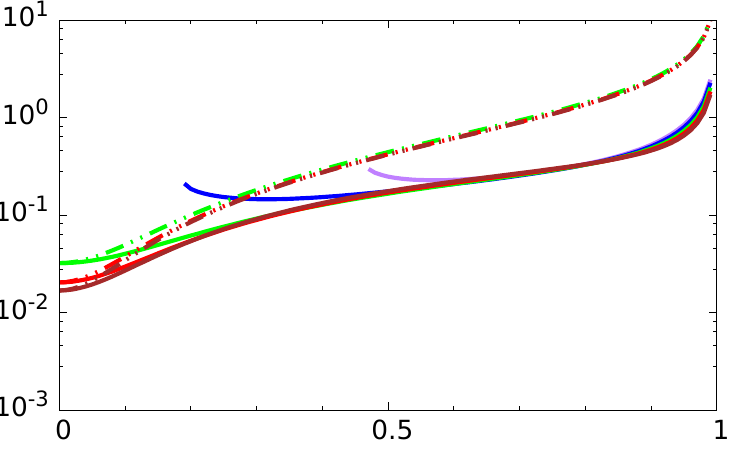}} &
            \rotatebox{90}{\hspace{1.2cm}Exp} \\
            & {~~$\epsilon$} & {~~$\epsilon$} &  
        \end{tabular}
        \caption{Relative error in $\theta_{max}$ ($|\theta_{max,a}-\theta_{max}|/\theta_{max}$) as a function of the frequency ratio $\epsilon$ for $r=R$. $q^*=-(0.50M)^2$. The purple, blue, green, red and brown lines correspond respectively to $R=3.5$, $4$, $5$, $6$ and $7$. The continuous lines were obtained from Eq. \eqref{eq:BeloPlas} (with plasma correction) while the dashed lines were obtained from Eq. \eqref{eq:Belo} (without plasma correction).}
        \label{fig:Ap_AA_TR_tdm}
    \end{figure}
    
    \begin{figure}
        \centering
        \begin{tabular}{cccc}
            & Sch & RN-like & \\
            \rotatebox{90}{\hspace{0.8cm} $\Delta\theta/\theta$} & 
            {\includegraphics[width=3.2cm,height=2.3cm,trim=0 0 0 0]{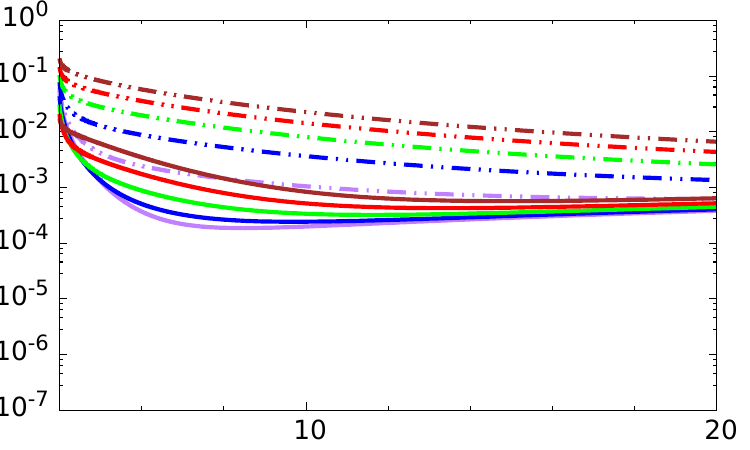}} &
            {\includegraphics[width=3.2cm,height=2.3cm,trim=0 0 0 0]{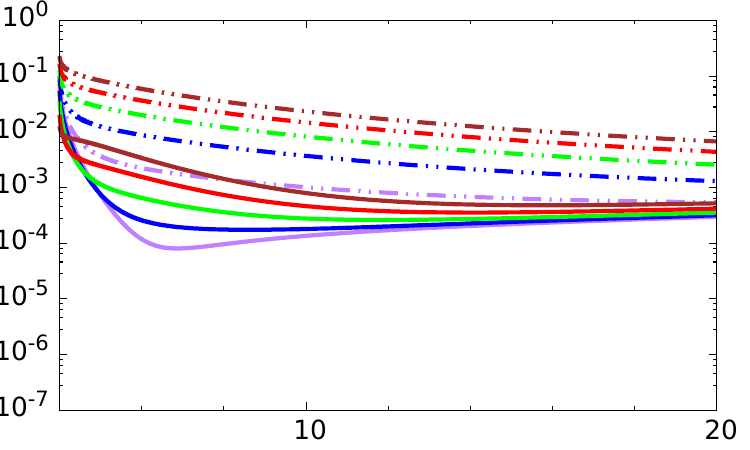}} &
            \rotatebox{90}{\hspace{0.85cm}$h=2$} \\
            \rotatebox{90}{\hspace{0.8cm} $\Delta\theta/\theta$} &  
            {\includegraphics[width=3.2cm,height=2.3cm,trim=0 0 0 0]{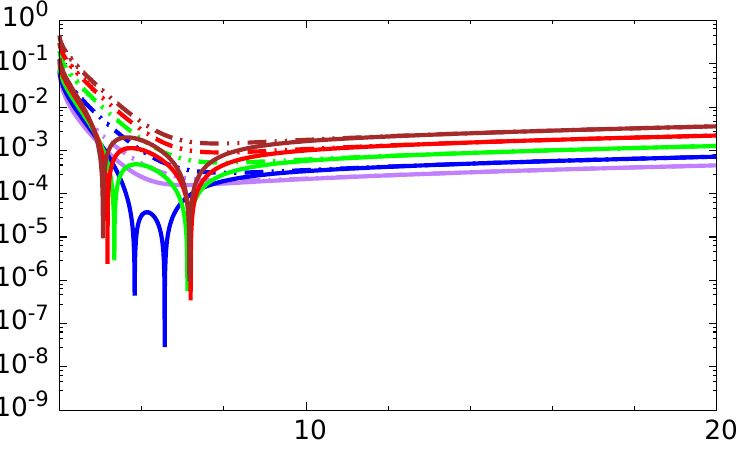}} &
            {\includegraphics[width=3.2cm,height=2.3cm,trim=0 0 0 0]{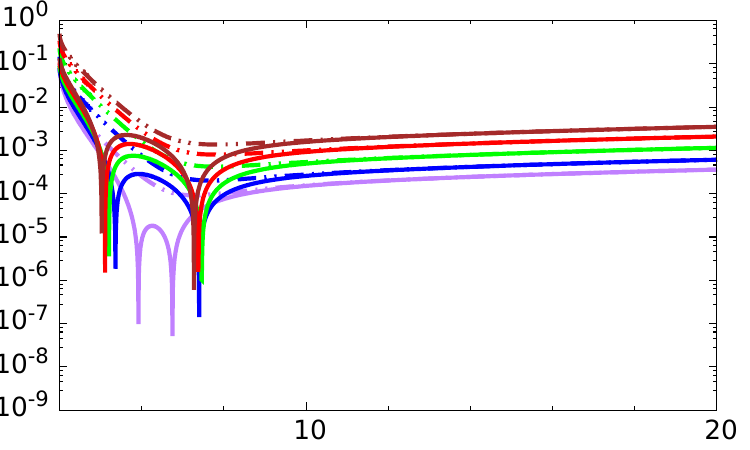}} &
            \rotatebox{90}{\hspace{1.2cm}Exp} \\
            & {~~$r$} & {~~$r$} &  
        \end{tabular}
        \caption{Relative error in $\theta$ ($|\theta_a-\theta|/\theta$) as a function of the radius $r$ for $\delta(R)=\pi/2$. $q^*=-(0.50M)^2$, $R=4M$. The purple, blue, green, red and brown lines correspond respectively to $\epsilon=0.1$, $0.2$, $0.3$, $0.4$ and $0.5$. The continuous lines were obtained from Eq. \eqref{eq:BeloPlas} (with plasma correction) while the dashed lines were obtained from Eq. \eqref{eq:Belo} (without plasma correction).}
        \label{fig:Ap_AA_TR_tvr}
    \end{figure}
    
    
    \begin{figure}
        \centering
        \begin{tabular}{cccc}
            & $h=2$ & Exp & \\
            \rotatebox{90}{\hspace{0.5cm}Flux ($\cdot100$)} &  
            {\includegraphics[width=3.1cm,height=2.2cm,trim=0 0 0 0]{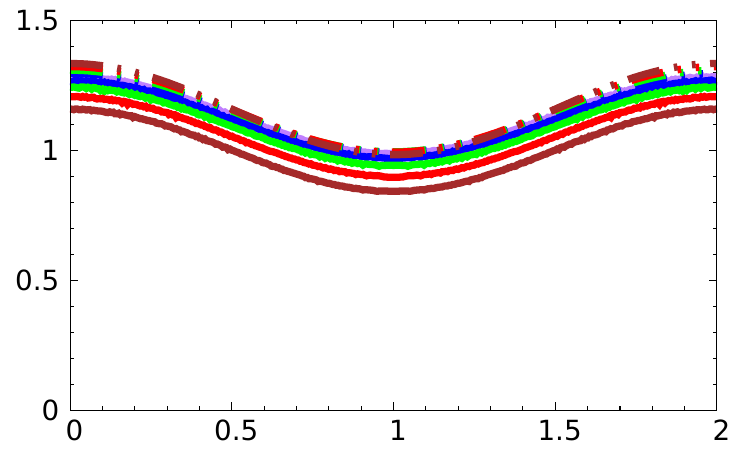}} &
            {\includegraphics[width=3.1cm,height=2.2cm,trim=0 0 0 0]{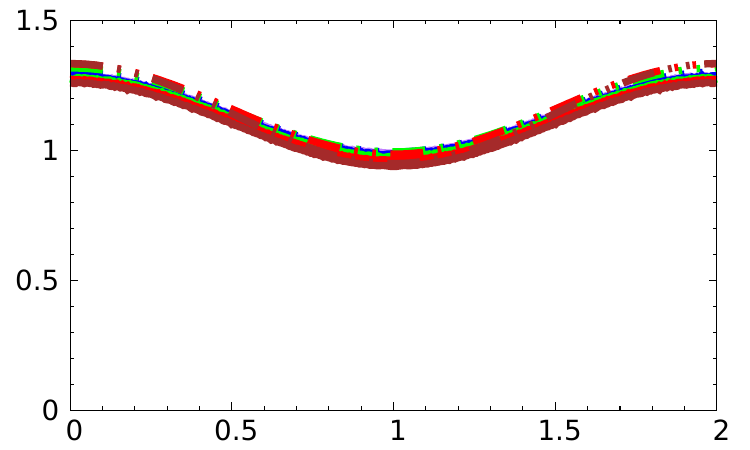}} &
            \rotatebox{90}{\hspace{0.3cm}$\chi=\frac{\pi}{9}$, $\xi=\frac{\pi}{6}$} \\
            \rotatebox{90}{\hspace{0.5cm}Flux ($\cdot100$)} &  
            {\includegraphics[width=3.1cm,height=2.2cm,trim=0 0 0 0]{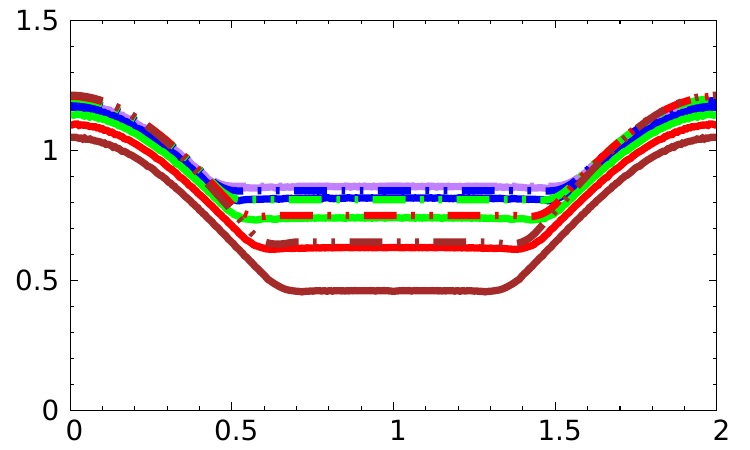}} &
            {\includegraphics[width=3.1cm,height=2.2cm,trim=0 0 0 0]{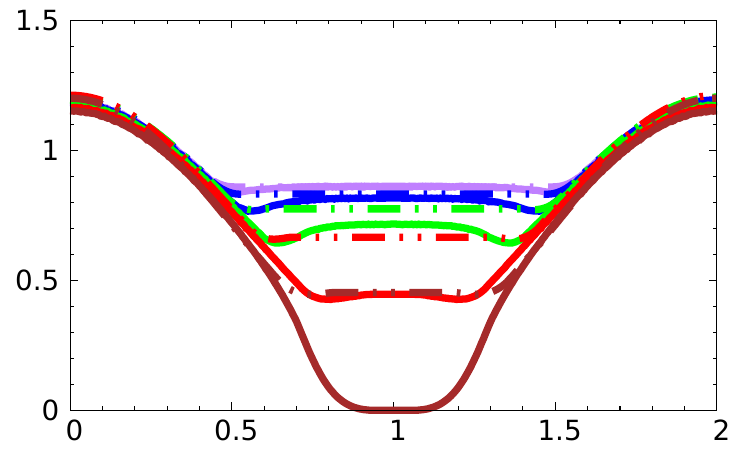}} &
            \rotatebox{90}{\hspace{0.3cm}$\chi=\frac{\pi}{6}$, $\xi=\frac{\pi}{3}$} \\
            \rotatebox{90}{\hspace{0.5cm}Flux ($\cdot100$)} &  
            {\includegraphics[width=3.1cm,height=2.2cm,trim=0 0 0 0]{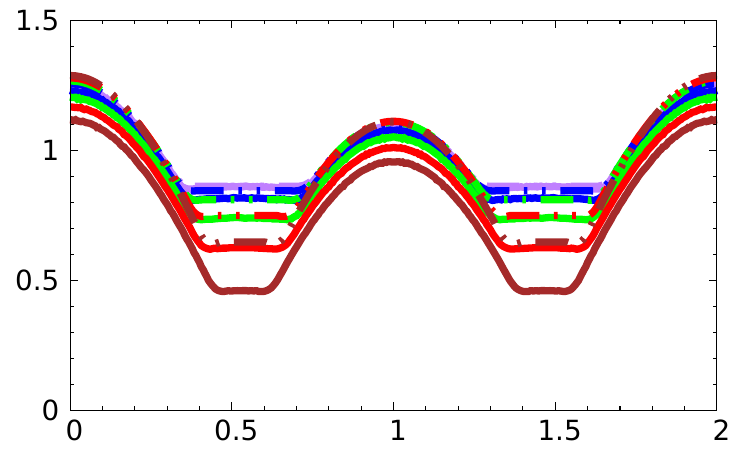}} &
            {\includegraphics[width=3.1cm,height=2.2cm,trim=0 0 0 0]{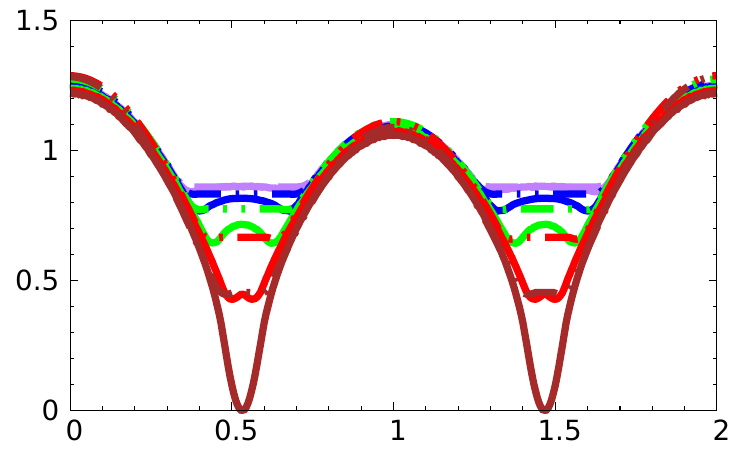}} &
            \rotatebox{90}{\hspace{0.3cm}$\chi=\frac{\pi}{3}$, $\xi=\frac{4\pi}{9}$} \\
            \rotatebox{90}{\hspace{0.5cm}Flux ($\cdot100$)} &  
            {\includegraphics[width=3.1cm,height=2.2cm,trim=0 0 0 0]{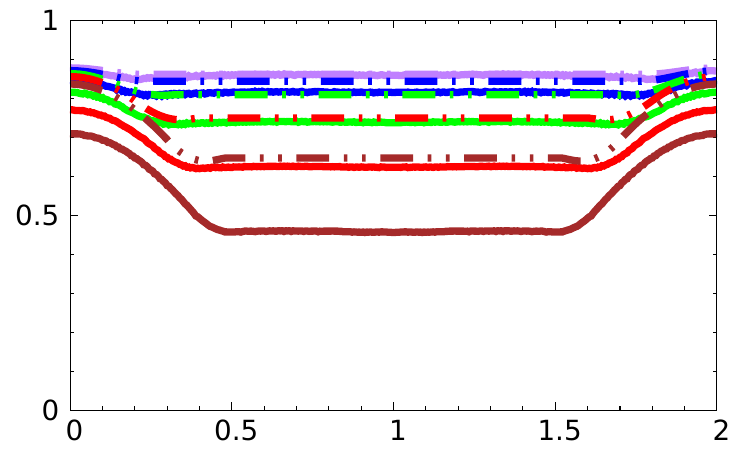}} &
            {\includegraphics[width=3.1cm,height=2.2cm,trim=0 0 0 0]{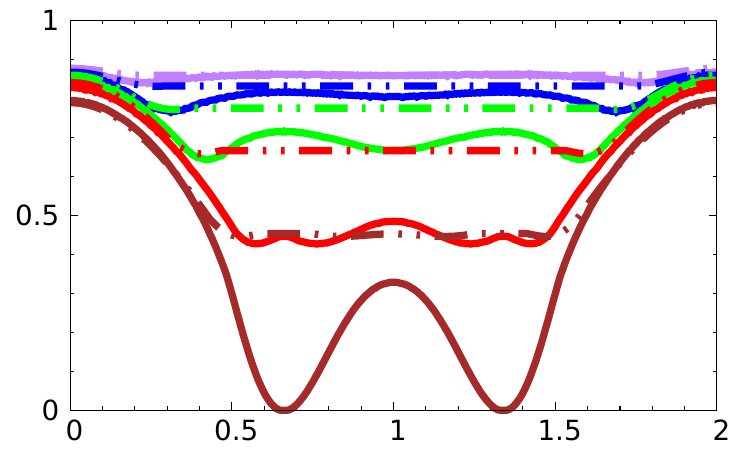}} &
            \rotatebox{90}{\hspace{0.3cm}$\chi=\frac{\pi}{9}$, $\xi=\frac{4\pi}{9}$} \\
            & {~~$\Omega t/\pi$} & {~~$\Omega t/\pi$} &
        \end{tabular}
        \caption{Analytical pulse profile for two identical, antipodal caps. Schwarzschild, $R=6M$, $q^*=-(0.50M)^2$, $\theta_c=5^o$. Purple, blue, green, green, red and brown lines correspond to $\epsilon=0.1$, $0.2$, $0.3$, $0.4$ and $0.5$ respectively. The solid lines correspond to the numerical result, while the dashed lines correspond to the analytical approximation.}
        \label{fig:Ap_AA_PP_SH}
    \end{figure}
    
    \begin{figure}
        \centering
        \begin{tabular}{cccc}
            & $h=2$ & Exp & \\
            \rotatebox{90}{\hspace{0.5cm}Flux ($\cdot100$)} &  
            {\includegraphics[width=3.1cm,height=2.2cm,trim=0 0 0 0]{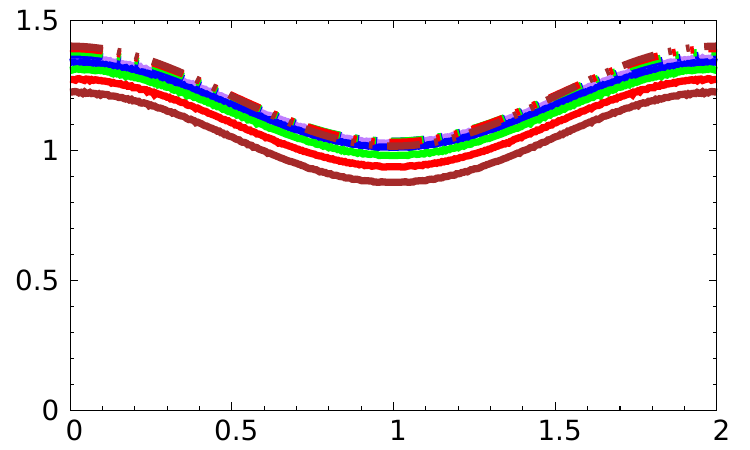}} &
            {\includegraphics[width=3.1cm,height=2.2cm,trim=0 0 0 0]{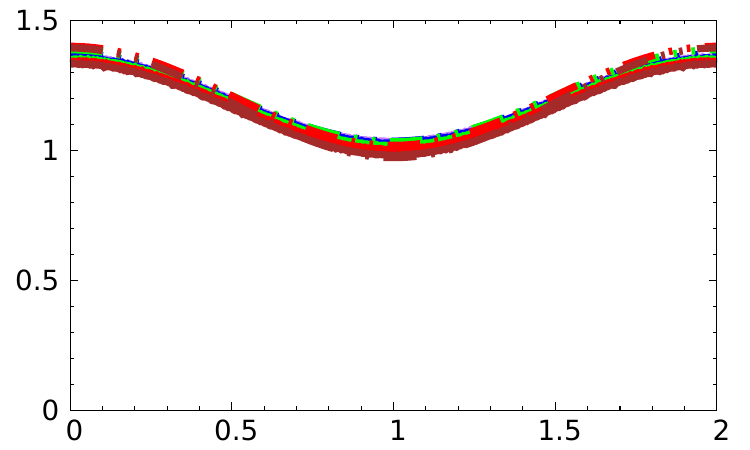}} &
            \rotatebox{90}{\hspace{0.3cm}$\chi=\frac{\pi}{9}$, $\xi=\frac{\pi}{6}$} \\
            \rotatebox{90}{\hspace{0.5cm}Flux ($\cdot100$)} &  
            {\includegraphics[width=3.1cm,height=2.2cm,trim=0 0 0 0]{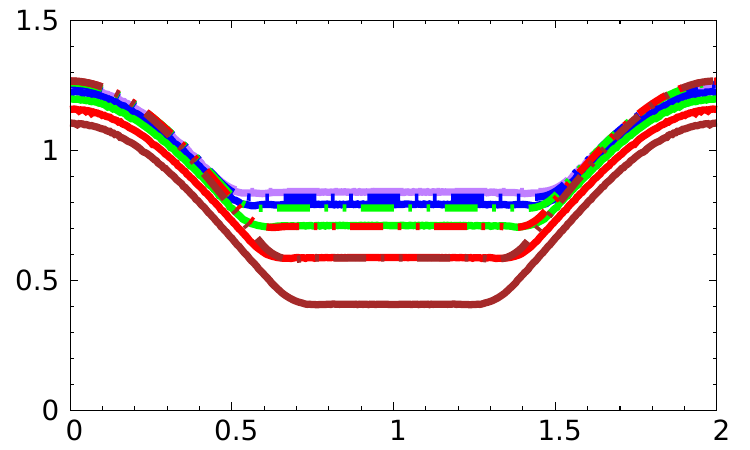}} &
            {\includegraphics[width=3.1cm,height=2.2cm,trim=0 0 0 0]{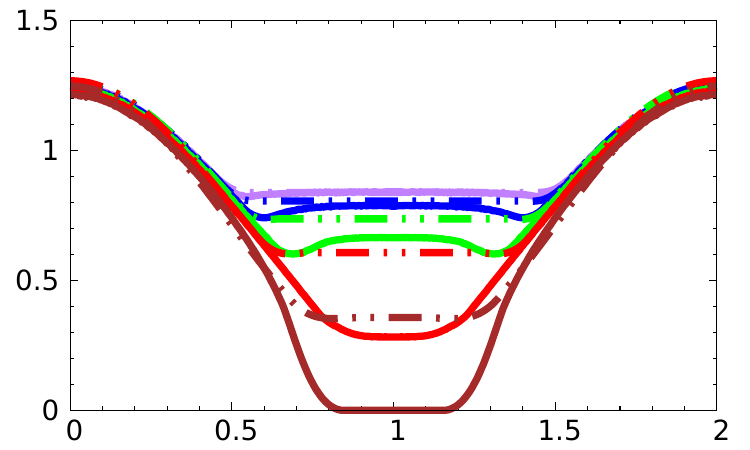}} &
            \rotatebox{90}{\hspace{0.3cm}$\chi=\frac{\pi}{6}$, $\xi=\frac{\pi}{3}$} \\
            \rotatebox{90}{\hspace{0.5cm}Flux ($\cdot100$)} &  
            {\includegraphics[width=3.1cm,height=2.2cm,trim=0 0 0 0]{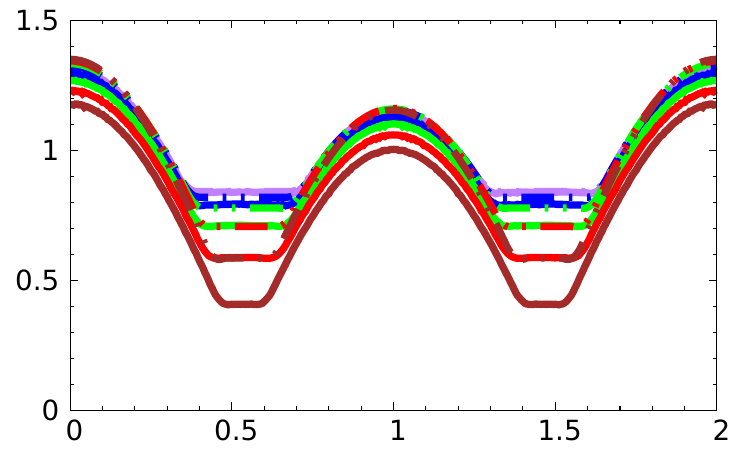}} &
            {\includegraphics[width=3.1cm,height=2.2cm,trim=0 0 0 0]{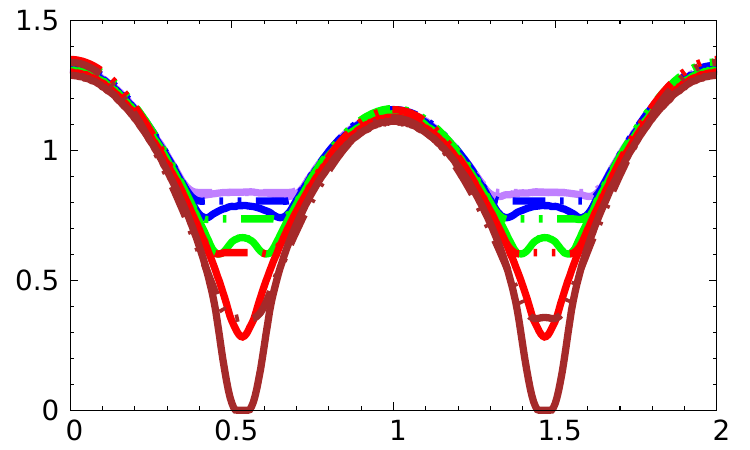}} &
            \rotatebox{90}{\hspace{0.3cm}$\chi=\frac{\pi}{3}$, $\xi=\frac{4\pi}{9}$} \\
            \rotatebox{90}{\hspace{0.5cm}Flux ($\cdot100$)} &  
            {\includegraphics[width=3.1cm,height=2.2cm,trim=0 0 0 0]{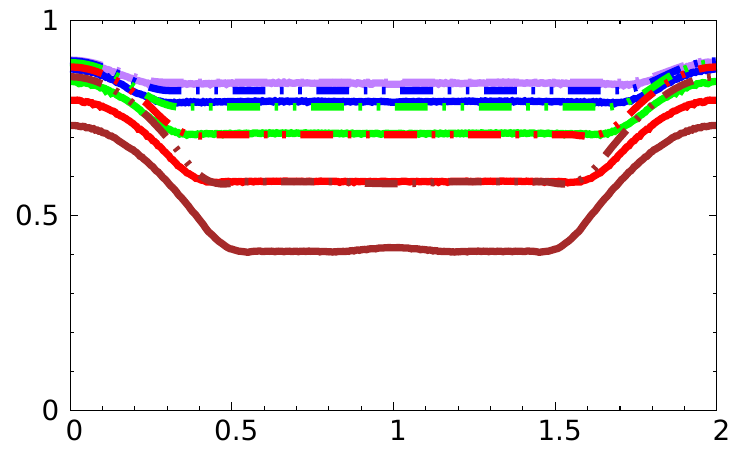}} &
            {\includegraphics[width=3.1cm,height=2.2cm,trim=0 0 0 0]{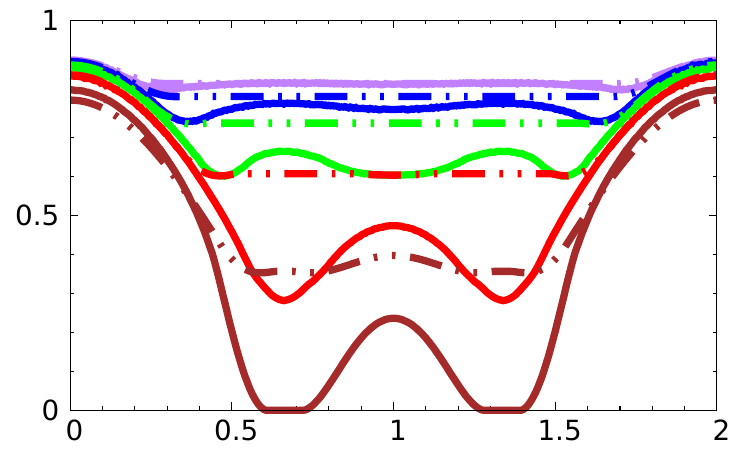}} &
            \rotatebox{90}{\hspace{0.3cm}$\chi=\frac{\pi}{9}$, $\xi=\frac{4\pi}{9}$} \\
            & {~~$\Omega t/\pi$} & {~~$\Omega t/\pi$} &
        \end{tabular}
        \caption{Analytical pulse profile for two identical, antipodal caps. Reissner-Nordström, $R=6M$, $q^*=-(0.50M)^2$, $\theta_c=5^o$. Purple, blue, green, green, red and brown lines correspond to $\epsilon=0.1$, $0.2$, $0.3$, $0.4$ and $0.5$ respectively. The solid lines correspond to the numerical result, while the dashed lines correspond to the analytical approximation.}
        \label{fig:Ap_AA_PP_RN}
    \end{figure}




\begin{thebibliography}{100}

\bibitem{Shapiro_1983}
S.~L. Shapiro and S.~A. Teukolsky.
\newblock {\em Black holes, white dwarfs, and neutron stars: \\ The physics of
  compact objects}.
\newblock John Wiley and Sons, 1983.

\bibitem{Harding_2013}
Alice~K Harding.
\newblock The neutron star zoo.
\newblock {\em Frontiers of Physics}, 8(6):679--692, 2013.

\bibitem{Ozel_2016}
Feryal {\"O}zel and Paulo Freire.
\newblock Masses, radii, and the equation of state of neutron stars.
\newblock {\em Annual Review of Astronomy and Astrophysics}, 54:401--440, 2016.

\bibitem{annala2018gravitational}
Eemeli Annala, Tyler Gorda, Aleksi Kurkela, and Aleksi Vuorinen.
\newblock Gravitational-wave constraints on the neutron-star-matter equation of
  state.
\newblock {\em Physical review letters}, 120(17):172703, 2018.

\bibitem{Lattimer_2021}
JM~Lattimer.
\newblock Neutron stars and the nuclear matter equation of state.
\newblock {\em Annual Review of Nuclear and Particle Science}, 71:433--464,
  2021.

\bibitem{Berti:2015itd}
Emanuele Berti et~al.
\newblock {Testing General Relativity with Present and Future Astrophysical
  Observations}.
\newblock {\em Class. Quant. Grav.}, 32:243001, 2015.

\bibitem{demorest2010two}
Paul~B Demorest, Tim Pennucci, SM~Ransom, MSE Roberts, and JWT Hessels.
\newblock A two-solar-mass neutron star measured using shapiro delay.
\newblock {\em nature}, 467(7319):1081--1083, 2010.

\bibitem{abbott2016observation}
Benjamin~P Abbott, Richard Abbott, TD~Abbott, MR~Abernathy, Fausto Acernese,
  Kendall Ackley, Carl Adams, Thomas Adams, Paolo Addesso, RX~Adhikari, et~al.
\newblock Observation of gravitational waves from a binary black hole merger.
\newblock {\em Physical Review Letters}, 116(6):061102, 2016.

\bibitem{Riley:2019yda}
Thomas~E. Riley et~al.
\newblock {A $NICER$ View of PSR J0030+0451: Millisecond Pulsar Parameter
  Estimation}.
\newblock {\em Astrophys. J. Lett.}, 887(1):L21, 2019.

\bibitem{Miller:2021qha}
M.~C. Miller et~al.
\newblock {The Radius of PSR J0740+6620 from NICER and XMM-Newton Data}.
\newblock {\em Astrophys. J. Lett.}, 918(2):L28, 2021.

\bibitem{Bogdanov:2021yip}
Slavko Bogdanov et~al.
\newblock {Constraining the Neutron Star Mass-Radius Relation and Dense Matter
  Equation of State with NICER. III. Model Description and Verification of
  Parameter Estimation Codes}.
\newblock {\em Astrophys. J. Lett.}, 914(1):L15, 2021.

\bibitem{petri2016theory}
J{\'e}r{\^o}me P{\'e}tri.
\newblock Theory of pulsar magnetosphere and wind.
\newblock {\em Journal of Plasma Physics}, 82(5), 2016.

\bibitem{Pechenick_1983}
K.R. Pechenick, C.~Ftaclas, and J.M. Cohen.
\newblock Hot spots on neutron stars - the near-field gravitational lens.
\newblock {\em apj}, 274:846--857, November 1983.

\bibitem{sotani2017pulse}
Hajime Sotani.
\newblock Pulse profiles from a pulsar in scalar-tensor gravity.
\newblock {\em Physical Review D}, 96(10):104010, 2017.

\bibitem{Sotani_2017}
Hajime Sotani and Umpei Miyamoto.
\newblock Sensitivity of pulsar light curves to spacetime geometry and efficacy
  of analytic approximations.
\newblock {\em Physical Review D}, 96(10), Nov 2017.

\bibitem{Sotani_2020}
Hajime Sotani.
\newblock Light curves from highly compact neutron stars with spot size effect.
\newblock {\em Physical Review D}, 101(6), Mar 2020.

\bibitem{Silva_2019}
Hector~O. Silva and Nicol\'{a}s Yunes.
\newblock Neutron star pulse profiles in scalar-tensor theories of gravity.
\newblock {\em Physical Review D}, 99(4), Feb 2019.

\bibitem{Dabrowski_1995}
Mariusz~P. D{{a}}browski and Janusz Osarczuk.
\newblock Light curves of relativistic charged neutron star.
\newblock {\em apss}, 229(1):139--155, July 1995.

\bibitem{Silva:2019leq}
Hector~O. Silva and Nicol\'as Yunes.
\newblock {Neutron star pulse profile observations as extreme gravity probes}.
\newblock {\em Class. Quant. Grav.}, 36(17):17LT01, 2019.

\bibitem{Xu:2020vbs}
Rui Xu, Yong Gao, and Lijing Shao.
\newblock {Strong-field effects in massive scalar-tensor gravity for slowly
  spinning neutron stars and application to X-ray pulsar pulse profiles}.
\newblock {\em Phys. Rev. D}, 102(6):064057, 2020.

\bibitem{Hu:2021tyw}
Zexin Hu, Yong Gao, Rui Xu, and Lijing Shao.
\newblock {Scalarized neutron stars in massive scalar-tensor gravity: X-ray
  pulsars and tidal deformability}.
\newblock {\em Phys. Rev. D}, 104(10):104014, 2021.

\bibitem{psaltis2014prospects}
Dimitrios Psaltis, Feryal {\"O}zel, and Deepto Chakrabarty.
\newblock Prospects for measuring neutron-star masses and radii with x-ray
  pulse profile modeling.
\newblock {\em The Astrophysical Journal}, 787(2):136, 2014.

\bibitem{bogdanov2016prospects}
Slavko Bogdanov.
\newblock Prospects for neutron star equation of state constraints using
  “recycled” millisecond pulsars.
\newblock {\em The European Physical Journal A}, 52(2):1--7, 2016.

\bibitem{poutanen2006pulse}
Juri Poutanen and Andrei~M Beloborodov.
\newblock Pulse profiles of millisecond pulsars and their fourier amplitudes.
\newblock {\em Monthly Notices of the Royal Astronomical Society},
  373(2):836--844, 2006.

\bibitem{2007ApJ...654..458C}
Coire {Cadeau}, Sharon~M. {Morsink}, Denis {Leahy}, and Sheldon~S. {Campbell}.
\newblock {Light Curves for Rapidly Rotating Neutron Stars}.
\newblock {\em \apj}, 654(1):458--469, January 2007.

\bibitem{psaltis2014pulse}
Dimitrios Psaltis and Feryal {\"O}zel.
\newblock Pulse profiles from spinning neutron stars in the hartle--thorne
  approximation.
\newblock {\em The Astrophysical Journal}, 792(2):87, 2014.

\bibitem{2010MNRAS.409..481N}
Kazutoshi {Numata} and Umin {Lee}.
\newblock {Light curves from rapidly rotating neutron stars}.
\newblock {\em MNRAS}, 409(2):481--490, December 2010.

\bibitem{Sotani:2018oad}
Hajime Sotani and Umpei Miyamoto.
\newblock {Systematical study of pulsar light curves with special relativistic
  effects}.
\newblock {\em Phys. Rev. D}, 98(10):103019, 2018.

\bibitem{Beloborodov_2002}
Andrei~M. Beloborodov.
\newblock Gravitational bending of light near compact objects.
\newblock {\em The Astrophysical Journal}, 566(2):L85–L88, Feb 2002.

\bibitem{Turolla_2013}
R.~Turolla and L.~Nobili.
\newblock Pulse profiles from thermally emitting neutron stars.
\newblock {\em The Astrophysical Journal}, 768(2):147, Apr 2013.

\bibitem{Hu:2022ehk}
Kun Hu, Matthew~G. Baring, Joseph~A. Barchas, and George Younes.
\newblock {Intensity and Polarization Characteristics of Extended Neutron Star
  Surface Regions}.
\newblock {\em Astrophys. J.}, 928(1):82, 2022.

\bibitem{Zyuzin:2021teu}
D.~A. Zyuzin, A.~V. Karpova, Y.~A. Shibanov, A.~Y. Potekhin, and V.~F.
  Suleimanov.
\newblock {Middle aged $\gamma$-ray pulsar J1957+5033 in X-rays: pulsations,
  thermal emission, and nebula}.
\newblock {\em Mon. Not. Roy. Astron. Soc.}, 501(4):4998--5011, 2021.

\bibitem{Petri:2019oqa}
J\'er\^ome P\'etri and Dipanjan Mitra.
\newblock {Joint radio and X-ray modelling of PSR J1136+1551}.
\newblock {\em Mon. Not. Roy. Astron. Soc.}, 491(1):80--91, 2020.

\bibitem{Giraud:2019iba}
Q.~Giraud and J.~P\'etri.
\newblock {Radio and high-energy emission of pulsars revealed by general
  relativity}.
\newblock {\em Astron. Astrophys.}, 639:A75, 2020.

\bibitem{Hu:2019zqy}
Chin-Ping Hu, C.~Y. Ng, and Wynn C.~G. Ho.
\newblock {A systematic study of soft X-ray pulse profiles of magnetars in
  quiescence}.
\newblock {\em Mon. Not. Roy. Astron. Soc.}, 485(3):4274--4286, 2019.

\bibitem{Gotthelf:2010um}
E.~V. Gotthelf, R.~Perna, and J.~P. Halpern.
\newblock {Modeling the Surface X-ray Emission and Viewing Geometry of PSR
  J0821-4300 in Puppis A}.
\newblock {\em Astrophys. J.}, 724:1316--1324, 2010.

\bibitem{Viironen:2004ze}
Kerttu Viironen and Juri Poutanen.
\newblock {Light curves and polarization of accretion- and nuclear-powered
  millisecond pulsars}.
\newblock {\em Astron. Astrophys.}, 426:985--997, 2004.

\bibitem{Melia:2011pq}
Fulvio Melia, Maurizio Falanga, and Andrea Goldwurm.
\newblock {Polarimetric Imaging of Sgr A* in its Flaring State}.
\newblock {\em Mon. Not. Roy. Astron. Soc.}, 419:2489, 2012.

\bibitem{Loktev:2021nhk}
Vladislav Loktev, Alexandra Veledina, and Juri Poutanen.
\newblock {Analytical techniques for polarimetric imaging of accretion flows in
  the Schwarzschild metric}.
\newblock {\em Astron. Astrophys.}, 660:A25, 2022.

\bibitem{EventHorizonTelescope:2021btj}
Ramesh Narayan et~al.
\newblock {The Polarized Image of a Synchrotron-emitting Ring of Gas Orbiting a
  Black Hole}.
\newblock {\em Astrophys. J.}, 912(1):35, 2021.

\bibitem{GJ_1969}
Peter Goldreich and William~H. Julian.
\newblock Pulsar electrodynamics.
\newblock {\em apj}, 157:869, 1969.

\bibitem{Petri:2016tqe}
J.~P\'etri.
\newblock {Theory of pulsar magnetosphere and wind}.
\newblock {\em J. Plasma Phys.}, 82(5):635820502, 2016.

\bibitem{Perlick:2021aok}
Volker Perlick and Oleg~Yu. Tsupko.
\newblock {Calculating black hole shadows: Review of analytical studies}.
\newblock {\em Phys. Rept.}, 947:1--39, 2022.

\bibitem{Perlick:2017fio}
Volker Perlick and Oleg~Yu. Tsupko.
\newblock {Light propagation in a plasma on Kerr spacetime: Separation of the
  Hamilton-Jacobi equation and calculation of the shadow}.
\newblock {\em Phys. Rev. D}, 95(10):104003, 2017.

\bibitem{Kimpson:2019mji}
Tom Kimpson, Kinwah Wu, and Silvia Zane.
\newblock {Spatial dispersion of light rays propagating through a plasma in
  Kerr space\textendash{}time}.
\newblock {\em Mon. Not. Roy. Astron. Soc.}, 484(2):2411--2419, 2019.

\bibitem{Huang:2018rfn}
Yang Huang, Yi-Ping Dong, and Dao-Jun Liu.
\newblock {Revisiting the shadow of a black hole in the presence of a plasma}.
\newblock {\em Int. J. Mod. Phys. D}, 27(12):1850114, 2018.

\bibitem{Zhang:2022osx}
Zhenyu Zhang, Haopeng Yan, Minyong Guo, and Bin Chen.
\newblock {Shadow of Kerr black hole surrounded by an angular Gaussian
  distributed plasma}.
\newblock {\em arXiv:2206.04430}, 6 2022.

\bibitem{Badia:2021kpk}
Javier Bad\'\i{}a and Ernesto~F. Eiroa.
\newblock {Shadow of axisymmetric, stationary, and asymptotically flat black
  holes in the presence of plasma}.
\newblock {\em Phys. Rev. D}, 104(8):084055, 2021.

\bibitem{CG_2018}
Gabriel Crisnejo and Emanuel Gallo.
\newblock Weak lensing in a plasma medium and gravitational deflection of
  massive particles using the gauss-bonnet theorem. a unified treatment.
\newblock {\em Physical Review D}, 97(12), Jun 2018.

\bibitem{CGR_2019}
Gabriel Crisnejo, Emanuel Gallo, and Adam Rogers.
\newblock Finite distance corrections to the light deflection in a
  gravitational field with a plasma medium.
\newblock {\em Physical Review D}, 99(12), Jun 2019.

\bibitem{CGV_2019}
Gabriel Crisnejo, Emanuel Gallo, and José~R. Villanueva.
\newblock Gravitational lensing in dispersive media and deflection angle of
  charged massive particles in terms of curvature scalars and energy-momentum
  tensor.
\newblock {\em Physical Review D}, 100(4), Aug 2019.

\bibitem{Crisnejo:2019ril}
Gabriel Crisnejo, Emanuel Gallo, and Kimet Jusufi.
\newblock {Higher order corrections to deflection angle of massive particles
  and light rays in plasma media for stationary spacetimes using the
  Gauss-Bonnet theorem}.
\newblock {\em Phys. Rev. D}, 100(10):104045, 2019.

\bibitem{Bisnovatyi-Kogan:2010flt}
G.~S. Bisnovatyi-Kogan and O.~Yu. Tsupko.
\newblock {Gravitational lensing in a non-uniform plasma}.
\newblock {\em Mon. Not. Roy. Astron. Soc.}, 404:1790--1800, 2010.

\bibitem{Bisnovatyi-Kogan:2015dxa}
G.~S. Bisnovatyi-Kogan and O.~Yu. Tsupko.
\newblock {Gravitational Lensing in Plasmic Medium}.
\newblock {\em Plasma Phys. Rep.}, 41:562, 2015.

\bibitem{Er:2019jkg}
Xinzhong Er and Adam Rogers.
\newblock {Two families of elliptical plasma lenses}.
\newblock {\em Mon. Not. Roy. Astron. Soc.}, 488(4):5651--5664, 2019.

\bibitem{Er:2017lue}
Xinzhong Er and Adam Rogers.
\newblock {Two families of astrophysical diverging lens models}.
\newblock {\em Mon. Not. Roy. Astron. Soc.}, 475(1):867--878, 2017.

\bibitem{Er:2021shs}
Xinzhong Er, Jenny Wagner, and Shude Mao.
\newblock {On the double-plane plasma lensing}.
\newblock {\em Mon. Not. Roy. Astron. Soc.}, 509(4):5872--5881, 2021.

\bibitem{Er:2013efa}
Xinzhong Er and Shude Mao.
\newblock {Effects of plasma on gravitational lensing}.
\newblock {\em Mon. Not. Roy. Astron. Soc.}, 437(3):2180--2186, 2014.

\bibitem{Tsupko:2019axo}
Oleg~Yu. Tsupko and Gennady~S. Bisnovatyi-Kogan.
\newblock {Hills and holes in the microlensing light curve due to plasma
  environment around gravitational lens}.
\newblock {\em Mon. Not. Roy. Astron. Soc.}, 491(4):5636--5649, 2020.

\bibitem{Er_2020}
Xinzhong Er, Yuan-Pei Yang, and Adam Rogers.
\newblock The effects of plasma lensing on the inferred dispersion measures of
  fast radiobursts.
\newblock {\em The Astrophysical Journal}, 889(2):158, feb 2020.

\bibitem{Er:2021pjc}
Xinzhong Er, Jiangchuan Yu, Adam Rogers, Shihang Liu, and Shude Mao.
\newblock {Bias in apparent dispersion measure due to de-magnification of
  plasma lensing on background radio sources}.
\newblock {\em Mon. Not. Roy. Astron. Soc.}, 510(1):197--204, 2021.

\bibitem{Rogers_2015}
Adam Rogers.
\newblock Frequency-dependent effects of gravitational lensing within plasma.
\newblock {\em Monthly Notices of the Royal Astronomical Society},
  451(1):17–25, May 2015.

\bibitem{Rogers:2017ofq}
Adam Rogers.
\newblock {Gravitational Lensing of Rays through the Levitating Atmospheres of
  Compact Objects}.
\newblock {\em Universe}, 3(1):3, 2017.

\bibitem{Rogers:2016xcc}
Adam Rogers.
\newblock {Escape and Trapping of Low-Frequency Gravitationally Lensed Rays by
  Compact Objects within Plasma}.
\newblock {\em Mon. Not. Roy. Astron. Soc.}, 465(2):2151--2159, 2017.

\bibitem{Battye:2021xvt}
R.~A. Battye, B.~Garbrecht, J.~I. McDonald, and S.~Srinivasan.
\newblock {Radio line properties of axion dark matter conversion in neutron
  stars}.
\newblock {\em JHEP}, 09:105, 2021.

\bibitem{Witte:2021arp}
Samuel~J. Witte, Dion Noordhuis, Thomas D.~P. Edwards, and Christoph Weniger.
\newblock {Axion-photon conversion in neutron star magnetospheres: The role of
  the plasma in the Goldreich-Julian model}.
\newblock {\em Phys. Rev. D}, 104(10):103030, 2021.

\bibitem{Main:2018kfc}
Robert Main, I-Sheng Yang, Victor Chan, Dongzi Li, Fang~Xi Lin, Nikhil Mahajan,
  Ue-Li Pen, Keith Vanderlinde, and Marten~H. van Kerkwijk.
\newblock {Pulsar emission amplified and resolved by plasma lensing in an
  eclipsing binary}.
\newblock {\em Nature}, 557:522--525, 2018.

\bibitem{Lin:2021epe}
F.~X. Lin, R.~A. Main, J.~P.~W. Verbiest, M.~Kramer, and G.~Shaifullah.
\newblock {Discovery and modelling of broad-scale plasma lensing in black-widow
  pulsar J2051~\ensuremath{-}~0827}.
\newblock {\em Mon. Not. Roy. Astron. Soc.}, 506(2):2824--2835, 2021.

\bibitem{Wang:2021cqk}
S.~Q. Wang et~al.
\newblock {Unusual Emission Variations Near the Eclipse of Black Widow Pulsar
  PSR J1720\ensuremath{-}0533}.
\newblock {\em Astrophys. J. Lett.}, 922(1):L13, 2021.

\bibitem{BW_2005}
Alikram~N. Aliev and A.~Emir Gumrukcuoglu.
\newblock Charged rotating black holes on a 3-brane.
\newblock {\em Physical Review D}, 71:104027, 2005.

\bibitem{Babichev:2017guv}
Eugeny Babichev, Christos Charmousis, and Antoine Leh\'ebel.
\newblock {Asymptotically flat black holes in Horndeski theory and beyond}.
\newblock {\em JCAP}, 04:027, 2017.

\bibitem{Moffat:2014aja}
J.~W. Moffat.
\newblock {Black Holes in Modified Gravity (MOG)}.
\newblock {\em Eur. Phys. J. C}, 75(4):175, 2015.

\bibitem{Reissner_1916}
Hans Reissner.
\newblock {\"U}ber die eigengravitation des elektrischen feldes nach der
  einsteinschen theorie.
\newblock {\em Annalen der Physik}, 355(9):106--120, 1916.

\bibitem{Nordstrom1918}
Gunnar Nordstr{\"o}m.
\newblock On the energy of the gravitation field in einstein's theory.
\newblock {\em Koninklijke Nederlandse Akademie van Wetenschappen Proceedings
  Series B Physical Sciences}, 20:1238--1245, 1918.

\bibitem{Dadhich:2000am}
Naresh Dadhich, Roy Maartens, Philippos Papadopoulos, and Vahid Rezania.
\newblock {Black holes on the brane}.
\newblock {\em Phys. Lett. B}, 487:1--6, 2000.

\bibitem{Germani:2001du}
Cristiano Germani and Roy Maartens.
\newblock {Stars in the brane world}.
\newblock {\em Phys. Rev. D}, 64:124010, 2001.

\bibitem{Kotrlova:2008xs}
Andrea Kotrlova, Zdenek Stuchlik, and Gabriel Torok.
\newblock {Quasiperiodic oscillations in a strong gravitational field around
  neutron stars testing braneworld models}.
\newblock {\em Class. Quant. Grav.}, 25:225016, 2008.

\bibitem{Morozova:2010gg}
V.~S. Morozova, B.~J. Ahmedov, A.~A. Abdujabbarov, and A.~I. Mamadjanov.
\newblock {Plasma Magnetosphere of Rotating Magnetized Neutron Star in the
  Braneworld}.
\newblock {\em Astrophys. Space Sci.}, 330:257--266, 2010.

\bibitem{Synge_1960}
John~Lighton Synge.
\newblock {\em Relativity: The General Theory}.
\newblock North-Holland, New York, 1 edition, 1960.

\bibitem{IntroPlasmaMHD_2003}
Marcel Goossens.
\newblock {\em An introduction to plasma astrophysics\\ and
  magnetohydrodynamics}.
\newblock Springer Science and Business Media, 2003.

\bibitem{kulsrud1992dynamics}
Russell Kulsrud and Abraham Loeb.
\newblock Dynamics and gravitational interaction of waves in nonuniform media.
\newblock {\em Physical Review D}, 45(2):525, 1992.

\bibitem{Dey:2016psn}
Rajat~K. Dey, Sabyasachi Ray, and Sandip Dam.
\newblock {Searching for PeV neutrinos from photomeson interactions in
  magnetars}.
\newblock {\em EPL}, 115(6):69002, 2016.

\bibitem{Dey:2021mwb}
Rajat~K. Dey, Animesh Basak, Sabyasachi Ray, and Tamal Sarkar.
\newblock {Newly Born Extragalactic Millisecond Pulsars as Efficient Emitters
  of PeV Neutrinos}.
\newblock {\em Braz. J. Phys.}, 51(5):1406--1415, 2021.

\bibitem{BH_1975}
J~Bicak and P~Hadrava.
\newblock General-relativistic radiative transfer theory in refractive and
  dispersive media.
\newblock {\em Astronomy and Astrophysics}, 44:389--399, 1975.

\bibitem{gralla2017inclined}
Samuel~E Gralla, Alexandru Lupsasca, and Alexander Philippov.
\newblock Inclined pulsar magnetospheres in general relativity: Polar caps for
  the dipole, quadrudipole, and beyond.
\newblock {\em The Astrophysical Journal}, 851(2):137, 2017.

\bibitem{lockhart2019x}
Will Lockhart, Samuel~E Gralla, Feryal {\"O}zel, and Dimitrios Psaltis.
\newblock X-ray light curves from realistic polar cap models: inclined pulsar
  magnetospheres and multipole fields.
\newblock {\em Monthly Notices of the Royal Astronomical Society},
  490(2):1774--1783, 2019.

\bibitem{Sotani_2018}
Hajime Sotani and Umpei Miyamoto.
\newblock Pulse profiles of highly compact pulsars in general relativity.
\newblock {\em Physical Review D}, 98(4), Aug 2018.

\bibitem{Sotani_2019}
Hajime Sotani, Hector~O. Silva, and George Pappas.
\newblock Finite size effects on the light curves of slowly-rotating neutron
  stars.
\newblock {\em Physical Review D}, 100(4), Aug 2019.

\bibitem{BE_1980}
RA~Breuer and J{\"u}rgen Ehlers.
\newblock Propagation of high-frequency electromagnetic waves through a
  magnetized plasma in curved space-time. i.
\newblock {\em Proceedings of the Royal Society of London. A. Mathematical and
  Physical Sciences}, 370(1742):389--406, 1980.

\bibitem{2003BroBla}
A.~Broderick and R.~Blandford.
\newblock Covariant magnetoionic theory -- i. ray propagation.
\newblock {\em Monthly Notices of the Royal Astronomical Society},
  342(4):1280–1290, Jul 2003.

\bibitem{Broderick:2003fc}
Avery Broderick and Roger Blandford.
\newblock {Covariant magnetoionic theory. 2. Radiative transfer}.
\newblock {\em Mon. Not. Roy. Astron. Soc.}, 349:994, 2004.

\bibitem{Beskin:2000si}
V.~S. Beskin and Roman~R. Rafikov.
\newblock {On the particle acceleration near the light surface of radio
  pulsars}.
\newblock {\em Mon. Not. Roy. Astron. Soc.}, 313:433--444, 2000.

\bibitem{gurevich_beskin_istomin_1993}
A.~V. Gurevich, V.~S. Beskin, and Ya.~N Istomin.
\newblock {\em Physics of the Pulsar Magnetosphere}.
\newblock Cambridge University Press, 1993.

\bibitem{2001Ap&SS.278...77I}
Ya. {Istomin}.
\newblock {Propagation of Electromagnetic Waves in Pulsar Magnetospheres}.
\newblock {\em Astrophysics and Space Science}, 278:77--80, October 2001.

\bibitem{PhysRevE.57.3399}
M.~Gedalin, D.~B. Melrose, and E.~Gruman.
\newblock Long waves in a relativistic pair plasma in a strong magnetic field.
\newblock {\em Phys. Rev. E}, 57:3399--3410, Mar 1998.

\bibitem{Mutka:2002ji}
P.~T. Mutka and P.~Mahonen.
\newblock {Approximation of light ray deflection angle and gravitational lenses
  in the Schwarzschild metric. 2. Lensing magnification in a binary system}.
\newblock {\em Astrophys. J.}, 581:1328--1336, 2002.

\bibitem{Frolov:2004cu}
Valeri~P. Frolov and Hyun~Kyu Lee.
\newblock {Observable form of pulses emitted from relativistic collapsing
  objects}.
\newblock {\em Phys. Rev. D}, 71:044002, 2005.

\bibitem{Amore:2006pi}
Paolo Amore and Santiago Arceo~Diaz.
\newblock {Analytical formulas for gravitational lensing}.
\newblock {\em Phys. Rev. D}, 73:083004, 2006.

\bibitem{Semerak:2014kra}
Oldrich Semer\'ak.
\newblock {Approximating light rays in the Schwarzschild field}.
\newblock {\em Astrophys. J.}, 800(1):77, 2015.

\bibitem{DeFalco:2016yox}
Vittorio De~Falco, Maurizio Falanga, and Luigi Stella.
\newblock {Approximate analytical calculations of photon geodesics in the
  Schwarzschild metric}.
\newblock {\em Astron. Astrophys.}, 595:A38, 2016.

\bibitem{LaPlaca:2019rjz}
Riccardo La~Placa, Pavel Bakala, Luigi Stella, and Maurizio Falanga.
\newblock {A New Approximation of Photon Geodesics in Schwarzschild Spacetime}.
\newblock {\em Res. Notes AAS}, 3(7):99, 2019.

\bibitem{Poutanen:2019tcd}
Juri Poutanen.
\newblock {Accurate analytic formula for light bending in Schwarzschild
  metric}.
\newblock {\em Astron. Astrophys.}, 640:A24, 2020.

\end{thebibliography}





\end{document}